\documentstyle[aps,preprint,floats,epsfig]{revtex}
\tightenlines

\makeatletter

\makeatletter

\tightenlines

\def\lsim{\mathrel{\rlap{\lower4pt\hbox{\hskip1pt$\sim$}}
    \raise1pt\hbox{$<$}}}         
\def\gsim{\mathrel{\rlap{\lower4pt\hbox{\hskip1pt$\sim$}}
    \raise1pt\hbox{$>$}}}         

\newcommand{\xslash}{\!\!\!\!/}

\newcommand{\lrope}{\nabla^{^{^{\!\!\!\!\!\!\!\!\!
    \Longleftrightarrow}}}}

\draft

\makeatother

\makeatother
\begin{document}

\title{Nuclear Anapole Moments}

\author{W. C. Haxton}

\address{Institute for Nuclear Theory, Box 351550, and Department of Physics,
\\
University of Washington, Seattle, WA 98195-1550, \\
and Department of Physics, University of California, Berkeley, CA 94720 \\
email: haxton@phys.washington.edu}

\author{C.-P. Liu}
\address{Institute for Nuclear Theory, Box 351550, and Department of Physics,
University of Washington, Seattle, WA 98195-1550, \\
email: cpliu@u.washington.edu}

\author{M. J. Ramsey-Musolf}

\address{
Kellogg Radiation Laboratory, California Institute of Technology,
Pasadena, CA 91125, \\
Department of Physics, University of Connecticut, Storrs, CN 
06269
\\
and Theory Group, Thomas Jefferson National Accelerator Facility, Newport
News, VA 23606
\\
email: mjrm@krl.caltech.edu
}

\date{\today{}}

\maketitle

\begin{abstract}
Nuclear anapole moments are parity-odd, time-reversal-even E1 moments of
the electromagnetic current operator. Although the existence of this
moment was recognized theoretically soon
after the discovery of parity nonconservation (PNC), its experimental isolation was
achieved only recently, when a new level of precision was reached in
a measurement of the hyperfine dependence of atomic PNC 
in \( ^{133}\)Cs.  An important anapole moment bound in \( ^{205}\)Tl
also exists. In this paper, we present the details of the first
calculation of these anapole moments in the framework commonly
used in other studies of hadronic PNC, a meson
exchange potential that includes long-range pion exchange and enough
degrees of freedom to describe the five independent $S-P$
amplitudes induced by short-range interactions.  The resulting
contributions of \( \pi - \),
\( \rho - \), and \( \omega - \)exchange to the single-nucleon 
anapole moment, to parity admixtures in the nuclear ground state,
and to PNC exchange currents are evaluated, using configuration-mixed
shell-model wave functions.  The experimental anapole moment constraints
on the PNC meson-nucleon coupling constants are derived and
compared with those from other tests of the hadronic weak interaction.
While the bounds obtained from the anapole moment results are
consistent with the broad ``reasonable ranges'' defined by theory,
they are not in good agreement with the constraints from the other
experiments.  We explore possible explanations for the discrepancy and
comment on the potential importance of new experiments.

\end{abstract}

\pagebreak

\section{Introduction}

The strangeness-conserving ($\Delta S = 0$) weak nucleon-nucleon interaction
is of considerable interest.  It provides the one experimentally 
accessible means of probing the neutral current component of the hadronic weak interaction, as
this component plays no role in flavor-changing reactions.
Furthermore the question of how long-range weak
forces between nucleons are connected to the underlying elementary
weak quark-boson couplings of the standard model is an important
strong-interaction question, one with potential connections to
poorly understood phenomena such as the $\Delta$ I = 1/2 rule.
One of the challenges in the field has been the experimental determination
of the various spin and isospin contributions to the low-energy weak
$NN$ interaction, as this interaction is dwarfed by much larger
strong and electromagnetic forces.  The weak effects can be isolated
only by precisely measuring tiny effects associated with the parity
nonconservation (PNC) accompanying this interaction.  Because the PNC effects are 
typically of relative size $\sim 10^{-7}$, only one class of 
elementary $NN$ scattering experiments, $\vec{p} + p$, 
has reached the requisite
sensitivity.  PNC effects have also been isolated in nuclear experiments,
but only a few nuclear systems are sufficiently well understood to
permit theorists to relate the observable to the underlying $NN$
interaction.  For these reasons there is interest in finding new
experimental constraints.

Shortly after Lee and Yang's proposal that weak interactions
violate parity, Vaks and
Zeldovich \cite{Ze58} noted independently that an elementary particle (as well
as composite systems like the nucleon or nucleus) could have
a new electromagnetic moment, the {}``anapole moment'',
corresponding to a PNC coupling to a virtual photon.  One contribition to
the anapole moments of hadrons would thus arise from 
PNC loop corrections to the electromagnetic vertex.
Despite some early work on the contribution of the
nucleon anapole moment to high-energy electron-nucleon scattering \cite{henley},
the interest in anapole moments might have been limited to theorists
had not Flambaum, Khriplovich, and collaborators \cite{FKS84} pointed out
their enhanced effects in atomic PNC experiments in heavy atoms.  As
the anapole moment is spin dependent, it contributes to the small
hyperfine dependence of atomic PNC.  (The dominant PNC effects
in such experiments arise from the coherent vector coupling of the
exchanged $Z^0$ to the nucleus, and are thus independent of nuclear
spin.)  While nuclear-spin-dependent effects do arise from Vector(electron)-
Axial(nucleus) $Z^0$ exchange, this nuclear coupling
does not grow systematically with the nucleon number $A$ of the nucleus:
naively, the axial coupling in an odd-$A$ nucleus is to the unpaired valence
nucleon.  Flambaum {\it et al.} \cite{FKS84} observed that the anapole
moment of a heavy nucleus grows as $A^{2/3}$, so that weak radiative 
corrections to spin-dependent atomic PNC associated with the anapole
moment would typically dominate over the corresonding tree-level $Z^0$
exchange for sufficiently large $A$ ($A \gsim$ 20).  This growth
means that spin-dependent atomic PNC effects should be dominated 
by the anapole moment -- a radiative ``correction'' -- and
measurable in heavy atoms.

Nevertheless, spin-dependent atomic PNC effects are still
exceedingly small, typically $\sim$ 1\% of the size of nuclear-spin-independent
atomic PNC effects.  Despite considerable effort, only limits existed on
the anapole contribution until very recently.   However, with the Colorado
group's measurement \cite{Colorado97} of atomic PNC in $^{133}$Cs at the level of 0.35\%,
a definitive (7$\sigma$) nuclear-spin-dependent effect emerged from
the hyperfine differences.  This measuement is the principal motivation
for the work presented here.  The goal of the present study is to carry
out an analysis of the $^{133}$Cs anapole moment that follows as closely as
possible the formalism developed and employed in other $NN$ and nuclear 
tests of the low-energy hadronic weak interaction \cite{Weak_Hadron}.  That formalism is based
on the finite-range PNC $NN$ potential of Desplanques, Donoghue, and
Holstein (DDH), a potential that contains sufficient freedom to describe the
long-range $\pi$ exchange and the short-range physics governing the five
independent PNC $S-P$ $NN$ amplitudes \cite{DDH}.  The resulting $\pi-$,
$\rho-$, and $\omega-$exchange PNC $NN$ potential is employed in estimating the
loop contributions to the single-nucleon anapole moment and the exchange
current and nuclear polarization contributions to the nuclear anapole moment
for $^{133}$Cs.  We also present results for Tl, where an interesting anapole
limit exists \cite{Seattle95,Oxford95}.  
  
The current work extends the treatment of Ref. \cite{HHM} by including
heavy-meson PNC contributions, thereby going beyond long-range
$\pi$ exchange to the full DDH potential.  This extension
is crucial in describing the isospin character of both the
single-nucleon and nuclear polarizability contributions to the
anapole moment.  The main results
of our study were recently presented in a letter \cite{hlrm}.  Here we give
the technical details of the heavy-meson current and polarizability calculations,
and discuss the associated shell-model calculations and their potential
shortcomings.
Our approach differs from most earlier
calculations \cite{FKS84,BoPi91,DKT94,DmTe97,WiBo98,AuBr99}
by avoiding one-body reductions of the currents and potentials:
exchange currents and polarizabilities are evaluated from shell-model
two-body densities matrices, modified by short-range correlation
functions that mimic the effects of missing high-momentum
components.  We also use a form for the anapole operator
in which components of the three-current constrained by current
conservation are rewritten in terms of a commutator with the
Hamiltonian, and thus explicitly removed.

The paper is organized as follows.  In Sec. II we define the 
anapole moment and the electron-nucleus interaction it induces,
and discuss connections with the generalized Siegert's theorm.
In Sec. III we describe the DDH PNC $NN$ interaction 
arising from $\pi-, \rho-$, and $\omega-$exchange and its connections
with the $S-P$ amplitudes.
The treatment of the one-body, exchange-current, and polarization
contributions to the anapole moment are given in Sec. IV.  The summation
over intermediate nuclear states in the polarizability is
performed by closure, after calibrating this approach in a series
of more complete shell-model calculations in lighter nuclei.
Other technical details -- particularly the rather complicated
heavy-meson exchange-current evaluations -- are presented in Appendices
A through D.  In Sec. V experimental values for the anapole
moments of $^{133}$Cs and $^{205}$Tl are deduced from the 
corresponding hyperfine PNC measurements.  Other tests of the 
low-energy PNC $NN$ interaction are discussed, and the constraints
they impose on various PNC meson-nucleon couplings described.
We address the issue of uncertainties in the shell-model nuclear
structure calculations and attempt to assess the effects
of missing correlations phenomenologically.
In the concluding Sec. VI we discuss the resulting discrepancies
and possible
future work that would help address some of the open questions.

\section{The Anapole Operator and Current Conservation}

In this section we describe the anapole moment in terms of a classical
current distribution \cite{Kh91,Le95}.  The corresponding operator for
a quantum mechanical current is obtained from a
multipole expansion that satisfies the generalized Siegert's
theorem.  We illustrate, in a simple one-body nuclear
model, the relationship between the anapole moment and the PNC $NN$ interaction
and the consequences of current conservation.

\subsection{Anapole Moments in Classical Electromagnetism\label{Subsec: Classical}}

Given classical charge and current distributions,
\( \rho (\vec{x}^{\, \, '}) \) and \( \vec{j}(\vec{x}^{\, \, '}) \),
the scalar and vector potentials, \( \Phi (\vec{x}) \) and \(
\vec{A}(\vec{x}) \),
are obtain from integrals over the Green's function.  After a Taylor
expansion around the source point \( \vec{x}^{\, \, '} \)
one obtains

\begin{eqnarray}
\Phi (\vec{x}) & = & \int d^{3}x^{\, \, '}\frac{\rho (\vec{x}^{\, \,
'})}{4\pi |\vec{x}-\vec{x}^{\, \, '}|}\nonumber \\
 & = & \int d^{3}x^{\, \, '}\rho (\vec{x}^{\, \, '})\{1-\vec{x}^{\, \,
'}\cdot \vec{\nabla }+\frac{1}{2}(\vec{x}^{\, \, '}\cdot \vec{\nabla
})^{2}+\cdots \}\frac{1}{4\pi |\vec{x}|},\\
\vec{A}(\vec{x}) & = & \int d^{3}x^{\, \, '}\frac{\vec{j}(\vec{x}^{\, \,
'})}{4\pi |\vec{x}-\vec{x}^{\, \, '}|}\nonumber \\
 & = & \int d^{3}x^{\, \, '}\vec{j}(\vec{x}^{\, \, '})\{1-\vec{x}^{\, \,
'}\cdot \vec{\nabla }+\frac{1}{2}(\vec{x}^{\, \, '}\cdot \vec{\nabla
})^{2}+\cdots \}\frac{1}{4\pi |\vec{x}|}.
\end{eqnarray}

In the scalar potential expansion, the first term inside the
curly bracket 
generates the total charge (monopole) moment; the second term, the electric
dipole moment; and the third term, a combination of the quadrupole
and monopole charge moments\cite{Ja75}.
For the vector potential, the first term
vanishes as there is no net current. After carefully taking the
constraints of current conservation and the boundedness of the current density
into account (which place six constraints on the bilinear products
$j(\vec{x}^{\, \, '})_i x^{\, \, '}_j$),
there remain three independent components in the second
term, corresponding to the magnetic dipole moment of a classical
current distribution.  Similarly,
the third term involves a symmetric product of two coordinates 
with the current, generating 18 independent trilinear combinations,
with 10 constraints.  The remaining eight independent components
comprise the static magnetic quadrupole moment and the $E1$ 
moment known as the ``anapole moment'' (AM).
 
One can extract the vector potential 
due to the AM explicitly,

\begin{equation}
\label{Eq: A'}
\vec{A}^{(anapole)}(\vec{x})=(-\vec{a}{\nabla^2 \over M^2}+{\vec{\nabla
} \over M}\vec{a}\cdot {\vec{\nabla} \over M})\frac{1}{4\pi |\vec{x}|},
\end{equation}
where
\begin{equation}
\label{Eq: abb}
\vec{a}=\frac{M^2}{6}\int d^{3}x^{\, \, '}\vec{x}^{\, \, '}\times
(\vec{x}^{\, \, '}\times \vec{j}(\vec{x}^{\, \, '})).
\end{equation}
(We multiply and divide by $M^2$ for consistency with the
definition of $\vec{a}$ we will later introduce via the Dirac equation.)
We can remove the second term in Eq. (\ref{Eq: A'})
by a gauge transformation, so that

\begin{equation}
\label{Eq: aaa}
\vec{A}^{(anapole)}(\vec{x})={\vec{a} \over M^2}\delta ^{(3)}(\vec{x}).
\end{equation}
Current conservation allows Eq. (\ref{Eq: abb})
to be rewritten as

\begin{equation}
\label{Eq: a}
\vec{a}=-\frac{M^2}{4}\int d^{3}x^{'}{x^{'}}^2\vec{j}(\vec{x}^{'}).
\end{equation}
(We use the Lorentz-Heaviside unit in which \( \alpha =e^{2}/4 \pi \hbar c=1/137
\).)
Eq. (\ref{Eq: a}) is often presented
as the definition of the 
AM \cite{FKS84,DKT94,DmTe97,WiBo98,AuBr99,Kh91,Le95,FlMu97}.
However it is important to note that this form is obtained only
after exploiting the constraints of current conservation.

It is apparent, for the ordinary electromagnetic current, that
the associated AM operator 
is odd under a parity transformation.  Therefore a
nonzero AM requires either the introduction of an axial vector
component into the current or a parity admixture in the ground
state (allowing the ordinary electromagnetic current to have a
nonvanishing expectation value).  This requirement of PNC
associates the AM with the weak interaction.

Another important property is the contact nature of the AM 
vector potential.  Thus an atomic electron interacts with the AM
of the nucleus only to the extent that its wave function 
penetrates the nucleus.

\begin{figure}
{\centering \resizebox*{0.6\columnwidth}{!}{\includegraphics{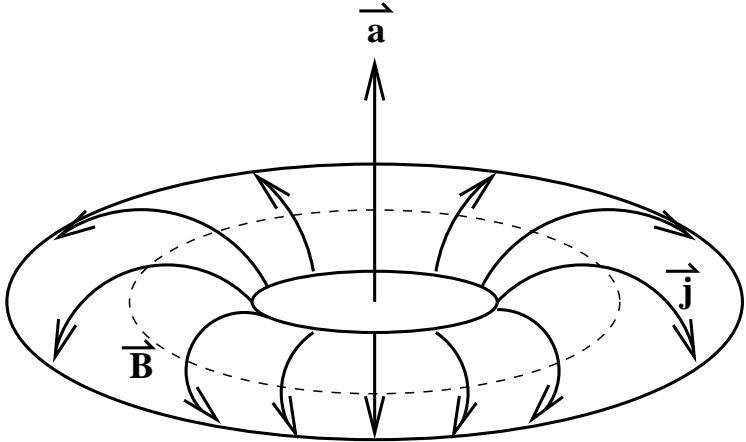}}
\par}

\caption{A toroidal current winding generates a nonzero anapole
moment.\label{Fig: Torus}}
\end{figure}

Figure \ref{Fig: Torus} gives a classical picture of the anapole
moment as a current winding.  Although the currents on the
inner and outer sides of the torus oppose one another, there is
a net contribution because of the $r^2$ weighting (in spherical
coordinates) of the current in the definition of the AM, 
leading to an AM that points upward.
The illustrated current distribution is odd under a parity 
reversal, as we have noted it must be for the ordinary
electromagnetic current.  If, however, the current has a 
chirality -- a small ``pitch'' corresponding to a left- or
right-handed winding that would signal PNC -- a parity-even
contribution to the operator would be induced.

\subsection{The Anapole Operator\label{Subsec: QM}}

Although one could quantize Eq. (\ref{Eq: a}) directly to generate the anapole
moment operator,
a better procedure is to avoid the assumption of current conservation,
as this is often violated in nuclear models.  Switching to a standard
spherical multipole decomposition yields the momentum-space charge and current 
operators \cite{FrWa66}

\begin{eqnarray}
\rho (\vec{q}) & = & \sum _{J,M}(-i)^{J}4\pi Y_{JM}^{*}(\Omega
_{q})M_{JM}^{coul}(q);\\
\vec{j}_\lambda(\vec{q}) & = & \sum _{J,M}(-i)^{J}\sqrt{2\pi (2J+1)}{\mathcal
D}_{M\lambda }^{(J)}(-\phi _{q},-\theta _{q},\phi
_{q})[T_{JM}^{el}(q)-\lambda T_{JM}^{mag}(q)],
\end{eqnarray}
and the associated charge, transverse electric, and transverse 
magnetic multipole projections of definite angular momentum and
(in the absence of PNC) parity

\begin{eqnarray}
M_{JM}^{coul}(q) & = & \int d^{3}x\, \, j_{J}(qx)Y_{JM}(\Omega _{x})\rho
(\vec{x});\\
T_{JM}^{el}(q) & = & \int d^{3}x\, \, \frac{1}{q}\vec{\nabla }\times
[j_{J}(qx)\vec{Y}^{M}_{JJ1}(\Omega _{x})]\cdot \vec{j}(\vec{x});\\
T_{JM}^{mag}(q) & = & \int d^{3}x\, \, j_{J}(qx)\vec{Y}^{M}_{JJ1}(\Omega
_{x})\cdot \vec{j}(\vec{x}),
\end{eqnarray}
where \( \vec{q} \) is the (outgoing) three-momentum transfer, \( j_{J} \)
the spherical Bessel function, \( Y_{JM} \) and \( \vec{Y}^{M}_{JJ1} \)
the ordinary and vector spherical harmonics, and \( {\mathcal
D}_{M\lambda }^{(J)}(-\phi _{q},-\theta _{q},\phi _{q}) \)
the rotation matrix.

The transformation properties of the possible multipole moments under
parity (P) and time-reversal (T) are listed in Table
\ref{Tab: Multipole PT}. Systems that are parity and time-reversal invariant
can have only even-rank Coulomb moments (charge, charge
quadrupole, etc.) and odd-rank transverse magnetic moments
(magnetic dipole, magnetic octupole, etc.).  The P- and T-odd
moments, which would arise in the standard model from small 
CP-violating contributions to the weak interaction,
correspond to the odd-rank Coulomb and even-rank transverse
magnetic multipoles (electric dipole, magnetic quadrupole, etc.).
The PNC but T-even moments, which would arise from the usual
weak interaction, correspond to the odd-rank transverse
electric multipoles, with the lowest of these being the dipole 
moment known as the anapole moment.

\begin{table}

\caption{Properties of multipole moments under parity and time reversal.
A slash (no slash) denotes odd (even) behavior.}
  
\label{Tab: Multipole PT}

\begin{center}

\begin{tabular}{cccc}
&
\( M^{coul} \)&
\( T^{el} \)&
\( T^{mag} \)\\
\hline
\( J=0 \)&
\( PT \)&
&
\\
\( J=1 \)&
\( P\xslash T\xslash  \)&
\( P\xslash T \)&
\( PT \)\\
\( J=2 \)&
\( PT \)&
\( PT\xslash  \)&
\( P\xslash T\xslash  \)\\
\( J=3 \)&
\( P\xslash T\xslash  \)&
\( P\xslash T \)&
\( PT \)\\
\( \vdots  \)&
\( \vdots  \)&
\( \vdots  \)&
\( \vdots  \)\\
\end{tabular}
\end{center}
\end{table}

For consistency with Eq. (\ref{Eq: aaa}),
we require

\begin{equation}
\nabla ^{2}\vec{A}(\vec{x})=-\vec{j}(\vec{x}),
\end{equation}
which then defines the anapole operator

\begin{equation}
\label{Eq: a_T^el}
a_{1 \lambda }= \lim_{\vec{q}^2 \rightarrow 0} \frac{-i\sqrt{6\pi } M^2}{_{\vec{q}^{2}}}T_{1\lambda }^{el}.
\end{equation}

The simplest case is the general expression for the matrix
element of a conserved
four-current for a free spin-${1 \over 2}$ particle
\begin{eqnarray}
\bar{U}(p') J^\mu(q) U(p) &=&
\bar{U}(p') (F_1(q^2) \gamma^\mu - i {F_2(q^2) \over 2 M} \sigma^{\mu \nu} q_\nu \nonumber \\
&& + {a(q^2) \over M^2} (\not q q^\mu - q^2 \gamma^\mu) \gamma_5
-i {d(q^2) \over M} \sigma^{\mu \nu} q_\nu \gamma_5 ) U(p).
\end{eqnarray}
from which the four moments of Table \ref{Tab: Multipole PT}
can be immediately identified.
The two vector terms define the charge $F_1(q^2)$ and 
magnetic $F_2(q^2)$ form factors.  The axial terms that
follow are the anapole and electric dipole terms, respectively.
The anapole term reduces
in the nonrelativistic limit to
\begin{eqnarray}
{a(q^2) \over M^2} (\not q q^\mu - q^2 \gamma^\mu) \gamma_5
&\rightarrow& {a(q^2) \over M^2} \vec{q}^{~2} (\vec{\sigma} - \hat{q} \hat{q}
\cdot \vec{\sigma}) \nonumber \\
&=& {a(q^2) \over M^2} \vec{q}^{~2} \vec{\sigma}_\perp,
\end{eqnarray}
showing that the current is transverse and spin-dependent.
From this current we then have the AM operator for
a nonrelativistic point particle
\begin{equation}
\hat{a}_{1 \lambda} = a(0) \sigma_{1 \lambda}.
\end{equation}

\subsection{Current Conservation and the Extended
Siegert's
Theorem\label{Subsec. j Conservation}}

The anapole operator $a_{1 \lambda}$ has been defined in terms
of $T^{el}_{1 \lambda}$, and it is well known that this operator
can be transformed into other forms by exploiting the continuity equation.
These forms are equivalent in calculations where 
consistent charge and current operators can be constructed
and exact matrix elements evaluated.
However we are interested in nuclear
calculations where, when one goes beyond the simplest descriptions
to models that treat the interactions among the nucleons, 
current conservation is not preserved.  We lack a
prescription for constructing the many-body currents 
consistently that are necessary for current conservation, and
for addressing the renormalizations that account for the limited
Hilbert spaces employed in nuclear models.  In such cases
there is a preferred form for $T^{el}_{1 \lambda}$, the form
in which all components of the three-current constrained by 
current conservation are reexpressed in terms of the charge
operator.
  
A familiar example is the case of $E1$ transitions generated by
the ordinary electromagnetic current.
Then $T^{el}_{1 \lambda}$ generates a one-body operator proportional
to $\vec{p}/M$, which is of order $v/c$, where $v$ is the
nucleon velocity.  It can be shown that the exchange current
contribution to $T^{el}_{1 \lambda}$ is also of this order.
As the exchange currents, in general, cannot be constructed faithfully,
it follows that errors will arise that are necessarily of 
leading order in the velocity.

Siegert \cite{Si37} showed that the situation could be greatly improved 
by exploiting the continuity equation
\begin{equation}
\vec{\nabla} \cdot \vec{j}(\vec{x}) = -i [H, \rho(\vec{x})]
\end{equation}
to write $T^{el}_J$, in the long-wavelength limit, entirely
in terms of the charge operator.  This generates the familiar
dipole form of the transverse electric operator, proportional to $\omega
\vec{r}$, where $\omega$ is the energy transfer.  The
importance of this rewriting is that the charge operator, which 
is of order $(v/c)^0$, has exchange current corrections only
of order $(v/c)^2$, or of relative size $\sim$ 1\%.  Thus
the Siegert's form of the $E1$ operator is a far more 
controlled operator in nuclear calculations.

A form of $T^{el}_{1 \lambda}$ consistent with Siegert's theorem
is in common use \cite{FrHa85}

\begin{eqnarray}
T_{JM}^{el'}(q) & \doteq  &
\frac{E_{i}-E_{f}}{q}(\frac{J+1}{J})^{1/2}M_{JM}^{coul}(q)\nonumber \\
 &  & -i (\frac{2J+1}{J})^{1/2}\int d^{3}x\, \,
j_{J+1}(qx)\vec{Y}^{M}_{JJ+11}(\Omega _{x})\cdot
\vec{j}(\vec{x}),\label{Eq: T^el'} 
\end{eqnarray}
where \( \doteq  \) means the equality holds after taking matrix
elements \( \langle f|\hat{O}|i\rangle  \).
This form has the correct leading-order behavior for transitions
due to the first term, with the second term vanishing as
$q \rightarrow 0$.  But for a static moment, the first term
vanishes; the leading-order behavior is then governed by the
second term, which the naive Siegert's theorem does not
properly constrain.

However the extension of Siegert's theorem to arbitrary $q$ 
was derived by Friar and Fallieros \cite{FrFa84,FrHa85}: at
every order in $q$ those components of the current constrained
by current conservation are identified and rewritten in terms of
the charge operator.  The result is

\begin{eqnarray}
T_{JM}^{el''}(q) & \doteq  &
\frac{E_{i}-E_{f}}{q}(\frac{J+1}{J})^{1/2}\int d^{3}x\, \,
\frac{(qx)^{J}}{(2J+1)!!}g_{J}(qx)Y_{JM}(\Omega _{x})\rho
(\vec{x})\nonumber \\
 &  & -\frac{q}{J+2}\int d^{3}x\, \,
\frac{(qx)^{J}}{(2J+1)!!}h_{J}(qx)\vec{Y}^{M}_{JJ+11}(\Omega _{x})\cdot
(\vec{x} \times \vec{j}(\vec{x})),\label{Eq: T^el''}
\end{eqnarray}
where \( g_{J} \) and \( h_{J} \) are polynomials in $q$ \cite{FrFa84}.
Combining Eq.
(\ref{Eq: T^el''}) and Eq. (\ref{Eq: a_T^el})
one finds \cite{HHM}

\begin{equation}
\label{Eq: a final}
a_{\lambda }=-\frac{M^2}{9}\int d^{3}x\, \, x^{2}[j_{\lambda
}(\vec{x})+\sqrt{2\pi }(Y_{2}(\Omega x)\otimes \vec{j}(\vec{x}))_{\lambda
}]
\end{equation}
This is the AM operator form used here and in our earlier work;
to our knowledge all other analyses have been based on the
naive form of Eq. (\ref{Eq: a}).

We stress that the three transverse electric operators
\( T^{el} \) ,\( T^{el'} \), and \( T^{el''} \)
are equivalent for simple one-body models
which ignore nucleon-nucleon interactions, 
provided the resulting one-body currents are properly generated
by minimal substitution.  The differences in these operators 
arise when they are used in more realistic calculations.

\subsection{Simple Examples\label{Subsec:1-body Calculation}}

In this section we illustrate how this equivalence is manifested
in non-interacting shell model calculations
of the nuclear AM of $^{133}$Cs.
The PNC interaction is also taken to be a one-body effective potential,
$H^{(1)}_{PNC}$.

The elements of the calculation include:

\begin{enumerate}
\item Extreme single-particle forms for the ground-state nuclear
wave function.  As $^{133}$Cs is an odd-even nucleus with $J$ = 
7/2$^+$, the odd proton is placed in the $1g_{7/2}$ shell, outside
an otherwise fully spin-paired closed core.
\item The strong Hamiltonian is a one-body harmonic oscillator
potential with spin-orbit interaction.  While this description is
primitive, it does yield the proper ground-state spin and parity
for the (nearly spherical) nucleus $^{133}$Cs.  The harmonic 
oscillator wave functions allow analytic calculations of 
polarizabilities, etc.
\item \( H^{(1)}_{PNC} \) is treated perturbatively: only linear terms are retained.
\end{enumerate}
Thus the resulting Hamiltonian is

\begin{equation}
H=H_{0}+H^{(1)}_{PNC}\, ,
\end{equation}
with

\begin{eqnarray}
H_{0} & = & \sum
^{A}_{i=1}\frac{\vec{p}(i)^{2}}{2M}+\frac{1}{2}M\omega
^{2}\vec{x}(i)^{2}-f\vec{s}(i)\cdot \vec{l}(i);\\
H^{(1)}_{PNC} & = & \sum ^{A}_{i=1}\frac{g_{S}+g_{V}\tau
_{3}(i)}{2M}\vec{\sigma }(i)\cdot \vec{p}(i),
\end{eqnarray}
where \( \omega  \) is related to the harmonic oscillator size parameter \( b \) by
\( \omega =1/Mb^{2} \), the spin-orbit strength \( f \) can
be determined from shell splittings near the $3s2d1g$ shell,
and \( g_{S} \) and \( g_{V} \), the
isoscalar and isovector strengths in the one-body PNC potential,
can be chosen to represent the average potential exerted by 
the core nucleons.
The analytic expressions we obtain illustrate the functional 
dependence on all of these parameters.
Thus we are not concerned here with specific numerical values.

By minimal substitution

\begin{equation}
H\rightarrow H+e\Phi \quad ;\quad \vec{p}\rightarrow \vec{p}-e\vec{A},
\end{equation}
one can derive the
charge and current densities to order \( 1/M \),
\begin{equation}
\rho (\vec{x})=e\sum ^{A}_{i=1}\frac{1+\tau _{3}(i)}{2}\delta
^{(3)}(\vec{x}-\vec{x}_{i}),
\end{equation}
and

\begin{mathletters}

\begin{eqnarray}
\vec{j}_{conv}(\vec{x}) & = & e\sum ^{A}_{i=1}\frac{1+\tau
_{3}(i)}{2M}\{\vec{p}(i),\delta ^{(3)}(\vec{x}-\vec{x}_{i})\}_{sym}\,
;\\
\vec{j}_{mag}(\vec{x}) & = & e\sum ^{A}_{i=1}\frac{\mu _{S}+\mu _{V}\tau
_{3}(i)}{2M}\vec{\nabla }\times (\vec{\sigma }(i)\delta
^{(3)}(\vec{x}-\vec{x}_{i}));\\
\vec{j}_{s.o.}(\vec{x}) & = & e\sum ^{A}_{i=1}\frac{1+\tau
_{3}(i)}{2}\frac{f}{2}\vec{x}(i)\times \vec{\sigma }(i)\delta
^{(3)}(\vec{x}-\vec{x}_{i});\\
\vec{j}_{PNC}(\vec{x}) & = & e\sum ^{A}_{i=1}\frac{1+\tau
_{3}(i)}{2M}\frac{g_{S}+g_{V}}{2}\vec{\sigma }(i)\delta
^{(3)}(\vec{x}-\vec{x}_{i}),
\end{eqnarray}
\end{mathletters}where the subscripts $conv, mag, s.o.$, and $PNC$ denote
the current densities arising from convection (kinetic
energy), magnetization (intrinsic nucleon spin), the spin-orbit
interaction, and the PNC potential, respectively.  The first three
are vector currents while the last is axial vector.  Current
conservation is then easily verified

\begin{equation}
\vec{\nabla }\cdot
(\vec{j}_{conv}(\vec{x})+\vec{j}_{mag}(\vec{x})+\vec{j}_{s.o.}(\vec{x})
+\vec{j}_{PNC}(\vec{x}))=-i[H_{0}+H^{(1)}_{PNC}\,
,\, \rho (\vec{x})].
\end{equation}

Contributions to the AM are generated by the axial-vector current
acting between the unperturbed nuclear ground state and by
vector currents that contribute because $H^{(1)}_{PNC}$ perturbs
the ground state,

\begin{eqnarray}
\langle \psi |T^{el}_{1}|\psi \rangle  & = & \langle \psi
_{0}|T^{el(A)}_{1}|\psi _{0}\rangle \nonumber \\
 &  & +\sum _{\chi _{0}}(\frac{\langle \psi _{0}|T^{el(V)}_{1}|\chi
_{0}\rangle \langle \chi _{0}|H^{(1)}_{PNC}|\psi _{0}\rangle }{E_{\psi
_{0}}-E_{\chi _{0}}}+H.c.),\label{Eq: Matrix Element} 
\end{eqnarray}
where \( \psi _{0} \) and \( \chi _{0} \)s are single particle unperturbed
eigenfunctions of definite (and opposite) parities, and the superscripts (A) and (V) 
label the components of $T^{el}_1$ generated by the axial-vector 
and vector currents, respectively.

The special case of no spin-orbit interaction is interesting
because the first-order perturbed wave function (in fact,
the result can be generalized to all orders) is given by the
Michel transformation \cite{Mi64}

\begin{eqnarray}
\psi _{0}(\vec{x})\rightarrow \psi (\vec{x}) & = & (1-ig\frac{\vec{\sigma
}}{2}\cdot \vec{x})\psi _{0}(\vec{x})\nonumber \\
 & \cong  & e^{-i\vec{s}\cdot (gx)\hat{x}}\psi
_{0}(\vec{x}),\label{Eq: Spin Helix} 
\end{eqnarray}
where \( g=g_{S}\pm g_{V} \) for a proton(+) or neutron(-). 
Eq. (\ref{Eq: Spin Helix}) shows that $H^{(1)}_{PNC}$ generates
a spin rotation
along the radial direction characterized by a small angle 
proportional to $g$ and to the distance to the center of the
nucleus.  
Consider an \( S_{1/2} \) state aligned along the \( +z \)
axis.  The spin probability around a ring, centered at the origin,
would be uniform and in the $+z$ direction: we visualize this as 
a uniform array of up-spinors.
When the weak interaction is turned on, the Michel
rotation will produce a spin helix \cite{Kh91} structure for this
chain of spinors as shown in Fig. \ref{Fig. Spin Hex}. If we picture
each spin as a small current loop, the combination of all horizontal
spin components \( S_{//} \) can be viewed as a toroidal current
winding producing an AM, as discussed in
Sec. \ref{Subsec: Classical}.

\begin{figure}
{\centering
\resizebox*{0.6\columnwidth}{!}{\includegraphics{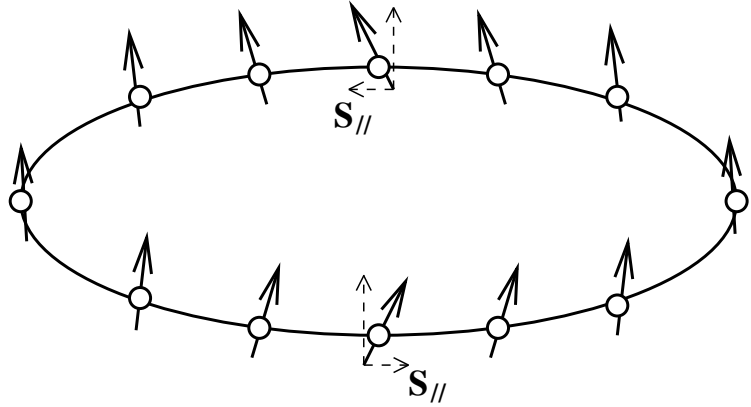}} \par}

\caption{Spin helix structure due to the parity mixing.\label{Fig. Spin
Hex}}
\end{figure}

Moreover, if the Michel transformed wave function is used in a 
calculation of the AM, one
finds that the contributions from \( \vec{j}_{PNC} \) and
\( \vec{j}_{conv} \) cancel exactly, so that \( \vec{j}_{mag} \)
is entirely responsible for the AM. Even with the inclusion of the spin-orbit interaction,
the magnetization current remains
the major contribution to the AM \cite{FKS84}.

The sum over intermediate states in Eq. (\ref{Eq: Matrix
Element})
simplies considerably in the harmonic oscillator since the 
momentum operator $\vec{p}$ only generates transitions of one
$\hbar \omega$.  Thus the transitions that must be consider in
the extreme single-particle limit are the simple $1p$ and $2p1h$ 
single-shell transitions of
Fig. \ref{Fig: Va & Co
Excitation} \cite{AuBr99}.

\begin{figure}
{\centering
\resizebox*{0.8\columnwidth}{!}{\includegraphics{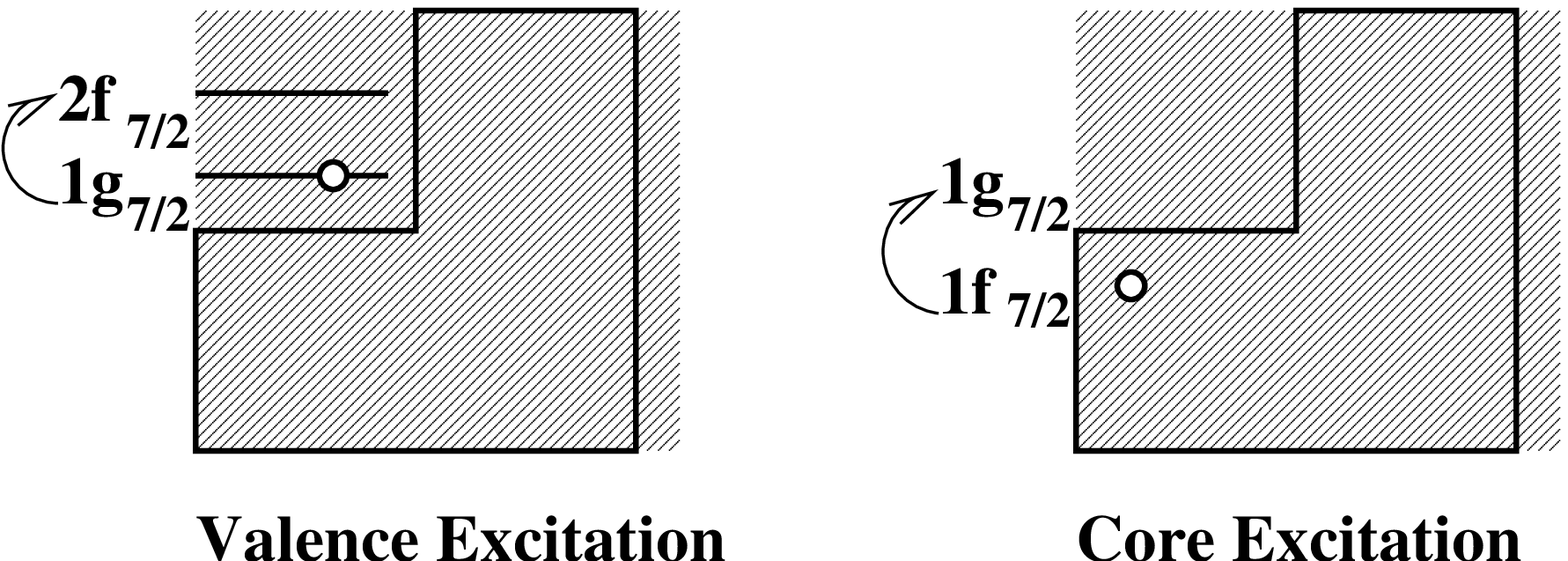}} \par}

\caption{Valence and core excitations produced by the PNC potential
acting on an extreme single-particle harmonic oscillation ground 
state.  The levels illustrated are appropriate for
\protect\( ^{133}\protect \)Cs.\label{Fig: Va &
Co Excitation}}
\end{figure}

Some preliminary algebraic manipulations are helpful.  Using the
commutation relation
\begin{equation}
[\frac{\vec{p}^{2}}{2M}+\frac{1}{2}M\omega
^{2}\vec{x}^{2},\vec{x}]=-i\frac{\vec{p}}{M},
\end{equation}
the PNC one-body potential can be rewritten as

\begin{eqnarray}
H^{(1)}_{PNC} & = & \frac{g}{2M}\vec{\sigma }\cdot \vec{p}\nonumber \\
 & = & i\frac{g}{2M}([H_{0},\vec{\sigma }\cdot
\vec{x}]+\frac{f}{2}[\vec{\sigma }\cdot \vec{l},\vec{\sigma }\cdot
\vec{x}]).
\end{eqnarray}
Using this result in the polarization sum yields

\begin{eqnarray}
a^{(pol)} & = & i\frac{g M^2}{2}\langle i|[\vec{\sigma }\cdot
\vec{x},T_{1}^{el(V)}]|i\rangle \nonumber \\
 &  & +\frac{g M^2}{2}f\sum _{n}\langle i|\vec{\sigma }\cdot (\vec{x}\times
\vec{l})|n\rangle \langle n|T^{el(V)}_{1}|i\rangle /(E_{i}-E_{n})+H.c.
\end{eqnarray}
As a typical value for the nuclear spin-orbit strength
is $f/\omega \equiv \alpha \sim 0.1$, one can work to first order in $f$,
yielding for the various AM contributions
\begin{eqnarray}
a^{(conv)} & \cong  & \frac{g M}{8}\langle
i|x^{2}\vec{\sigma }|i\rangle \nonumber \\
 &  & -\frac{g M}{4}f\sum _{n}\langle i|\vec{\sigma }\cdot
(\vec{x}\times \vec{l})|n\rangle \langle
n|\frac{1}{2}(x^{2}\vec{p}-i\vec{x})|i\rangle /(E_{i}-E_{n})+H.c.\, ,\\
a^{(mag)} & \cong  & \frac{g M}{4} \mu\Big \{\langle
i|(\vec{\sigma }\cdot \vec{x})\vec{x}-x^{2}\vec{\sigma }|i\rangle
\nonumber \\
 &  & -f\sum _{n}\langle i|\vec{\sigma }\cdot (\vec{x}\times
\vec{l})|n\rangle \langle n|\frac{1}{2}\vec{\sigma }\times
\vec{x}|i\rangle /(E_{i}-E_{n})+H.c.\Big \}\, ,\\
a^{(s.o.)} & \cong  & -\frac{g M^2}{8} f\langle i|x^{2}((\vec{\sigma
}\cdot \vec{x})\vec{x}-x^{2}\vec{\sigma })|i\rangle \, ,\\
a^{(PNC)} & \cong  & -\frac{g M}{8}\langle
i|x^{2}\vec{\sigma }|i\rangle \, .
\end{eqnarray}
As the \( O(f^{0}) \) terms from \( a^{(conv)} \) and \( a^{(PNC)} \)
exactly cancel, \( a^{(mag)} \) determines the leading-order (LO) contribution

\begin{equation}
a_{LO}=\frac{g M}{4} \mu\langle i|(\vec{\sigma }\cdot
\vec{x})\vec{x}-x^{2}\vec{\sigma }|i\rangle .
\end{equation}
For the next-to-leading-order (NLO) \( O(f^{1}) \) contributions, we
approximate \( E_{i}-E_{n}\simeq -\hbar \omega  \) 
(as the spin-orbit correction to this are higher order) and invoke closure

\begin{eqnarray}
a^{(conv)}_{NLO} & \simeq  & \frac{g M}{4} \alpha\langle
i|\{\vec{\sigma }\cdot (\vec{x}\times
\vec{l}),\frac{1}{2}(x^{2}\vec{p}-\vec{x})\}|i\rangle ,\\
a^{(mag)}_{NLO} & \simeq  & \frac{g M}{4} \alpha \mu\langle
i|\{\vec{\sigma }\cdot (\vec{x}\times \vec{l}),\frac{1}{2}(\vec{\sigma
}\times \vec{x})\}|i\rangle .
\end{eqnarray}
Therefore, assuming these matrix elements are of the same order of
magnitude, one obtains for the relative sizes

\begin{eqnarray}
|a^{(conv)}_{NLO}/a_{LO}| & \sim  & \left| \frac{\alpha}{\mu } \right| ,\\
|a^{(mag)}_{NLO}/a_{LO}| & \sim  & \left| \alpha \right|,\\
|a^{(s.o.)}/a_{LO}| & \sim  & \left| \frac{\alpha}{2\mu} \right|\frac{\langle x^{2}\rangle
}{b^{2}}\sim A^{1/3} \left| \frac{\alpha}{2\mu } \right|\sim \left| \alpha \right|,
\end{eqnarray}
where in the last line we assume an odd-proton nucleus with
$A \sim 100$, similar to Cs.

In Table \ref{Tab: 1-body H_PNC Result} we present our results for
the AM of \( ^{133} \)Cs in this single-particle scheme using the four different
\( T_{1}^{el} \)s discussed previously (\( T_{1}^{el} \),\( T_{1}^{el\, \, '} \),\(
T_{1}^{el\, \, ''} \),
and the one from the Cartesian decomposition using Eq. (\ref{Eq: a_T^el})
and Eq. (\ref{Eq: a})) to define our anapole operator.
Agreement is achieved only
when: i) all the currents--$conv, mag, s.o.$, and $PNC$--and ii) a complete set
of excitations--valence and core--are considered. This illustrates
a point made earlier, that the use of incomplete current operators
or Hilbert spaces breaking current conservation will in general
lead to difficulties.

\begin{table}

\caption{Matrix elements corresponding to four choices of the anapole
operator.
Note the dimensionless parameter 
\protect\( \alpha\equiv f/\omega =fMb^{2}\protect \),
where \protect\( b \protect \) is the oscillator parameter.
The results are in units of \protect\(
C\equiv 4\pi \sqrt{\frac{2}{7}}(g_{S}+g_{V}) M b^{2}\protect \),
another dimensionless quantity.}
  
\label{Tab: 1-body H_PNC Result}

\begin{center}

\begin{tabular}{ccccc}
\( \langle ||a||\rangle/e \)&
 \( T_{1}^{el} \)&
 \( T_{1}^{el'} \)&
 \( T_{1}^{el''} \)&
 \( Cartesian \)\\
\hline
\( conv \)&
 \( \frac{-1.1}{1-4\alpha}+\frac{-5.4}{1+4\alpha} \)&
 \( \frac{-2.8}{1-4\alpha}+\frac{-2.7}{1+4\alpha} \)&
 \( \frac{-4/3}{1-4\alpha}+\frac{-6}{1+4\alpha} \)&
 \( \frac{0.5}{1-4\alpha}+\frac{-10.125}{1+4\alpha} \)\\
\( mag \)&
 \( \frac{11.16}{1-4\alpha}+\frac{50.22}{1+4\alpha} \)&
 \( \frac{11.16}{1-4\alpha}+\frac{50.22}{1+4\alpha} \)&
 \( \frac{11.16}{1-4\alpha}+\frac{50.22}{1+4\alpha} \)&
 \( \frac{11.16}{1-4\alpha}+\frac{50.22}{1+4\alpha} \)\\
\( s.o. \)&
 \( \frac{-15.6\alpha}{1-4\alpha}+\frac{-30.6\alpha}{1+4\alpha} \)&
 \( \frac{-8.8\alpha}{1-4\alpha}+\frac{-19.9\alpha}{1+4\alpha} \)&
 \( \frac{-44\alpha/3}{1-4\alpha}+\frac{-33\alpha}{1+4\alpha} \)&
 \( \frac{-22\alpha}{1-4\alpha}+\frac{-49.5\alpha}{1+4\alpha} \)\\
\( PNC \)&
 \( 13/2 \)&
 \( 11/2 \)&
 \( 22/3 \)&
 \( 77/8 \)\\
\( Tot. \)&
 \( \frac{11.16-20\alpha}{1-4\alpha}+\frac{50.22-9\alpha}{1+4\alpha} \)&
 same&
 same&
 same\\
\end{tabular}
\end{center}
\end{table}
The table also shows that the contribution of the magnetization current, which is
separately conserved (\( \vec{\nabla }\cdot \vec{j}^{(mag)}=0) \),
is independent of the choice of the anapole operator.  
This term is entirely responsible for the leading $O(\alpha^0)$
result (given the $conv$--$PNC$ cancellation in this order).  It
is also apparent that the NLO contribution attributed to a given
current depends on the anapole operator chosen: it is the sum
over all contributions, not individual contributions, that is
kept constant in calculations satisfying current conservation.
The numerical value of the LO contribution
(61.4) is reduced by $\sim$ 20\% to 50.5 when the NLO contributions
are included ($\alpha$ = 0.1), consistent with our earlier
assertion that these corrections are perturbations.

\section{The PNC Nucleon-Nucleon Potential}
  
The AM calculations presented here are the first to
employ an $NN$ weak potential sufficiently general to describe
long-range $\pi$ exchange and all five short-range $S-P$ $NN$ amplitudes.
This section summarizes the isospin structure of the 
$\Delta S$ = 0 hadronic weak interaction and its description in terms of
$\pi-$, $\rho-$, and $\omega-$exchange.

\subsection{The Isospin Structure of the Hadronic Weak Interaction
\label{Subsec: H_PNC Quark}}
  
The standard model specifies the weak charged and neutral
currents, $J_W$ and $J_Z$, associated with the absorption/emission
of weak bosons by quarks \cite{Weak_Quark}.  The couplings to the light quarks
(u,d,s) are
\begin{eqnarray}
J^{\mu }_{W} & = & \cos \theta _{C}\bar{u}\gamma ^{\mu }(1-\gamma
_{5})d+\sin \theta _{C}\bar{u}\gamma ^{\mu }(1-\gamma _{5})s\, ;\\
J^{\mu }_{Z} & = & \frac{1}{\sqrt{2}\cos \theta _{W}}\{\bar{u}\gamma ^{\mu
}(1-\frac{8}{3}\sin ^{2}\theta _{W}+\gamma _{5})u-\bar{d}\gamma ^{\mu
}(1-\frac{4}{3}\sin ^{2}\theta _{W}+\gamma _{5})d\nonumber \\
 &  & -\bar{s}\gamma ^{\mu }(1-\frac{4}{3}\sin ^{2}\theta _{W}+\gamma
_{5})s\},
\end{eqnarray}
where \( \theta _{C} \) is the Cabbibo angle, with \( \sin \theta _{C}\sim
.22 \) and \( \theta _{W} \) is the Weinberg angle, with \( \sin ^{2}\theta _{W}\sim
0.23 \).
The effective quark-quark weak interaction at low
energies can be described by a phenomenological current-current
Lagrangian
\begin{equation}
{\mathcal L}_{Weak}=\frac{G_{F}}{\sqrt{2}}(J^{\dag }_{W}J_{W}+J_{W}J_{W}^{\dag}+J^{\dag
}_{Z}J_{Z}) ,
\end{equation}

By assigning proper isospin and strangeness quantum number to each
quark field, we can decompose these hadronic currents 

\begin{eqnarray}
J_{W} & = & \cos \theta _{C}J^{(1,0)}_{W}+\sin \theta
_{C}J^{(\frac{1}{2},1)}_{W};\\
J_{Z} & = & J^{(1,0)}_{Z}+J^{(0,0)}_{Z},
\end{eqnarray}
where the first superscript denotes the change in isospin (\( \Delta I
\)),
and the second in strangeness (\( \Delta S \)). 
The current $J^{(1,0)}_W$ drives the $u \rightarrow d$ transition,
while $J^{(\frac{1}{2},1)}_W$ drives the $u \rightarrow s$ 
transition.
We can construct
the strangeness conserving (\( \Delta S \)=0) hadronic weak interaction
Lagrangian density

\begin{eqnarray}
{\mathcal L}^{(\Delta S=0)}_{Weak} & = & \frac{G_{F}}{\sqrt{2}} \bigg ( (\cos
^{2}\theta _{C}J^{(1,0)\dag }_{W}J^{(1,0)}_{W}+\sin ^{2}\theta
_{C}J^{(\frac{1}{2},1)\dag }_{W}J^{(\frac{1}{2},1)}_{W} + H.c.) \nonumber \\
 & + & J^{(1,0)\dag }_{Z}J^{(1,0)}_{Z}+J^{(1,0)\dag
}_{Z}J^{(0,0)}_{Z}+J^{(0,0)\dag }_{Z}J^{(1,0)}_{Z}+J^{(0,0)\dag
}_{Z}J^{(0,0)}_{Z} \bigg ).
\end{eqnarray}

An important aspect of this Lagrangian density is its isospin content.
The symmetric product of two \( J^{(1,0)}_{W} \)
currents forms \( \Delta I=0 \) (isoscalar) and \( \Delta I=2
\)
(isotensor) interactions, while the symmetric product of two \(
J^{(\frac{1}{2},0)}_{W} \)
currents forms a \( \Delta I=1 \) (isovector) interaction. Therefore
the charged current weak $NN$ interaction in the \( \Delta I=1
\)
channel is suppressed by \( \tan ^{2}\theta _{C} \) relative to
the \( \Delta I=0 \) or \( 2 \) contributions. As there is no
isovector suppression for the neutral current, one 
concludes that the $\Delta I = 1$ $NN$ channel provides 
experimentalists their best opportunity for studying the neutral
current component of the hadronic weak interaction.

The physical states are strongly interacting composites, nucleons
and mesons.  The strong interaction dresses the underlying 
quark-boson couplings, and we have not yet developed the 
theoretical tools needed to evaluate the strong effects
quantitatively.  The physical couplings associated with the
effective operators for nucleons and mesons are thus expected
to differ -- perhaps substantially -- from the underlying bare
couplings.  One famous example of this is the $\Delta I = 1/2$
rule in strangeness-changing weak decays: in experiments one finds a strong
enhancement of $\Delta I =1/2$ over $\Delta I =3/2$ amplitudes,
relative to expectations based on the underlying standard-model
couplings and efforts to evaluate strong renormalizations.
One reason for the interest in PNC is the hope that we can
learn more about such strong effects by adding precise
data on $\Delta S = 0$ weak hadronic interactions.

\subsection{Meson Exchange and the Long-range PNC $NN$ Potential
\label{Subsec: H_PNC OPEP}}

The most straightforward contribution to the PNC nuclear potential
is from the direct exchange of \( W^{\pm } \)and \( Z^{0} \) between
nucleons.  Because of the small Compton wavelengths of these bosons
($\sim$ 0.002 fm) direct exchanges effectively occur only when two
nucleons overlap.  We do not yet have an adequate understanding
of such short-range contributions to either the PNC or 
parity-conserving (PC) $NN$ interactions.  Fortunately, for energies
characteristic of bound nucleons, the $NN$ interaction takes
place primarily at distances large compared to the nucleon size.
This is due in part to the strong repulsion in the $NN$ interaction
at short distances, and in part because nuclei are moderately dilute
Fermi systems.  Thus we expect long-range contributions,
which can be described without explicit reference to the 
structure of the nucleon, to dominate the PNC interaction at
low energies.

The strong PC $NN$ interaction at low energies
($\lsim$ 400 MeV) has been quite successfully modeled in terms
of meson-exchange potentials.  The complicated short-distance quark and gluon
dynamics governing this interaction are parameterized by various
meson-nucleon couplings and phenomenological form factors
constrained by experiment.
This meson-exchange strong interaction model
can be enlarged to include the weak PNC $NN$
interaction by replacing one of the strong meson-nucleon couplings
by a weak coupling.  All of the physics of $W$ and $Z$ exchange
between quarks -- and the attendant strong interaction dressing --
is buried inside the weak meson-nucleon vertices.
As in the case of the strong $NN$ interaction, the weak vertices
depend on momentum-independent meson-nucleon couplings and
phenomenological form factors.  For this model to make sense,
one should, at a minimum, be able to derive a consistent and 
reliable set of meson-nucleon couplings from PNC observables.
Should such a set emerge, the longer term goal would be to
develop a first-principles understanding of the relationship
between the effective hadronic couplings and the underlying 
standard model bare couplings, dressed by a complicated soup
of strong quark-quark interactions.

In developing a sensible meson-exchange model for the PNC $NN$
force, one must first truncate the tower of possible dynamical
mesons, effectively ``integrating out'' those which do not
contribute explicitly to the interaction.
At small center-of-mass energies light mesons
dominate the PNC potential because they have longer ranges.
Candidates below the chiral symmetry breaking scale of $\sim$ 1 GeV
include the pseudoscalar mesons \( \pi
\)(140MeV),
\( \eta  \)(549MeV), and \( \eta ^{'} \)(958MeV); the scalar mesons
S(975 MeV) and \( \delta  \)(983MeV); and the vector mesons \( \rho
\)(769MeV),
\( \omega  \)(783MeV) and \( \phi  \)(1020MeV).  One could
also consider crossed two-pion exchanges, etc.  Barton's
theorem \cite{Ba61}, which states that CP invariance forbids
any coupling between neutral $J$=0 mesons and on-shell nucleons,
helps to restrict the possibilities, eliminating exchanges
of $\pi^0$, $\eta$, $\eta^{'}$, $S$, and $\delta^0$ (to the extent
that CP violation can be ignored).  
Furthermore McKellar and Pick have
argued that \( \delta ^{\pm } \) exchange can be regarded as a form
factor correction to \( \pi ^{\pm } \) exchange \cite{McPi72}, and
\( \phi  \) is strongly suppressed relative to \( \rho  \) and \( \omega
\).  This motivates a PNC potential based on $\pi^{\pm}$,
\( \rho ^{0} \), \( \rho ^{\pm } \), and \( \omega ^{0} \) exchanges.
(We will present below another argument that will make this 
potential seem less arbitrary.)

The PC and PNC meson-nucleon interaction Lagrangian density in the
\( \pi - \), \( \rho - \), and \( \omega - \)exchange model is

\begin{eqnarray}
{\mathcal L}_{PC} & = & ig_{\pi NN}\overline{N}^{\, \, '}\gamma
_{5}\vec{\tau }\cdot \vec{\pi }N-g_{\rho NN}\overline{N}^{\, \, '}(\gamma
_{\mu }-i\frac{\mu _{V}}{2M}\sigma _{\mu \nu }q^{\nu })\vec{\tau
}\cdot \vec{\rho }^{\mu }N\nonumber \\
 &  & -g_{\omega NN}\overline{N}^{\, \, '}(\gamma _{\mu }-i\frac{\mu
_{S}}{2M}\sigma _{\mu \nu }q^{\nu })\omega ^{\mu }N,\\
{\mathcal L}_{PNC} & = & -\frac{f_{\pi }}{\sqrt{2}}\overline{N}^{\, \,
'}(\vec{\tau }\times \vec{\pi })_{3}N+\overline{N}^{\, \, '}\Big
(h^{0}_{\rho }\vec{\tau }\cdot \vec{\rho }^{\mu }+h_{\rho }^{1}\rho ^{\mu
}_{3}+\frac{h_{\rho }^{2}}{2\sqrt{6}}(3\tau _{3}\rho ^{\mu }_{3}-\vec{\tau
}\cdot \vec{\rho }^{\mu })\Big )\gamma _{\mu }\gamma _{5}N\nonumber \\
 &  & +\overline{N}^{\, \, '}(h^{0}_{\omega }\omega ^{\mu }+h^{1}_{\omega
}\tau _{3}\omega ^{\mu })\gamma _{\mu }\gamma _{5}N,
\end{eqnarray}
where \( g_{\pi NN} \), \( g_{\rho NN} \), and \( g_{\omega NN} \)
are the strong \( \pi - \), \( \rho - \), and \( \omega - \)nucleon
coupling constants and \( f_{\pi } \), \( h^{(0,1,2)}_{\rho } \),
and \( h^{(0,1)}_{\omega } \) (the superscripts denote the rank of
isospin) are the weak \( \pi - \), \( \rho - \), and \( \omega - \)nucleon
coupling constants.  (In the literature $f_\pi$ is also frequently
called $h_\pi$ or $h_\pi^1$.)  Note that the $\gamma_5$ convention is that
of Bjorken and Drell, and that $q$ is the outgoing momentum of the
produced meson.  (Both of these conventions are opposite in sign to
those of \cite{DDH}.)  Evaluating the one-boson exchange diagrams,
where one of the vertices is PC and the other PNC, and making a
non-relativistic
reduction, one obtains the PNC $NN$ potential

\begin{eqnarray}
H^{(2)}_{PNC}(\vec{r}) & = & \frac{iF_{\pi }}{M}[\vec{\tau }(1)\times
\vec{\tau }(2)]_{3}[\vec{\sigma }(1)+\vec{\sigma }(2)]\cdot \vec{u}_{\pi
}(\vec{r})\nonumber \\
 &  & +\frac{1}{M}\Bigg (\bigg \{F_{0}\vec{\tau }(1)\cdot \vec{\tau
}(2)+\frac{F_{1}}{2}[\tau (1)_{3}+\tau
(2)_{3}]+\frac{F_{2}}{2\sqrt{6}}[3\tau_{3}(1)\tau_{3}(2)-\vec{\tau
}(1)\cdot \vec{\tau }(2)]\bigg \}\nonumber \\
 &  & \times \{(1+\mu _{V})i[\vec{\sigma }(1)\times \vec{\sigma }(2)]\cdot
\vec{u}_{\rho }(\vec{r})+[\vec{\sigma }(1)-\vec{\sigma }(2)]\cdot
\vec{v}_{\rho }(\vec{r})\}\nonumber \\
 &  & +\bigg \{G_{0}+\frac{G_{1}}{2}[\tau_{3}(1)+\tau_{3}(2)]\bigg
\}\times \{(1+\mu _{S})i[\vec{\sigma }(1)\times \vec{\sigma }(2)]\cdot
\vec{u}_{\omega }(\vec{r})\nonumber \\
 &  & +[\vec{\sigma }(1)-\vec{\sigma }(2)]\cdot \vec{v}_{\omega
}(\vec{r})\}+\frac{1}{2}[\tau_{3}(1)-\tau_{3}(2)][\vec{\sigma
}(1)+\vec{\sigma }(2)]\cdot [G_{1}\vec{v}_{\omega
}(\vec{r})-F_{1}\vec{v}_{\rho }(\vec{r})]\Bigg ), 
\end{eqnarray}
where \( \vec{r}=\vec{r}_{1}-\vec{r}_{2} \), \(
\vec{u}(\vec{r})=[\vec{p},e^{-mr}/4\pi r] \),
\( \vec{v}(\vec{r})=\{\vec{p},e^{-mr}/4\pi r\} \), and \(
\vec{p}=\vec{p}_{1}-\vec{p}_{2} \).
The various coefficients in this potential are products of
PC and PNC couplings: \( F_{\pi }=g_{\pi NN}f_{\pi }/\sqrt{32} \),
\( F_{0}=-g_{\rho NN}h^{0}_{\rho }/2 \), \( F_{1}=-g_{\rho NN}h^{1}_{\rho
}/2 \),
\( F_{2}=-g_{\rho NN}h^{2}_{\rho }/2 \), \( G_{0}=-g_{\omega
NN}h^{0}_{\omega }/2 \), and
\( G_{1}=-g_{\omega NN}h^{0}_{\omega }/2 \).
We use the strong couplings \( g_{\pi NN}=13.45 \), \( g_{\rho NN}=2.790
\),
and \( g_{\omega NN}=8.37 \).  Vector dominance fixes
the strong scalar and vector magnetic moments, \( \mu
_{S}=-0.12 \)
and \( \mu _{V}=3.70 \). 
Note that the $\pi$-exchange channel is I=1; numerically it 
dominates the isovector $NN$ weak interaction.  
This is the channel which tests the strength of the neutral current component
of the hadronic weak interaction.

\begin{table}

\caption{Weak meson-nucleon coupling {}``best values'' and {}``reasonable
ranges'' (in units of \protect\( 10^{-7}\protect \)) from the standard
model calculations of Desplanques, Donoghue, and Holstein.
For comparison, the last two columns give the corresponding
results of Dubovik and Zenkin (DZ) and
Feldman, Crawford, Dubach, and Holstein (FCDH).}

\label{Tab: DDH}

\begin{center}

\begin{tabular}{ccccc}
Coupling&
{}``Reasonable Range{}``&
{}``Best Value{}``&
DZ\cite{DZ86}&
FCDH\cite{FCDH91}\\
\hline
\( \textrm{f}_{\pi } \)&
0.0\( \leftrightarrow  \)11.4&
4.6&
1.1&
2.7\\
\( h^{0}_{\rho } \)&
-30.8\( \leftrightarrow  \)11.4&
-11.4&
-8.4&
-3.8\\
\( h^{1}_{\rho } \)&
-0.38\( \leftrightarrow  \)0.0&
-0.19&
0.4&
-0.4\\
\( h^{2}_{\rho } \)&
-11.0\( \leftrightarrow  \)-7.6&
-9.5&
-6.8&
-6.8\\
\( h^{0}_{\omega } \)&
-10.3\( \leftrightarrow  \)5.7&
-1.9&
-3.8&
-4.9\\
\( h^{1}_{\omega } \)&
-1.9\( \leftrightarrow  \)-0.8&
-1.1&
-2.3&
-2.3\\
\end{tabular}
\end{center}
\end{table}

While the field has seen considerable experimental progress in constraining
the PNC meson-nucleon couplings, the theoretical situation has hardly
advanced beyond the benchmark analysis of Desplanques, Donoghue, and
Holstein  (DDH) \cite{DDH}, carried out twenty years ago.  Using SU(6)$_W$
symmetry, current algebra, and the constituent quark model, DDH related
charged current components of $f_\pi$ and the $h_V^i$ to experimental PNC
amplitudes for $\Delta S=1$ nonleptonic hyperon decays. Portions of the
neutral current contributions were also related to hyperon decays,
while the remaining pieces -- unaccessible through symmetry techniques --
were computed using explicit quark model calculations.
Uncertainties associated with the latter imply considerable lattitude in
the theoretical predictions. The resulting ``best values'' and
``reasonable ranges" are given in Table \ref{Tab: DDH}. The case of $f_\pi$ is
particularly acute, as this coupling is nominally dominated by neutral
current interactions.

Subsequent to the DDH work, other approaches, such as soliton models
\cite{meissner} and QCD sum rules \cite{henley2}, have been applied to the weak
meson-nucleon couplings. None of these approaches, however, has yielded a
sharper theoretical picture. Part of the difficulty may lie in the
assumption of valence quark dominance for the hadronic weak interaction. In
particular, it has recently been shown, in the context of chiral
perturbation theory, that chiral corrections to the leading-order PNC $\pi NN$
interaction may be large \cite{zhu}.
These corrections, which have no analog in constituent quark models,
reflect the presence of ``disconnected'' light $\bar{q} q$ sea
contributions.
Given the present interest of hadron structure physicists in the sea quark
structure of light hadrons, the possibility of important sea quark
contributions makes $f_\pi$ a particularly interesting object of study.
Achieving agreement among all determinations of this coupling is, thus,
important.  As we observe below, the current interpretation of the
Cs and Tl AMs in terms of DDH couplings shows that such agreement
is not yet in hand.

In can be argued that an analysis in terms of meson-exchange
PNC couplings is in fact quite general, if limited to low-energy observables:
the DDH couplings are a shorthand for
another representation of the low energy PNC $NN$ interaction, 
one based on the five independent $S-P$ amplitudes.   
The DDH description in terms of $\pi$, $\rho$,
and $\omega$ exchange can be viewed as an effective theory,
valid at momentum scales much below the inverse range of the
vector mesons.  At low momentum the detailed short-range 
behavior of the potential is not resolvable: thus one could
characterize the vector-meson contribution to the weak $NN$ interaction by five strengths
describing the five $S-P$ amplitudes.  A sixth parameter 
would be needed to describe $\pi$ exchange, as this interaction
is long ranged.  The six DDH couplings thus are equivalent to
such a description of the weak potential.

In an ideal world one would determine the low-energy $NN$ $S-P$
amplitudes, or equivalently the six weak meson-nucleon couplings,
by a series of $NN$ scattering experiments.  Such experiments
require measurements of asymmetries $\sim 10^{-8}$, the
natural scale for the ratio of weak and strong amplitudes,
$4 \pi G_F m_\pi^2/g^2_{\pi N N}$.  As we will detail later, 
only a single $NN$ measurement, the longitudinal analyzing 
power for $A_L$ for $\vec{p} + p$, has produced a definitive
result.  This result has been supplemented by PNC measurements in
few-body nuclei and in some special nuclear systems where 
nuclear structure uncertainties can be largely circumvented,
allowing the experiments to be interpretted reliably.
An analysis of these results, which have been in hand for some
time, suggests that the isoscalar PNC interaction -- which is
dominated by $\rho$ and $\omega$ exchange -- is comparable to or
slightly larger than the DDH ``best value,'' while the isovector
interaction -- dominated by $\pi$ exchange -- is significantly
weaker \cite{Weak_Hadron}.
As the isovector channel is expected to be enhanced by neutral
currents, there is great interest in confirming this result.
One reason for the interest in the $^{133}$Cs AM
is the hope that spin-dependent atomic PNC measurements can
provide such a cross check.

\section{Contributions to Nuclear Anapole Moments}

The DDH meson exchange model -- which we have argued provides a 
very general description of the PNC $NN$ interaction at low energies --
has become the standard formalism for discussing low-energy
properties of the weak $NN$ interaction.  
We now extend this formalism to nuclear AMs, 
discussing the various PNC meson-exchange mechanisms
by which a virtual $E1$ photon can be absorbed by the nucleus.

\begin{figure}
{\centering \resizebox*{0.5\columnwidth}{!}{\includegraphics{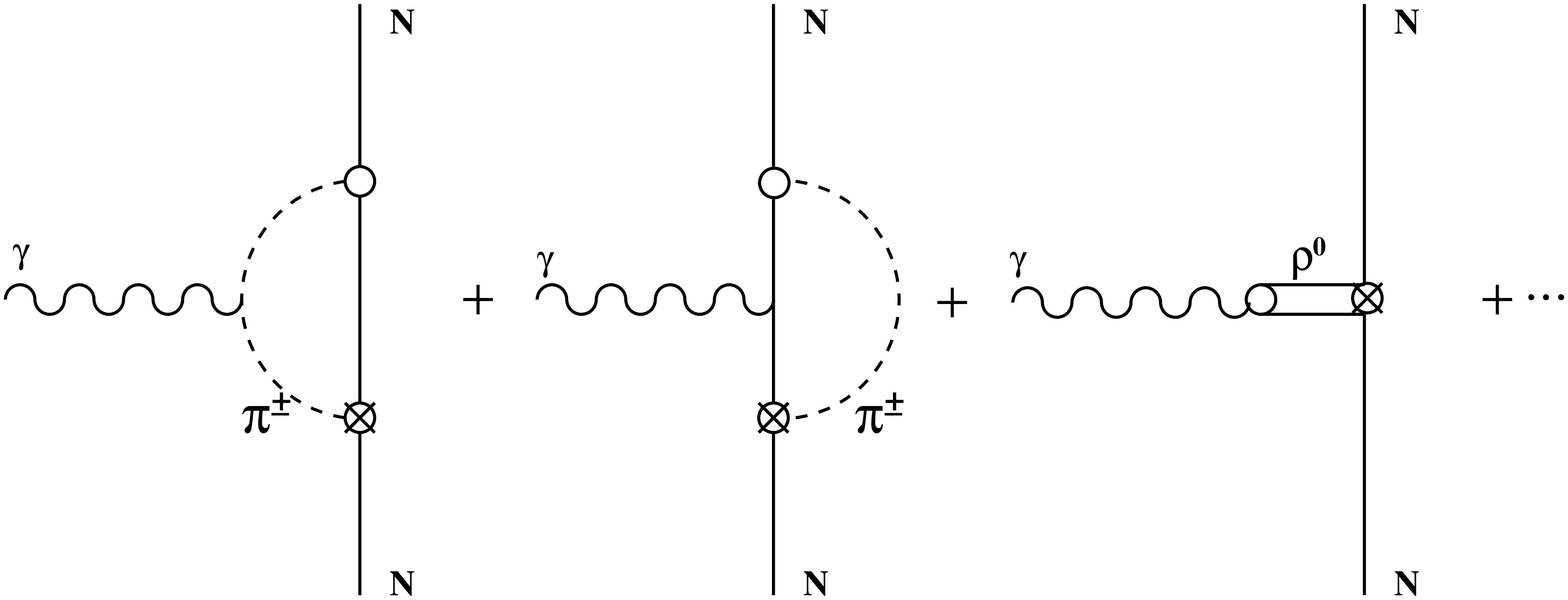}}
\par}

\caption{One-body axial-vector currents contributing to nucleonic
anapole moments are generated by pion loop diagrams
and by vector meson dominance diagrams.
\label{Fig: 1-body}}
\end{figure}

\begin{figure}
{\centering \resizebox*{0.5\columnwidth}{!}{\includegraphics{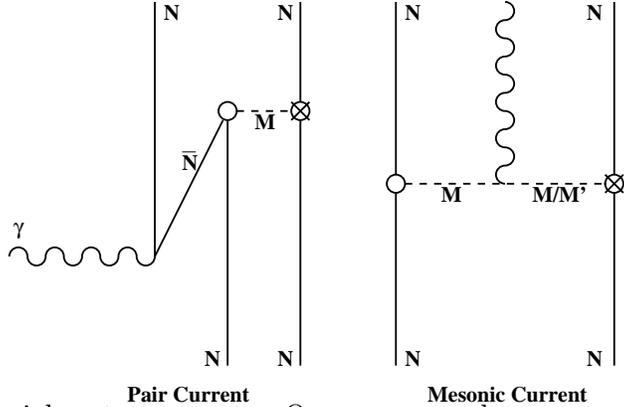}}
\par}

\caption{Two-body axial-vector currents. One meson-nucleon vertex
is strong, the second is weak.
\label{Fig: 2-body}}
\end{figure}

\begin{figure}
{\centering \resizebox*{0.3\columnwidth}{!}{\includegraphics{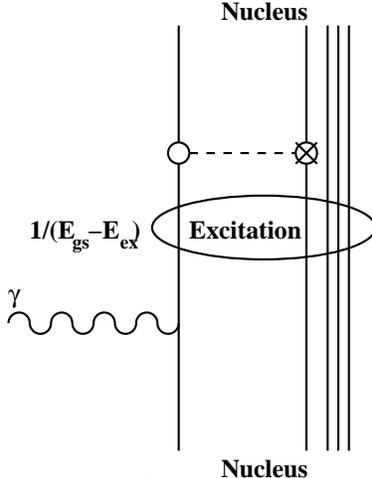}}
\par}

\caption{An opposite-parity polarization of the nuclear ground state
induced by the PNC weak $NN$ interaction.\label{Fig: Pol}}
\end{figure}

\begin{enumerate}
\item Fig. \ref{Fig: 1-body} illustrates a PNC pion cloud
dressing of a nucleon (one pion-nucleon coupling is PNC and 
one PC) and a vector-meson pole graph, leading to $E1$
photon absorption by a nucleon.
The axial currents corresponding to such pion-loop
and vector meson dominance diagrams generate
nucleonic AMs \cite{HHM,MuHo91,KaSa93,ZPHM00,MaKo00}, which we discuss
in more detail in Sec. \ref{Subsec: 1-body}.
\item The two-body \( H^{(2)}_{PNC} \) also generates two-body axial vector
exchange currents (See Fig. \ref{Fig: 2-body}).  The diagrams
we evaluate include: i) pair currents, where E1 photons couple
to the \( N\bar{N} \) pairs excited by the two-body potential, and
ii) transition currents, where E1 photons couple to the exchanged mesons.
Detailed calculations are described in Sec. \ref{Subsec: 2-body}.
\item The two-body \( H^{(2)}_{PNC} \) polarizes the
nucleus, producing an opposite-parity ground state component.  This
component then couples back to the unperturbed ground state via
the amplitude for absorbing a virtual $E1$ photon.  The resulting
polarizability requires one to sum over a complete set of 
opposite-parity intermediate states (Fig. \ref{Fig: Pol}).
This is discussed in Sec. \ref{Subsec: Pol}.
\end{enumerate}

The dependence of these contributions on nucleon number $A$ is
important.  As the one-body anapole contribution involves a 
coupling to spin, it is easy to see that the nucleonic contribution
acts very much like a nuclear magnetic moment: in a naive picture
of an odd-$A$ nucleus as an unpaired nucleon outside of a 
spin-paired core, the core contribution cancels, leaving only
the valence nucleon contribution.  While that contribution will
depend on the quantum labels of the valence orbital, there is
no general growth of the nucleonic contribution with $A$.
In contrast, it was the important observation that that polarization
contribution grows as $A^{2/3}$ \cite{FKS84} that led atomic
experimentalists to realize that AMs might be
measureable.  This growth not only leads to larger AMs
in heavy nuclei, but guarantees that the AM will
dominate over other sources of spin-dependent PNC, such as 
direct V(electron)-A(nucleus) $Z^0$ exchange (another nucleonic
coupling that effectively sees only the unpaired valence spin).
Similarly, it was shown in \cite{HHM} that the exchange current 
contribution also grows like $A^{2/3}$.  Note that the
polarization contribution could be additionally enhanced 
if the ground state is a member a fortuitous parity doublet.
There has been some discussion of anapole (and electric dipole)
moment enhancements because of such accidental near-degeneracies
\cite{edm}.

In Figs. \ref{Fig: 1-body} - \ref{Fig: Pol}
the AM is shown interacting with an external photon.
Yet the illustrated processes are not physical, as the anapole
coupling vanishes for on-shell photons.  The underlying physical
processes involve a scattering particle -- e.g., an atomic
electron, the source of the virtual photon.  It follows that the
AM need not be a gauge invariant quantity:
instead it is one of a larger class of weak radiative corrections --
corrections naively of \( O(G_{F}\alpha ) \) -- that together form
a gauge invariant physical amplitude.
Included in this larger set of radiative corrections would be
various ``box'' diagrams corresponding to simultaneous exchange
between the electron and nucleus of a photon and $Z^0$, etc.
However the long-distance contributions to the AM
of a nucleus -- the meson cloud contributions and many-body
contributions due to wave function polarization and exchange
currents discussed here -- are both dominant numerically and
separately gauge invariant \cite{MuHo91}.  This is one reason the set of 
anapole contributions associated with $H_{PNC}^{(2)}$ discussed
here are of such interest.

The calculations require wave functions for the nuclear ground
state and one- and two-body transition density matrices for
evaluating the effects of one- and two-body operators on the
ground state.  The wave functions were derived from shell model (SM) diagonalizations
with harmonic oscillator Slater determinants and
with suitable residual two-body interactions. For \( ^{133}\)Cs,
the oscillator parameter is \( b=2.27\)f and the canonical SM space
is between the magic shells of 50 and 82, i.e., 1\( g_{7/2} \)-2\( d_{5/2}
\)-1\( h_{11/2} \)-2\( d_{3/2} \)-3\( s_{1/2} \).
Calculations were performed with the five valence protons restricted to the
first two of these shells and four neutron holes to the last three.
This produced an m-scheme basis of about 200,000. Two interactions were employed,
the Baldridge-Vary potential \cite{BaVa} and a recent potential developed
by the Strasbourg group \cite{Stras}, both of which are based on
the addition of multipole terms to g-matrix interactions and are designed
for the \( ^{132}\)Sn region. As the results are very similar, here
we only quote results from the Baldridge-Vary calculation.  For \( ^{205}\)Tl, an oscillator parameter \( b=2.54\)f
was chosen.  The ground state was described as a proton hole in the orbits immediately below
the Z=82 closed shell, i.e., 3\( s_{1/2} \)-2\( d_{3/2} \)-2\( d_{5/2} \)
(though the \( 1h_{11/2} \) lies between two d shells, we omitted
this opposite-parity shell to keep the SM space manageable), and
the two neutron holes are in the space between magic shells of 126
and 82, i.e., 3\( p_{1/2} \)-2\( f_{5/2} \)-3\( p_{3/2} \)-1\( i_{13/2}
\)-2\( f_{7/2} \)-1\( h_{9/2} \).
A simple Serber-Yukawa force was used as the residual interaction.

\subsection{Nucleonic Anapole Moments\label{Subsec: 1-body}}

As illustrated in Fig. \ref{Fig: 1-body}, the one-body
PNC electromagnetic currents (parity even) can be derived from
pion loop diagrams, where one meson-nucleon vertex is weak and PNC
and the other strong and PC, and from vector meson dominance.
After plugging these one-body PNC currents into
Eq. (\ref{Eq: a final}), the one-body anapole operator takes the form

\begin{equation}
a_{\lambda }^{1-body}=\sum _{i=1}^{A}\left[ a_{s}(0)+a_{v}(0)\tau
_{3}(i)\right] \sigma _{\lambda }(i).
\end{equation}
This form makes it clear that the contributions of spin-paired
core nucleons cancel, leaving only the valence nucleon AM.  The results from
\cite{HHM}, where only the pion contribution was considered, are

\begin{eqnarray}
a_{s}(0) & \cong  & -0.193f_{\pi } e,\\
a_{v}(0) & \cong  & -0.048f_{\pi } e.
\end{eqnarray}
Thus the pion loops generate
an isoscalar coupling that is about four times larger than the isovector
one. Later this calculation was extended to include the \( \rho ^{0}
\)-pole
contribution by vector meson dominance \cite{MuHo91}. This work was
further extended to included the full set of one-loop contributions
involving the DDH vector meson PNC couplings \cite{ZPHM00}, using
the framework of heavy baryon chiral perturbation theory (HB\( \chi  \)PT)
and retaining contributions through \( O(1/\Lambda ^{2}_{\chi }) \),
where \( \Lambda _{\chi }=4\pi F_{\pi }\sim 1\)GeV is the chiral
symmetry breaking scale and \( F_{\pi }\cong 93\)MeV is the pion
decay constant.  This yielded the nucleonic AM couplings

\begin{eqnarray}
{a_{s}(0) \over e} & \cong  & -0.24f_{\pi }-0.37h^{1}_{\rho }-0.11h^{0}_{\omega
}+0.07h^{0}_{\phi }-1.43h^{1}_{A}+0.0051h_{n\Sigma ^{-}K}\nonumber \\
 &  & +0.047(h^{n\bar{\Sigma }K^{+}}_{V}+\frac{h^{p\Sigma
^{0}K^{+}}_{V}}{\sqrt{2}})-0.3(h^{pK}_{A}+h^{nK}_{A})+0.009h_{p\Lambda
K}-0.125h^{p\Lambda K^{+}}_{V},\\
{a_{v}(0) \over e} & \cong  & -0.37(h^{0}_{\rho }+\frac{h^{2}_{\rho
}}{\sqrt{6}})-0.12h^{1}_{\omega }+0.07h^{1}_{\phi
}-1.43h^{1}_{A}+0.9(h^{0}_{V}+\frac{3}{4}h^{2}_{V})+0.0051h_{n\Sigma
^{-}K}\nonumber \\
 &  & +0.047(-h^{n\bar{\Sigma }K^{+}}_{V}+\frac{h^{p\Sigma
^{0}K^{+}}_{V}}{\sqrt{2}})-0.3(h^{pK}_{A}-h^{nK}_{A})+0.009h_{p\Lambda
K}-0.125h^{p\Lambda K^{+}}_{V}.
\end{eqnarray}
The HB$\chi$PT result for the pionic contribution is consistent with
the earlier pion loop estimates: the isoscalar coupling is 1.3 times
the pion loop value, while the isoscalar coupling is zero to this
order in $\chi$PT.  However, the vector mesons
greatly enhance the isovector AM.  An evaluation using DDH best
values shows that \( a_{v}(0)\sim 7a_{s}(0) \). That is, the inclusion
of the vector mesons enhances the AM and qualitatively changes
its isospin character, with the proton and neutron AMs opposite in sign.
The HB$\chi$PT calculation included
non-Yukawa type \( \pi NN \) couplings (defined as \( h^{i}_{v} \)s
and \( h^{i}_{A} \)s in \cite{ZPHM00}) associated with derivative
interactions.  Here we include only the standard DDH contributions,
omitting the rest.  Using ``best values'' for the neglected terms
\cite{ZPHM00}, this omission is estimated to generate a 3\%
error in the dominant isovector coupling and 100\% in $a_s(0)$.
The reason for the omission is consistency: such derivative couplings
are absent in the DDH PNC $NN$ potential, the parameters of which 
are constrained by experiment.  A consistent treatment of the 
derivative coupling would require not only their propagation through
the polarization and exchange current calculations for the AM,
but also redoing the DDH potential fits to all other low-energy
$NN$ and nuclear PNC observables.  We leave this ambitious task
to future work.

Folding these expressions with our SM matrix elements (\( \langle
I||\sum _{i=1}^{A}\sigma (i)||I\rangle  \)
= -2.372 and 2.532, \( \langle I||\sum _{i=1}^{A}\sigma (i)\tau
(i)||I\rangle  \)
= -2.305 and 2.282, for Cs and Tl, respectively) yields the results
in Table \ref{Tab: 2-body SM}.

\subsection{Exchange Currents\label{Subsec: 2-body}}
  
The virtual $E1$ photon can also be absorbed on a pair of nucleons
coupled by the PNC potential.
Such PNC exchange currents are evaluated in the standard way.  The
transition matrix is derived and reduced nonrelativistically,
retaining terms through $1/M$.  This resulting momentum-space
current is then Fourier transformed to produce a coordinate-space
two nucleon current,

\begin{eqnarray}
j^{\mu }(\vec{x},\vec{x}_{1},\vec{x}_{2}) & = & \int \frac{d\vec{k}}{(2\pi
)^{3}}e^{i\vec{k}\cdot \vec{x}}\int \frac{d(\vec{p}_{1}^{\, \,
'}-\vec{p}_{1})}{(2\pi )^{3}}e^{i(\vec{p}_{1}^{\, \, '}-\vec{p}_{1})\cdot
\vec{x}_{1}}\int \frac{d(\vec{p}_{2}^{\, \, '}-\vec{p}_{2})}{(2\pi
)^{3}}e^{i(\vec{p}_{2}^{\, \, '}-\vec{p}_{2})\cdot \vec{x}_{2}}\nonumber
\\
 &  & \times j^{\mu }(\vec{k},\vec{p}_{1}^{\, \,
'}-\vec{p}_{1},\vec{p}_{2}^{\, \, '}-\vec{p}_{2}),
\end{eqnarray}
where \( \vec{x} \) is the field point, \( \vec{x}_{1} \) and \(
\vec{x}_{2} \)
the source points. 

In Appendix A we give the two-body charge and 
current operators in momentum space.  In Appendix B we give the
nonvanishing three-current coordinate-space operators to $O(1/M)$,
the forms needed for the AM calculation.  The $\pi$ contribution,
which turns out to dominate numerically, is
\begin{eqnarray}
\vec{j}^{(\pi )}(\vec{x},\vec{x}_{1},\vec{x}_{2}) & = & \vec{j}_{\gamma
-PC}^{(\pi \, \, pair)}(\vec{x},\vec{x}_{1},\vec{x}_{2})+\vec{j}_{\gamma
-PNC}^{(\pi \, \, pair)}(\vec{x},\vec{x}_{1},\vec{x}_{2})+\vec{j}^{(\pi \pi
\gamma )}(\vec{x},\vec{x}_{1},\vec{x}_{2})\nonumber \\
 & = & \frac{-eg_{\pi NN}f_{\pi }}{8\sqrt{2}\pi M}(\vec{\tau }(1)\cdot
\vec{\tau }(2)-\tau (1)_{3}\tau (2)_{3})\bigg \{\vec{\sigma }(1)\delta
^{(3)}(\vec{x}-\vec{x}_{1})+\vec{\sigma }(2)\delta
^{(3)}(\vec{x}-\vec{x}_{2})\nonumber \\
 &  & -\frac{1}{2}(\vec{\sigma }(1)\cdot \vec{\nabla }_{1}-\vec{\sigma
}(2)\cdot \vec{\nabla }_{2})\Big [(\delta
^{(3)}(\vec{x}-\vec{x}_{1})+\delta
^{(3)}(\vec{x}-\vec{x}_{2}))(\vec{x}_{1}-\vec{x}_{2})\nonumber \\
 &  & +\frac{1}{2}\frac{|\vec{x}_{1}-\vec{x}_{2}|}{m_{\pi }}\vec{\nabla
}(\delta ^{(3)}(\vec{x}-\vec{x}_{1})+\delta
^{(3)}(\vec{x}-\vec{x}_{2}))\Big ]\bigg \}\frac{e^{-m_{\pi
}|\vec{x}_{1}-\vec{x}_{2}|}}{|\vec{x}_{1}-\vec{x}_{2}|}.
\end{eqnarray}

Even with the complete exchange currents in hand, evaluating their
shell model matrix elements is a formidable task.  (The one previous AM 
exchange current calculation treated only $\pi$ exchange \cite{HHM}.)
For example,
the form of \( \vec{j}^{\rho \rho \gamma } \) is far more involved
than any of the pionic contributions.  The procedure we follow is
to first identify which currents are numerically significant by
averaging the currents over the nuclear core.  Once 
identified, full two-body evaluations are then performed
for these cases.

The one-body average, first performed for PNC potentials by 
Michel \cite{Mi64}, involves direct and exchange terms

\begin{equation}
\langle \alpha |O^{(1)}|\beta \rangle \equiv \sum _{\gamma }\langle \alpha
\gamma |O^{(2)}|\beta \gamma \rangle -\langle \alpha \gamma
|O^{(2)}|\gamma \beta \rangle ,
\end{equation}
where the sum extends over all single-particle core states.
The averages are done in a Fermi gas, a simple choice because
spin, isospin, and spatial averages can be performed independently.
The nucleus is viewed as a single particle outside a spin-paired
(but isospin asymmetric) Fermi sea.  The one-body average operators
are obtained in closed form, though the average done over
the spatial functions produces, in general, a complicated but smooth
function of the single-particle initial and final momenta (the $Y$ and $W$ functions below).
The smoothness allows us to replace this function with an average
value, with little loss of accuracy.  Appendix C contains an 
example of this averaging procedure, while the full results for
the various currents are listed in Appendix D.  In the case of $\pi$
exchange the result is

\begin{eqnarray}
\vec{j}^{(\pi )}(\vec{x},\vec{x}_{i}) & = & \vec{j}_{\gamma -PC}^{(\pi
-pair)}(\vec{x},\vec{x}_{i})+\vec{j}_{\gamma -PNC}^{(\pi
-pair)}(\vec{x},\vec{x}_{i})+\vec{j}^{(\pi \pi \gamma
)}(\vec{x},\vec{x}_{i})\nonumber \\
 & = & \frac{e g_{\pi NN}f_{\pi }}{2\sqrt{2}Mm^{2}_{\pi }}((\theta
_{n}+\theta _{p})+(\theta _{n}-\theta _{p})\tau _{3})\times \rho \bigg
\{\langle W^{'(\pi )}\rangle \vec{\sigma }\delta
^{(3)}(\vec{x}-\vec{x}_{i})\nonumber \\
 &  & -\frac{2}{m^{2}_{\pi }}\Big [\langle Y_{3}\rangle
p^{2}_{F}\vec{\sigma }\delta ^{(3)}(\vec{x}-\vec{x}_{i})-\langle
Y_{1}\rangle [\vec{\sigma }\cdot \vec{\nabla },[\vec{\nabla },\delta
^{(3)}(\vec{x}-\vec{x}_{i})]]\nonumber \\
 &  & -\frac{1}{4}\langle Y_{2}\rangle \{\vec{\sigma }\cdot \vec{\nabla
}_{i},\{\vec{\nabla }_{i},\delta ^{(3)}(\vec{x}-\vec{x}_{i})\}\}\Big ]
\end{eqnarray}

The one-body estimate of the exchange current contributions to the
AM can be obtained by plugging the averaged currents into
Eq. (\ref{Eq: a final}).
The Fermi-gas-averaged AM results are tabulated in the FGA columns
of Table \ref{Tab: FG Vs SM}.
The results are given as a fraction of the $\pi$ pair current
contribution, as this is the dominant term.
These results are compared to full two-body SM results, similarly normalized
to the SM $\pi$ pair current AM value.  The absolute $\pi$ pair current
results are also given for both calculations.
  
We see from the table that, while the Fermi gas average tends
to overestimate the AM contribution by a factor of $\sim$ 2-3,
compared to the SM, the Fermi gas and SM agree very well
on the relative values of the various contributions. 
(The comparison is less impressive for Tl
than for Cs, but the Fermi gas parameters used for both nuclei
were tailored to Cs.)  This suggests that the one-body average
AM values should be reliable indicators of which exchange 
current contributions are important.

\begin{table}

\caption{A comparison of anapole moment estimates from a one-body
Fermi gas average with full two-body shell model results.  DDH
best-value couplings are used, and no short-range correlation function corrections
are included in either set of results.  The labels PC and
PNC denote whether the nucleon absorbing the photon has a PC
or PNC meson-nucleon coupling.}

\label{Tab: FG Vs SM}

\begin{center}

\begin{tabular}{ccccccccccc}
&
\( \langle ||a||\rangle /e\times 10^{7} \)&
\( \pi \, \, pair \)&
\( \rho \, \, pair_{PC} \)&
\( \rho \, \, pair_{PNC} \)&
\( \omega \, \, pair_{PC} \)&
\( \omega \, \, pair_{PNC} \)&
\( \pi \pi \gamma  \)&
\( \rho \rho \gamma  \)&
\( \rho \pi \gamma _{PC} \)\\
\hline
\( ^{133}\)Cs&
FGA&
110&
13.0\%&
-19.0\%&
-0.4\%&
8.1\%&
-34.9\%&
6.6\%&
0.5\%\\
&
SM&
67&
12.9\%&
-18.2\%&
&
8.6\%&
-24.0\%&
&
\\
\( ^{205}\)Tl&
FGA&
-75&
12.8\%&
-18.2\%&
-0.3\%&
7.8\%&
-35.5\%&
7.8\%&
0.0\%\\
&
SM&
-27&
15.4\%&
-21.5\%&
&
12.8\%&
-29.4\%&
&
\\
\end{tabular}
\end{center}
\end{table}

The Fermi gas model is an independent particle model.  The SM, while
incorporating certain correlations, omits the high-momentum
components of the Hilbert space necessary for describing the 
short-range hard core.  While the SM (and associated Fermi
gas) shortcomings could in principle be corrected by introducing
effective operators and wave function renormalizations, in practice
this is never done.  Instead, most frequently the omitted short-range
physics is mocked up by a correlation function 
which, in SM PNC studies, is often taken from Miller and Spencer \cite{MiSp76},
\begin{equation}
f(r_{12})=1-(1-br^{2}_{12})e^{-ar^{2}_{12}},
\end{equation}
with \( a=1.1\, fm^{-2} \) and \( b=0.68\, fm^{-2} \).
This correlation function reduces two-body matrix elements by $\sim$
25-30\% for $\pi$ currents, 75-80\% for $\rho$ and $\omega$
currents, $\sim$ 80\% for $\pi \pi$ currents, and $\sim$ 90-95\%
for $\rho \rho$ and $\rho \pi$.
 
No short-range correlation corrections have been included in the results of
Table \ref{Tab: FG Vs SM}.  It is thus apparent that the true
\( \vec{j}_{\gamma-PC}^{\omega \, \, pair} \)
, \( \vec{j}^{\rho \rho \gamma } \) (the most complicated current), and
\( j^{\rho \pi \gamma } \) 
exchange current contributions (with short-range correlations
included) would be $\sim$ 1\% of the dominant
$\pi$ pair result.
It is then reasonable to ignore these unimportant but complicated
exchange currents, evaluating all others with the full two-body
SM density matrix, modified by the Miller-Spencer correlation
function.  While a complete list of the two-body AM operators
is too long to list here, the dominant $\pi$ operator is 
found to be

\begin{eqnarray}
\vec{a}^{(\pi )} & = & \frac{eg_{\pi NN}f_{\pi } M}{72\sqrt{2}\pi
}(\vec{\tau }(1)\cdot \vec{\tau }(2)-\tau_{3}(1)\tau
_{3}(2))\nonumber \\
 &  & \times \bigg [x_{1}^{2}\vec{\sigma }(1)+x_{2}^{2}\vec{\sigma
}(2)+\sqrt{2\pi }\{x_{1}^{2}[Y_{2}(\Omega _{1})\otimes \vec{\sigma
}(1)]_{1}+x_{2}^{2}[Y_{2}(\Omega _{2})\otimes \vec{\sigma
}(2)]_{1}\}\nonumber \\
 &  & -\frac{1}{2}(\vec{\sigma }(1)\cdot \vec{\nabla }_{1}-\vec{\sigma
}(2)\cdot \vec{\nabla }_{2})\nonumber \\
 &  & \times \Big ((x^{2}_{1}+x^{2}_{2})\vec{x}+\sqrt{2\pi
}\{x_{1}^{2}[Y_{2}(\Omega _{1})\otimes \vec{x}]_{1}+x_{2}^{2}[Y_{2}(\Omega
_{2})\otimes \vec{x}]_{1}\}+\frac{3}{2}\frac{x}{m_{\pi }}\vec{x}\Big
)\bigg ]\frac{e^{-m_{\pi }x}}{x},
\end{eqnarray}
where \( \vec{x}=\vec{x}_{1}-\vec{x}_{2} \).
The numerical results for the sum of all exchange current contributions
to the Cs and Tl AMs is given in Table \ref{Tab: 2-body SM}.

\subsection{Nuclear Polarization Contributions\label{Subsec: Pol}}

As illustrated in Fig. \ref{Fig: Pol}, the two-body PNC $NN$ potential 
perturbs the ground state, mixing it with excited states of opposite parity.
The resulting odd-parity ground state component allows the 
ordinary (vector) $E1$ current to couple to the ground state.
The first-order perturbation theory AM is thus

\begin{equation}
\sum _{n}\frac{\langle I|a^{V}|n\rangle \langle n|H^{(2)}_{PNC}|I\rangle
}{E_{g.s.}-E_{n}}+\frac{\langle I|H^{(2)}_{PNC}|n\rangle \langle
n|a^{V}|I\rangle }{E_{g.s.}-E_{n}},
\end{equation}
where
\( |I\rangle \)
is the unperturbed ground state of good parity and the sum extends over a complete
set of nuclear states $n$ of angular momentum $I$ and opposite parity.
The operator \( a^{V} \) is obtained by plugging the ordinary
electromagnetic current into Eq. \ref{Eq: a final}, 
\begin{eqnarray}
\vec{a}_1^V &=& -{M e \over 6 \sqrt{2}} \sum^A_{i=1} \left \{ {1 \over \sqrt{2}}
\vec{r}_1(i) \tau_3(i)+[\vec{r}(i) \otimes \vec{l}(i)]_1[1+\tau_3(i)]
\right. \nonumber \\
&+& \left. {3 \over 2} [\vec{r}(i) \otimes \vec{\sigma}(i)]_1 
[\mu_s + \mu_v\tau_3(i)] \right \} 
\end{eqnarray}
where $\mu_s=0.88$ and $\mu_v=4.706$.

The summation over a complete set of intermediate SM states
for \( ^{133}\)Cs or \( ^{205}\)Tl is impractical either directly
or by the summation-of-moments method discussed in Ref. \cite{HHM}
and below. However, because no nonzero $E1$ transition exists among
the valence orbits (e.g., the $h_{11/2}$ and $g_{7/2}$ orbitals
have opposite parity but cannot be connected by a dipole operator),
an alternative
of completing the sum by closure,
after replacing \( 1/\Delta E_{n} \)
by an average value \( \langle 1/\Delta E\rangle  \) is quite attractive 

\begin{eqnarray}
-\sum _{n}\frac{\langle I|a^{V}|n\rangle \langle n|H^{(2)}_{PNC}|I\rangle
+\langle I|H^{(2)}_{PNC}|n\rangle \langle n|a^{V}|I\rangle }{\Delta E_{n}}
&  & \nonumber \\
\rightarrow -\langle \frac{1}{\Delta E}\rangle \sum _{n}\langle
I|a^{V}|n\rangle \langle n|H^{(2)}_{PNC}|I\rangle +\langle
I|H^{(2)}_{PNC}|n\rangle \langle n|a^{V}|I\rangle  & = & -\langle
\frac{1}{\Delta E}\rangle \langle I|\{a^{V},H^{(2)}_{PNC}\}|I\rangle 
\end{eqnarray}
The resulting product of \( a^{V} \) and \( H_{PNC}^{(2)} \)
contracts to a two-body operator, so that only the two-body ground
state density matrix is needed, a considerable simplification.
(No three-body terms arise because
the absence of $E1$ valence transitions guarantees they vanish 
for our SM spaces.)

The closure approximation can be considered as an identity, clearly,
if one knows the correct \( \langle 1/E\rangle  \), 
that is, how to parameterize the relationship between the 
$1/E$-weighted and non-energy-weighted sums.
In practical terms, this means demonstrating that a systematic
relationship exists between $ \langle 1/E \rangle $ and some
experimentally known quantity, such as the position of the $E1$
giant resonance.  Note that the
$E1$ operator is closely related to the anapole operator $a^V$. 

To investigate the systematics we completed a series of exact calculations
in \( 1p- \) and light \( 2s1d- \)shell nuclei (\( ^{7} \)Li, \( ^{11}
\)B,
\( ^{17,19,21} \)F, \( ^{21,23} \)Na), evaluating both the
$\langle 1/E \rangle$ and non-energy-weighted sums.  First, the ground states
are determined from full \( 0\hbar \omega  \) diagonalizations. The
polarization sum involves the complete set of 1\( \hbar \omega  \)
states that connect to the ground state through the anapole operator.
The summation was performed by exploiting a variation of the Lanczos algorithm
to evaluate the effect of the nuclear propagator \( 1/E_{g.s.}-H \)
(see Sec. V.D).
The algorithm efficiently completes the sum via moments,
even though the dimensions of the $1 \hbar \omega$ bases
ranged up to $\sim$
500,000.  The appropriate closure energies were found not only
for the anapole polarization sum, but also for $E1$ operator.
This allowed us to compare the $\langle 1/E \rangle$ appropriate
for the AM calculation with that appropriate for photoexcitation.
As photoexcitation response functions have been mapped in many 
nuclei, this in turn allows us to relate the anapole $\langle 1/E \rangle$
to an experimental observable.
  
The results show that the anapole and photoexcitation 
average excitation energies
track each other very well, provided one takes into account the
three isospins contributing to \( H_{PNC}^{(2)} \). Measured as a
fraction of the \( 1/E \)-weighted giant dipole average excitation
energy, which is \( \langle 1/E\rangle ^{-1}\sim  \) (22-26) MeV
for these nuclei, the appropriate effective energies for the anapole
closure approximation are \( 0.604\pm 0.056 \) for \( h_{\rho }^{0} \)
and \( h_{\omega }^{0} \)(isoscalar channel), \( 0.899\pm 0.090 \)
for \( f_{\pi } \) (isovector channel), and \( 1.28\pm 0.14 \) for
\( h_{\rho }^{2} \) (isotensor channel). The larger \( \langle 1/E\rangle
\)
for \( h_{\rho }^{0} \) and \( h_{\omega }^{0} \) enhances the isoscalar
contribution to the anapole polarizability. The small variation in
\( \langle 1/E\rangle  \), once the isospin dependence is recognized,
supports the notion that we can connected the closure result to the
true polarization sum. 

Inspired by the nuclear systematics we found above, we estimate
T=0,1,2 closure energies from known \( E1 \) distributions, that
is, we fix the anapole closure energy as 0.6, 0.9, and 1.28 of the \( E1 \)
closure energy evaluated from the experimental dipole distribution.
For \( ^{133} \)Cs \cite{Cs_E1}, this gives 9.5, 14.1, and 20.2 MeV,
respectively. The corresponding \( ^{205} \)Tl values are 8.7, 12.9,
and 18.5 MeV.  The ground-state expectation values for the contracted two-body effective operator
\{$a^V,H^{(2)}_{PNC}$\} are then evaluated from the SM two-body density
matrices for Cs and Tl.  The Miller-Spencer correlation function is again included
in the two-nucleon matrix elements of $H^{(2)}_{PNC}$.
The resulting polarization contributions are given in
Table \ref{Tab: 2-body SM}. 

\begingroup

\squeezetable

\begin{table}

\caption{Nuclear systematics found in light odd-proton nuclei: The second
column shows the functional dependences of SM results for the direct
anapole polarization sums and the third column shows the forms for
sums by closure approximation using the closure energy \protect\( \langle
1/E\rangle _{E1}\protect \)
which is derived from the \protect\( 1/E\protect \)-weighted $E1$ sum
rule (also evaluated in the SM).
The same normalization has been applied to the second and third
columns.  By comparing
these two columns, it is apparent that in order for the closure
approximation
to be correct, the anapole closure energies \protect\( \langle 1/E\rangle
_{AM(T=0,1,2)}\protect \)
should be different from \protect\( \langle 1/E\rangle _{E1}\protect \).
In columns 4-6 we express \protect\( \langle 1/E\rangle
^{-1}_{AM(T=0,1,2)}\protect \)
in units of \protect\( \langle 1/E\rangle ^{-1}_{E1}\protect \):
thus a value less than one means that the appropriate anapole 
average excitation energy is lower than the corresponding average over the
photoexcitation peak.
Note the closure result faithfully reproduces the correct \protect\(
h_{\rho }^{0}-h_{\omega }^{0}\protect \)
combination.  We omit the dependence on $h_\rho^1$ and $h_\omega^1$
because the net isovector contribution is almost entirely from $f_\pi$.
In the case of \protect\( ^{19}\protect \)F, the lowest, nearly degenerate
\protect\( 1/2^{-}\protect \) state was removed from all sums.}

\begin{center}

\begin{tabular}{cccccc}
 Nucleus&
Direct Pol. Sum&
Closure with \( \langle 1/E\rangle _{E1} \)&
\( \langle 1/E\rangle _{AM(0)}^{-1} \)&
\( \langle 1/E\rangle ^{-1}_{AM(1)} \)&
\( \langle 1/E\rangle _{AM(2)}^{-1} \)\\
\hline
\( ^{7} \)Li&
\( f_{\pi }-0.34(h_{\rho }^{0}+0.58h_{\omega }^{0})+0.05h^{2}_{\rho } \)&
\( 0.80f_{\pi }-0.20(h_{\rho }^{0}+0.63h_{\omega }^{0})+0.05h^{2}_{\rho }
\)&
0.59&
0.80 &
1.0\\
 \( ^{11} \)B&
\( f_{\pi }-0.53(h_{\rho }^{0}+0.52h_{\omega }^{0})+0.05h^{2}_{\rho } \)&
\( 0.89f_{\pi }-0.37(h_{\rho }^{0}+0.52h_{\omega }^{0})+0.07h^{2}_{\rho }
\)&
0.70&
0.89 &
1.4\\
 \( ^{17} \)F&
\( f_{\pi }-0.60(h_{\rho }^{0}+0.48h_{\omega }^{0})+0.04h^{2}_{\rho } \)&
\( 1.02f_{\pi }-0.40(h_{\rho }^{0}+0.46h_{\omega }^{0})+0.05h^{2}_{\rho }
\)&
0.66&
1.02 &
1.2\\
 \( ^{19} \)F&
\( f_{\pi }-0.33(h_{\rho }^{0}+0.56h_{\omega }^{0})+0.02h^{2}_{\rho } \)&
\( 0.90f_{\pi }-0.19(h_{\rho }^{0}+0.59h_{\omega }^{0})+0.03h^{2}_{\rho }
\)&
0.58&
0.90 &
1.5\\
 \( ^{21} \)F&
\( f_{\pi }-0.41(h_{\rho }^{0}+0.55h_{\omega }^{0})+0.03h^{2}_{\rho } \)&
\( 0.97f_{\pi }-0.24(h_{\rho }^{0}+0.54h_{\omega }^{0})+0.04h^{2}_{\rho }
\)&
0.60&
0.97 &
1.3\\
 \( ^{21} \)Na&
\( f_{\pi }-0.57(h_{\rho }^{0}+0.51h_{\omega }^{0})+0.02h^{2}_{\rho } \)&
\( 0.77f_{\pi }-0.31(h_{\rho }^{0}+0.49h_{\omega }^{0})+0.03h^{2}_{\rho }
\)&
0.54&
0.77 &
1.5\\
 \( ^{23} \)Na&
\( f_{\pi }-0.67(h_{\rho }^{0}+0.53h_{\omega }^{0})+0.05h^{2}_{\rho } \)&
\( 0.95f_{\pi }-0.38(h_{\rho }^{0}+0.52h_{\omega }^{0})+0.07h^{2}_{\rho }
\)&
0.57&
0.95 &
1.4\\
\end{tabular}
\end{center}
\end{table}

\endgroup

\begin{table}

\caption{Decomposition of the SM estimates of the anapole matrix element
\protect\( \langle I||A_{1}||I\rangle /e\protect \)
into its weak coupling contributions.}

\label{Tab: 2-body SM}

\begin{center}

\begin{tabular}{ccrrrrrr}
 Nucleus&
Source&
\( f_{\pi } \)&
\( h_{\rho }^{0} \)&
\( h_{\rho }^{1} \)&
\( h_{\rho }^{2} \)&
\( h_{\omega }^{0} \)&
\( h_{\omega }^{1} \)\\
\hline
\( ^{133} \)Cs&
nucleonic&
0.59&
0.87&
0.90&
0.36&
0.28&
0.29 \\
&
ex. cur.&
8.58&
0.02&
0.11&
0.06&
-0.57&
-0.57 \\
&
polariz.&
51.57&
-16.67&
-4.88&
-0.06&
-9.79&
-4.59 \\
&
total&
60.74&
-15.78&
-3.87&
0.36&
-10.09&
-4.87 \\
 \( ^{205} \)Tl&
nucleonic&
-0.63&
-0.86&
-0.96&
-0.35&
-0.29&
-0.29 \\
&
ex. cur.&
-3.54&
-0.01&
-0.06&
-0.03&
0.28&
0.28 \\
&
polariz.&
-13.86&
4.63&
1.34&
0.08&
2.77&
1.27 \\
&
total&
-18.03&
3.76&
0.33&
-0.30&
2.76&
1.26 \\
\end{tabular}
\end{center}
\end{table}

\section{Experimental Constraints, Results, and Uncertainties}

In this sections we discuss atomic PNC experiments that determined (or limited) the 
AMs of $^{133}$Cs and $^{205}$Tl,
other experimental tests of the PNC hadronic weak interaction,
and the consistency of the AM results with these
other tests.  We also discuss nuclear structure uncertainties in
the interpretation of the AM measurements.

\subsection{Constraints from the Nuclear Anapole Moments of \protect\protect\(
^{133} \)Cs \protect \protect 
and \protect\protect\( ^{205} \)Tl \protect \protect \label{Subsec: Atomic
PNC}}

A thirty-year program to study atomic PNC \cite{BoBo} has yielded in the past
few years exquisitely precise (sub 1\%) results.
The primary focus of these studies has been to obtain accurate
values of the strength of direct $Z^0$ exchange between electrons
and the nucleus.  The PNC effects are dominated by the exchange
involving an axial $Z^0$ coupling to the electron and a vector
coupling to the nucleus.  The nuclear coupling is thus coherent,
proportional to the weak vector charge, 
$Q_W \sim Z(1-4\sin^2\theta_W) - N \sim -N$, and independent of the
nuclear spin direction.  It is widely recognized that these atomic
measurements are important tests of the standard electroweak
model and its possible extensions, complementing what has been
learned at high energy accelerators that directly probe physics
near the $Z^0$ pole \cite{erler,ramsey}.

\begin{figure}
{\centering
\resizebox*{0.5\columnwidth}{!}{\includegraphics{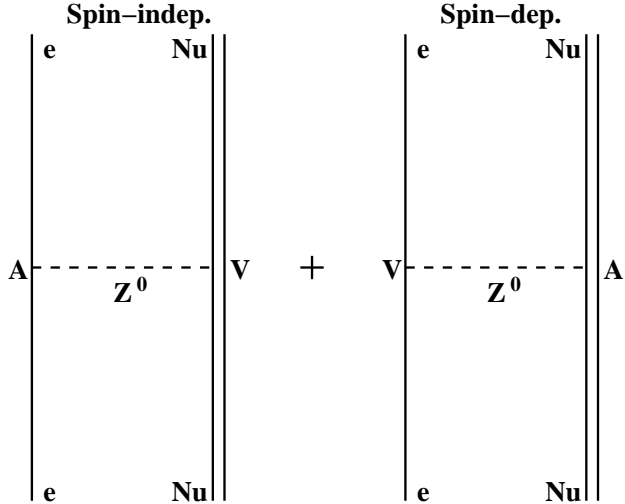}} \par}

\caption{Atomic parity mixing induced by \protect\( Z^{0}\protect
\) exchange.\label{Fig: Atomic PNC Z}}
\end{figure}

In heavy atoms the weak electron-nucleus interaction will induce a small $P-$wave 
parity admixture in an atomic $S$ orbital on the order of parts in
$10^{11}$.  This will produce, in a transition that
is normally $M1$, a small $E1$ component.  The PNC signal will 
be easier to detect if the parity-allowed $M1$ transition
is hindered, as the observable depends on the $E1/M1$ ratio.  The forbidden
M1 transitions of \( 6S_{1/2}\rightarrow 7S_{1/2} \) in Cs
and \( 6P_{1/2}\rightarrow 7P_{1/2} \) in Tl are two examples of
this sort.  Moreover, the structure of these atoms is comparatively
simple, allowing theorists to extract the underlying weak couplings
from the PNC observables.

One popular atomic technique exploits the linear Stark response
to an applied static electric field.
A coordinate system in the atom is established by mutually 
perpendicular Stark, magnetic (for producing the Zeeman spectrum
of states that can be populated by optical pumping), and laser
(stimulating the E1 transition) fields.  The ``parity transformation''
is accomplished by inverting these fields.  The PNC signal is  
associated with any difference seen in 
the interference between the Stark, PNC $E1$, and
hindered $M1$ amplitudes 
after various reversals of the coordinate system.  The
elimination of spurious signals associated with imperfect field reversals 
and other sources of systematic error is a
tedious task.  A recent review of the Cs and Tl experiments
can be found in \cite{haxwie}.

The dominant axial(electron)-vector(nucleus) atomic PNC interaction is independent of the
nuclear spin (see Fig. \ref{Fig: Atomic PNC Z}).  There is also a
tree-level contribution to atomic PNC that is nuclear-spin-dependent,
where the $Z^0$ exchange is vector(electron)-axial(nucleus).
This contribution is highly suppressed because the 
vector electron
weak coupling is small, \( g^{(e)}_{V}=-(1-4\sin ^{2}\theta _{W})\approx -0.1 \),
and the nuclear coupling is no longer coherent.
But, given sufficiently accurate ($\lsim$ 1\%) measurements,
this suppressed signal can be cleanly extracted by studying 
the hyperfine (and thus nuclear spin) dependence of the PNC
measurements.

\begin{figure}
{\centering
\resizebox*{0.5\columnwidth}{!}{\includegraphics{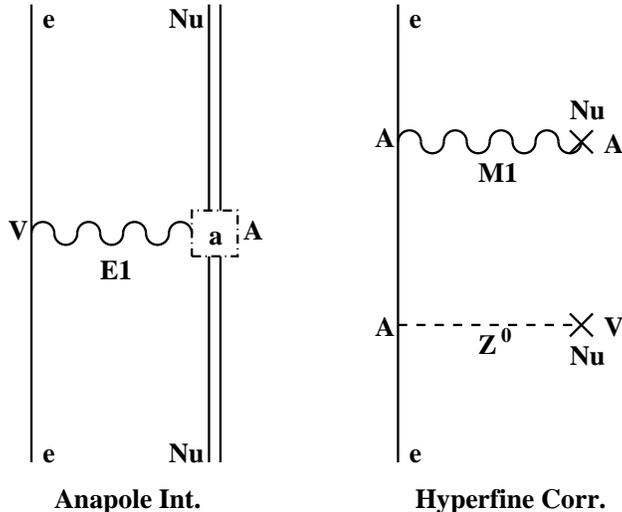}}
\par}

\caption{Radiative corrections in atomic parity mixing due to the nuclear
AM and hyperfine interactions.\label{Fig: Atomic PNC Corr.}}
\end{figure}

In Sec. II we noted that the nuclear AM will also 
generate a nuclear-spin-dependent weak interaction between 
the electron and the nucleus, thus contributing in combination
with tree-level V(electron)-A(nucleus) $Z^0$ exchange.
Furthermore other $O(G_F \alpha)$ radiative corrections also
contribute to that spin dependence, with the hyperfine interaction
between the electron and nucleus (see Fig.
\ref{Fig: Atomic PNC Corr.})
of particular importance because of the coherent $Z^0$ coupling.
While the naive expectation is that radiative corrections will
indeed be corrections of strenth $\sim$ $\alpha$ relative to the
tree-level contribution, the small vector coupling of the 
$Z^0$ to the electron
combined with the $A^{2/3}$ growth of the anapole 
moment leads to a surprise.  The AM becomes the
dominant source of nuclear-spin-dependent atomic PNC for
$A \gsim 20$ \cite{FKS84,HHM}.  This guarantees not only that the
nuclear spin dependence is signifcant for heavy atoms, but 
also that the AM contribution might be
deduced from the measurements.

The nuclear-spin-dependent(NSD) PNC electron-nucleus
contact interaction which generates the parity mixing can be expressed
as
\begin{eqnarray}
H^{NSD}_{PNC} & = & \frac{G_{F}}{\sqrt{2}}\kappa
_{tot}\vec{\alpha }\cdot \vec{I}\rho (r),\\
\kappa _{tot} & = & \kappa _{Z^{0}}+\kappa _{hf}+\kappa _{AM},
\end{eqnarray}
where \( \vec{I} \) and \( \rho (r) \) are the nuclear spin and
density, \( \vec{\alpha } \) the usual Dirac matrix of the electron, and
\( \kappa  \) a dimensionless constant which characterizes the
strength of the PNC.
(Note that our definition of \( \kappa  \) is different from the one
given by Khriplovich and others by a factor \(
(-1)^{I+1/2+l}(I+1/2)/(I(I+1)) \),
where \( l \) is a single-particle orbital angular momentum.
The Khriplovich definition thus assumes a single-particle picture,
though there are examples of nuclei where the dominant 
single-particle orbital is characterized by an $l$ that is
naively inconsistent with the many-body $I$, e.g.,
$l \neq I \pm 1/2$.)
The $\kappa$ subscripts denote contributions from \( Z^{0} \)
exchange, the hyperfine interaction correction, and the AM.
From the \( ^{133}\)Cs (extracted by Flambaum and Murray
\cite{FlMu97})
and \( ^{205}\)Tl results \cite{Seattle95,Oxford95}, one finds

\begin{eqnarray}
\kappa _{tot}(^{133}\mathrm{Cs}) & = & 0.112\pm 0.016,\nonumber\\
\kappa _{tot}(^{205}\mathrm{Tl}) & = & 0.29\pm 0.40~~\mathrm{Seattle},\nonumber \\
\kappa _{tot}(^{205}\mathrm{Tl}) & = & -0.08\pm 0.40~~\mathrm{Oxford}.
\end{eqnarray}
Henceforth we will focus on the Seattle Tl result, as this 
proves to be more restrictive than the Oxford result 
in the parameter space of PNC hadronic couplings favored by other
experiments.  (The Oxford AM result is quoted with opposite signs
in different sections of \cite{Oxford95} and the accuracy of the
spin-independent measurement is considerably less than that of
the corresponding Seattle measurement.  These observations contributed to our
decision to focus on the result of \cite{Seattle95}.)
We treat
the Tl constraint as one on the principal isotope $^{205}$Tl
(70.5\%).  The other stable isotope, $^{203}$Tl (29.5\%),
differs in structure only by a pair of neutrons, and thus 
should have very similar properties.

The \( Z^{0} \) contribution is

\begin{equation}
\kappa _{Z^{0}}=-\frac{g_{A}}{2}(1-4\sin ^{2}\theta _{W})\frac{\langle
I||\sum ^{A}_{i=1}\sigma (i)\tau _{3}(i)||I\rangle }{\langle
I||\hat{I}||I\rangle },
\end{equation}
with the axial vector coupling \( g_{A}=1.267 \)  and \( \sin ^{2}\theta
_{W}=0.2230 \).  Here
\( || \) denotes a matrix element reduced in angular momentum.
The reduced matrix element of \( \hat{I} \) is \( \sqrt{I(I+1)(2I+1)} \).
The Gamow-Teller matrix elements, taken from the SM studies, are \( -2.305
\)
\( (^{133}\)Cs) and \( 2.282 \) \( (^{205}\)Tl), not too different
from the corresponding single-particle (s.p.) values of \( -2.494 \) (unpaired \( 1g_{7/2}
\)
proton) and \( 2.449 \) (\( 3s_{1/2} \) proton). This yields: 

\begin{eqnarray}
\kappa _{Z^{0}}(^{133}\mathrm{Cs}) & = & 0.0140,\\
\kappa _{Z^{0}}(^{205}\mathrm{Tl}) & = & -0.127.
\end{eqnarray}
Note that the inclusion
of one-loop standard model electroweak radiative corrections
modify these results, reducing the isovector contribution 
substantially and inducing a small isoscalar component.

For the hyperfine correction, from the measured nuclear weak charge
and magnetic moment, Bouchiat and Piketty \cite{BoPi91} find 
\begin{eqnarray}
\kappa _{hf}(^{133}\mathrm{Cs}) & = & 0.0078,\\
\kappa _{hf}(^{205}\mathrm{Tl}) & = & 0.044.
\end{eqnarray}
Note that the conversion of the notation of Ref. \cite{BoPi91}
to ours is
\begin{equation}
\kappa _{hf}=c^{(2)}_{p}(hf)\frac{\langle I||\sigma _{p}||I\rangle
_{s.p.}}{\langle I||\hat{I}||I\rangle }.
\end{equation}
By subtracting \( \kappa _{Z^{0}} \) and \( \kappa _{hf} \) from
\( \kappa _{tot} \) we obtain the AM contribution
\begin{eqnarray}
\kappa _{AM}(^{133}\mathrm{Cs}) & = & 0.090\pm 0.016,\\
\kappa _{AM}(^{205}\mathrm{Tl}) & = & 0.376\pm 0.400.
\end{eqnarray}
These values are related to the nuclear AMs by
\begin{equation}
\kappa _{AM}=\frac{4\pi \alpha \sqrt{2}}{G_{F} M^2}\frac{\langle
I||\hat{a}||I\rangle /e}{\langle I||\hat{I}||I\rangle },
\end{equation}
where \( \hat{a} \) is the anapole operator.   As our results for \( \langle
I||\hat{a}||I\rangle /e \)
are expressed in terms of the PNC meson-nucleon couplings in Table
\ref{Tab: Exp Results}, 
we have the needed AM coupling constraints.

\subsection{Constraints from Nuclear PNC
Experiments\label{Subsec: Nuclear PNC}}

The nuclear experiments measuring an interference between PC and
PNC amplitudes generally fall into four types

\begin{enumerate}
\item Measurement of the longitudinal asymmetry \( A_{L} \) in a scattering
experiment (e.g., 
\( \vec{p}p \), \( \vec{p}d
\),
or \( \vec{p} \alpha \)).
\item Measurement of the circular polarization $P_\gamma$ of
photons emitted in a nuclear decay (e.g., $^{18}$F, $^{21}$Ne) or
reaction (e.g., $np \rightarrow d \gamma$).
\item Measurement of the asymmetry $A_\gamma$ of photons emitted
in the decay of a polarized nucleus (e.g., $^{19}$F) or in a
polarized nuclear reaction (e.g., $\vec{n} p \rightarrow d \gamma,
\vec{n} d \rightarrow t \gamma$).
\item Measurement of the degree of spin rotation for polarized
neutrons through various targets (e.g., $p,d,^4$He).
\end{enumerate}

It is unfortunate that only a single $NN$ PNC scattering
observable, the longitudinal analyzing power $A_L$
for $\vec{p}+p$, has been successful~\cite{balzer,potter,ramsay}.  (Experiments have been
done at 13.6, 45, and 221 MeV.)  These results have been
supplemented by a number of PNC measurements in nuclear
systems, where accidental degeneracies between pairs of
opposite-parity states
can produce, in some cases, large enhancements in the 
PNC signal.  Unfortunately not all of these results are readily
interprettable because of nuclear structure uncertainties.  
Those that can be analyzed with confidence \cite{Weak_Hadron} include
$A_L$ for $\vec{p} + \alpha$
at 46 MeV~\cite{lang}, the circular polarization $P_\gamma$ of the $\gamma$-ray
emitted from the 1081 keV state in $^{18}$F~\cite{bini}, and $A_\gamma$ for
the decay of the 110 keV state in polarized $^{19}$F~\cite{elsener}.
These examples involve either
few-body systems, where quasi-exact structure calculations 
can be done, or special nuclei in which the PNC mixing
matrix elements can be calibrated from axial-charge $\beta$ 
decay~\cite{haxton81}.  An analysis of these results, which have been in hand 
for some time, suggests that the isoscalar PNC $NN$ interaction --
which is dominated by $\rho$ and $\omega$ exchange -- is 
comparable to or slightly stronger than the DDH ``best value,''
whereas the isovector interaction -- dominated by $\pi$ exchange --
is significantly weaker ($\lsim$ 1/3)~\cite{Weak_Hadron}.  Because one expects the
isovector channel to be governed by neutral currents and to
receive potentially significant light sea-quark contributions,
there is considerable interest in testing
this result.  The Cs and Tl
AM results provide one possible cross check.
\begin{table}

\caption{PNC observables and corresponding theoretical predictions,
decomposed
into the designated weak-coupling combinations.}

\label{Tab: Exp Results}

\begin{center}

\begin{tabular}{cccccccc}
 Observable&
Exp.(\( \times 10^{7} \))&
\( f_{\pi }-0.12h_{\rho }^{1}-0.18h_{\omega }^{1} \)&
\( h_{\rho }^{0}+0.7h_{\omega }^{0} \)&
\( h_{\rho }^{1} \)&
\( h_{\rho }^{2} \)&
\( h_{\omega }^{0} \)&
\( h_{\omega }^{1} \)\\
\hline
\( A_{L}^{pp}(13.6) \)&
-0.93 \( \pm  \) 0.21&
&
0.043&
0.043&
0.017&
0.009&
0.039 \\
 \( A_{L}^{pp}(45) \)&
-1.57 \( \pm  \) 0.23&
&
0.079&
0.079&
0.032&
0.018&
0.073 \\
 \( A_{L}^{pp}(221) \)&
0.84 \( \pm \) 0.34&
&
-0.030&
-0.030&
-0.012&
0.021&
\\
 \( A_{L}^{p\alpha }(46) \)&
-3.34 \( \pm  \) 0.93&
-0.340&
0.140&
0.006&
&
-0.039&
-0.002 \\
 \( P_{\gamma }(^{18} \)F)&
1200 \( \pm  \) 3860&
4385&
&
34&
&
&
-44 \\
 \( A_{\gamma }(^{19} \)F)&
-740 \( \pm  \) 190&
-94.2&
34.1&
-1.1&
&
-4.5&
-0.1 \\
 \( \langle ||A_{1}||\rangle /e, \) Cs&
800 \( \pm  \) 140&
60.7&
-15.8&
3.4&
0.4&
1.0&
6.1 \\
 \( \langle ||A_{1}||\rangle /e, \) Tl&
370 \( \pm  \) 390&
-18.0&
3.8&
-1.8&
-0.3&
0.1&
-2.0 \\
\end{tabular}
\end{center}
\end{table}

\subsection{Results\label{Subsec: Res}}

\begin{figure}
{\centering
\resizebox*{0.8\columnwidth}{!}{\includegraphics{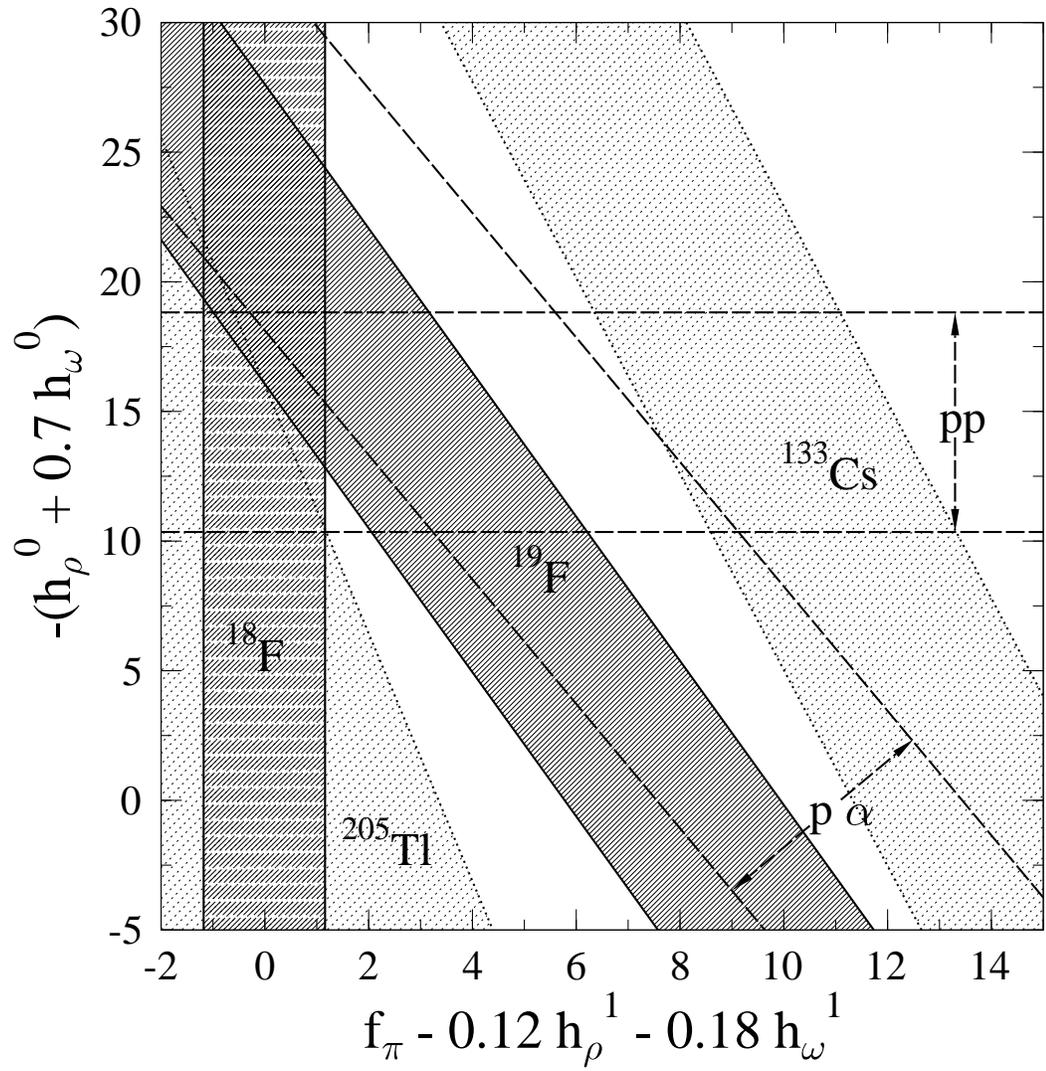}} \par}

\caption{Constraints on the PNC meson couplings (\protect\( \times
10^{7}\protect \))
that follow from the results in Table \ref{Tab: Exp Results}. The
error bands are one standard deviation. The \protect\( \vec{p}p\protect \)
band is the union of \protect\( 13.6\protect \), \protect\( 45\protect \),
and \protect\( 221\protect \)MeV results.\label{Fig: Comparison}}
\end{figure}

The constraints on PNC meson-nucleon couplings of Table \ref{Tab: Exp Results}
are displayed graphically in Fig. \ref{Fig: Comparison}.
Although there are six independent couplings, two combinations 
of these, one isoscalar and one isovector,
dominate the observables: \( f_{\pi
}-0.12h^{1}_{\rho }-0.18h^{1}_{\omega } \)
and \( h^{0}_{\rho }+0.7h^{0}_{\omega } \).
The decomposition of Table \ref{Tab: Exp Results} thus uses these two
degrees of freedom along with $h_\rho^2$ and the residual contributions
in $h_\rho^1, h_\omega^0$, and $h_\omega^1$.
The $1\sigma$ error bands of Fig. \ref{Fig: Comparison}
are generated from the experimental uncertainties, broadened
somewhat by allowing uncorrelated variations in each of the 
four minor degrees of freedom (that is, $h_\rho^2$ and the residuals in
in $h_\rho^1$, $h_\omega^0$, and $h_\omega^1$) over the DDH broad
``reasonable ranges.''  Note that only a fraction of the region
allowed by the Seattle Tl constraint is shown: the total width 
of the Tl band is an order of magnitude broader than the width
of the Cs allowed band, with most of the Tl allowed region 
lying outside the DDH ``reasonable ranges'' (i.e., in the region
of negative $f_\pi - 0.12h_\rho^1 - 0.18 h_\omega^1$ and
positive $h_\rho^0 + 0.7h_\omega^0$).  That is, the
bulk of the Seattle Tl band corresponds to an AM value opposite
in sign to that expected theoretically, given what we know
experimentally about PNC meson-nucleon couplings.  The corresponding Oxford
Tl band (not illustrated) includes almost all of the parameter
space in Fig. \ref{Fig: Comparison}, as well as a substantial 
region outside the bounds of the figure, to the lower left.
 
The weak coupling ranges covered by Fig. \ref{Fig: Comparison}
correspond roughly to the DDH broad ``reasonable ranges.''
Thus the anapole constraints are not inconsistent with the
theoretical ``ball-park'' estimates.  However, the detailed
lack of consistency among the various measurements is disconcerting.
Before the anapole results are included, the indicated solution
is a small $f_\pi$ and an isoscalar coupling somewhat larger
than, but consistent with, the DDH best value,
$-(h_\rho^0+0.7h_\omega^0)_{b.v.}^{DDH} \sim 12.7$.  But
the AM results agree poorly with this solution, as well as with
each other.  In particular, the precise result for $^{133}$Cs
tests a combination of PNC couplings quite similar to those
measured in $A_\gamma(^{19}$F) and in $A_L^{p \alpha}$, but
requires larger values for the weak couplings.

Despite substantial differences between our work and that of
Flambaum and Murray \cite{FlMu97},
the predicted AMs from these two calculations are in
relatively good agreement.  The corresponding interpretations, however,
are quite different.  Flambaum and Murray adopted the 
viewpoint that the Cs AM result could be accommodated by a value
$f_\pi \sim 9.5$, about twice the DDH best value, $f_{\pi~b.v.}^{DDH}
\sim 4.6$.  (The DDH reasonable range is 0-11.4, in units of $10^{-7}.$)
The difficulty with this suggestion is its inconsistency with
$P_\gamma(^{18}$F), a measurement that has been performed by
five groups.  The constraint from this measurement is almost 
devoid of theoretical uncertainty
\begin{equation}
-0.6 \lsim f_\pi - 0.11h_\rho^1 - 0.19h_\omega^1 \lsim 1.2.
\end{equation}
If one allows $h_\rho^1$ and $h_\omega^1$ to vary throughout their
DDH reasonable ranges, one finds $-1.0 \lsim f_\pi \lsim$ 1.1,
clearly ruling out $f_\pi \sim 9$.  There is also some tension 
between the Cs band and those for $p+\alpha$ and $A_\gamma(^{19}$F).

Thus, unfortunately, the hint of a consistent pattern of weak
meson-nucleon couplings that was emerging from nuclear tests of
the weak hadronic current is disturbed when the Cs and Tl 
results are added.

\subsection{Operator Renormalization and other Nuclear Structure Issues\label{Subsec: Issues}}

It thus appears that the calculated value of the Cs AM, using 
weak meson-nucleon couplings determined from $NN$ and nuclear
experiments, is significantly smaller than the measured
value.  While there are
several questions that could be raised about this conclusion,
perhaps the most difficult one is the quality of the nuclear
structure calculations for Cs and Tl: what error bar should we
assign because of the inherent uncertainties in such calculations?

Despite the rather extensive theoretical literature on AMs, 
it would be fair to characterize the general quality of the 
associated nuclear structure work as unsophisticated.  Much
of the previous work is based on extreme single particle 
models and employs effective one-body PNC potentials, a choice
that tends to obscure the discrepancies apparent in Fig. 
\ref{Fig: Comparison}.
Only a few attempts have been made to estimate the effects of
correlations, even in schematic ways.  In \cite{BoPi91} quenching
factors were introduced as a phenomenological correction to
single-particle estimates.  Solid motivation for this approach
can be found in classic studies of magnetic moments and Gamow-Teller
transitions in nuclear physics.  In \cite{AuBr99} single-particle
calculations were corrected for core polarization effects, 
employing a realistic g-matrix interaction but a very simple
set of particle-hole excitations.
Despite the highly truncated model space, this may be the
only paper, other than our work here and in earlier papers
\cite{HHM,hlrm}, to use a realistic interaction in calculations
of the Cs and Tl AMs.
Finally, in Ref. \cite{DmTe97} core polarization effects were evaluated
in the random phase approximation, but with a schematic 
zero-range spin-spin residual interaction.
  
One factor limiting what can be done is the challenge of completing
the polarization sum: apart from \cite{HHM,hlrm}, the work referenced above performed
this sum state-by-state.  Such a summation technique rules out 
a sophisticated ground-state wave function: the number of 
opposite-parity eigenstates connecting to the ground state by
the $E1$ operator would be enormous.  The two attempts to move
beyond direct summation have come from our studies.  In
\cite{HHM} summation to a complete set of $1\hbar \omega$ 
states for $^{19}$F was carried out by a Lanczos algorithm
moments method.  In this approach one recognizes that the 
quantity of interest is the distribution of the vector 
$a^V |I \rangle$ over the full set of $1\hbar \omega$ eigenstates:
if that distribution is known, it can be weighted by
$1/(E_{g.s.}-E_n)$ and dotted with $\langle I| H_{PV}^{(2)}$
to generate the polarization sum.  Instead of diagonalizing
a very large matrix of dimension $N$, where $N$ is the number
$1 \hbar \omega$ eigenstates, 
to get the eigenvalues $E_n$ and eigenstates needed to do this
sum state-by-state, the Lanczos method maps the large matrix
into a series of smaller matrices of dimension $N' = 1,2,3...,$
where $N' \ll N$.
This mapping extracts exact information from 
the original large matrix, the $2N'-1$ lowest 
moments of the vector $a^V |I \rangle$ over the $1 \hbar \omega$
eigenspectrum.  It is readily seen that the distribution must be
very well determined after a modest number of iterations,
$N' \sim 50$.  There is a variation of this algorithm that uses
the information in the Lanczos matrix to construct the effect
of the Green's function \cite{HHM}: it is obvious physically
that one can obtain the Green's function from the detailed moments
construction.  (The algorithm develops the Green's function
acting on a vector as an expansion in the Lanczos vectors, with the
the coefficients of the vectors updated with each iteration \cite{engel}.
The method is thus exact in a numerical sense, allowing one to
evaluate the convergence.)  This was the method used in the present
study of $p-$ and $sd-$shell nuclei, to assess average excitation
energies.  We have applied this method in cases where $N \sim 10^6$,
and it is possible with modern machines to tackle problems of
dimension $\sim 10^8$ in this way.  Unfortunately, 
given the complexity of our $^{133}$Cs ground state wave function,
the dimension of the negative-parity space required to saturate the
$E1$ sum is substantially larger than $10^8$.  Thus this technique,
while exceedingly powerful, cannot be applied to a case 
like $^{133}$Cs, at least at the present time.
 
Because we felt it was important to use a realistic large-scale
SM wave function in describing the $^{133}$Cs ground state,
another method was needed to evaluate the polarization sum.
We did this by closure, which was tractable in part
because of an attractive property of the canonical $^{133}$Cs
SM space, no nonzero matrix elements of $a^V$.  In our view there
are two worrisome features of this calculation.  The first is
the reliability of the average excitation energy estimate,
which we defined as the ratio of the non-energy-weighted to
$1/E_n$-weighted sums.  We performed a large set of calculations
in lighter nuclei, using the exact Lanczos Green's function
method described above, to calibrate the method.  The average
excitation energies,
normalized to the photoexcitation $E1$ peak 
and evaluated for each isospin channel, proved to be very
stable.  One cannot prove that the extrapolation to heavy nuclei
like Cs and Tl is valid, clearly: perhaps there is some
systematic evolution with neutron excess.  On the other hand,
the naive expectation is that the method should improve with 
$A$, as the $E1$ profile tends to become more collective in
heavier nuclei, and as the spin-orbit force tends to remove 
$E1$ strength from low excitations: the closure approximation
is clearly exact in the limit of an infinitely narrow $E1$
resonance.  Because the measured Cs AM is large,  
one would need a substantial amount of strength quite low
in the Cs spectrum to enhance the $1/E_n$ sum and thus "fix"
the SM calculation: this is unexpected and, while the $a^V$
and photoexcitation $E1$ operators are somewhat different,
there is no evidence in the photoexcitation distribution for
such strength \cite{Cs_E1}.

\begin{table}

\caption{Magnetic moments of \protect\( ^{133}\protect \)Cs and \protect\(
^{205}\protect \)Tl
measured in nuclear magnetons.}

\label{Tab: M1 Moment}

\begin{center}

\begin{tabular}{cccc}
&
s.p.&
SM&
exp.\\
\hline
\( ^{133}\)Cs&
1.72&
1.65&
2.58\\
\( ^{205}\)Tl&
2.79&
2.58&
1.64\\
\end{tabular}
\end{center}
\end{table}

The second question is the adequacy of our ground-state wave
function: though the Cs and Tl SM calculations are serious efforts,
numerical limitations forced restrictions on the proton
and neutron occupation numbers.  The unrestricted
$1g_{7/2}-2d_{5/2}-3s_{1/2}-2d_{3/2}-1h_{11/2}$ SM calculation was not
attempted.  Furthermore, it is well known that even full-shell
calculations often must be renormalized phenomenologically.
Two operators closely related to the AM, the Gamow-Teller
and $M1$ operators, are well-studied examples \cite{arima}.
In Table \ref{Tab: M1 Moment} our Cs and Tl SM magnetic moment values
are compared to the experimental and s.p. values.  The SM
and uncorrelated s.p. values are not that different, and both
differ significantly from experiment.  The conclusion is that 
potential important physics is absent in our truncated SM 
calculations.

The deviations of magnetic moments from the Schmidt line (or s.p. values) around the
Pb region have been extensively studied by Arima \emph{et al.}
\cite{arima}.  The deviations from the s.p. predictions can be
described as a set of corrections to the bare gyromagnetic factors

\begin{eqnarray}
\langle I||\mu ||I\rangle /\mu _{N} & = & (\frac{1}{2}+\delta
g^{(0)}_{l})\langle ||\vec{l}||\rangle _{s.p.}+(\frac{1}{2}+\delta
g^{(1)}_{l})\langle ||\vec{l}\tau _{3}||\rangle _{s.p.}+(0.88+\delta
g^{(0)}_{s})\langle ||\vec{\sigma }||\rangle _{s.p.}\nonumber \\
 &  & +(4.70+\delta g^{(1)}_{s})\langle ||\vec{\sigma }\tau _{3}||\rangle
_{s.p.}\, ,
\end{eqnarray}
These factors represent the operator and wave function normalization
corrections that would result from a faithful treatment of the 
omitted parts of the Hilbert space.
Equivalently (and perhaps more appropriately) one can quote this 
result in terms of renormalized matrix elements

\begin{equation}
\langle I||\mu ||I\rangle /\mu _{N}=\frac{1}{2}\langle
I||\vec{l}||I\rangle _{ren}+\frac{1}{2}\langle I||\vec{l}\tau
_{3}||I\rangle _{ren}+0.88\langle I||\vec{\sigma }||I\rangle
_{ren}+4.70\langle I||\vec{\sigma }\tau _{3}||I\rangle _{ren}\, .
\end{equation}
The fit of \cite{arima} gives the following quenching for the spin
matrix elements near Pb
\begin{eqnarray}
\langle I||\vec{\sigma }||I\rangle _{ren} & = & 0.86\langle ||\vec{\sigma
}||\rangle _{s.p.}\, ,\\
\langle I||\vec{\sigma }\tau _{3}||I\rangle _{ren} & = & 0.54\langle
||\vec{\sigma }\tau _{3}||\rangle _{s.p.}\, .
\end{eqnarray}

Although there exists no such large body of data on the anapole
moment operator, we now explore whether some tentative 
conclusions can be drawn about effects of missing correlations
on that operator.  We begin with the observation that the 
effects of correlations on a many-body operator are expected
to be quite similar to their effects on the one-body equivalent of
that operator.  (One specific illustration of this is detailed
in \cite{haxton81}.)  Thus we start by looking for the one-body
equivalent of the anapole polarization operator.  The most general
spin-isospin form for a rank-one operator is
\begin{equation}
\vec{a}_{pol}^{equiv}=\frac{e}{\langle E\rangle }(a^{(0)}_{l}\vec{l}+a^{(1)}_{l}\tau _{3}
\vec{l}+a^{(0)}_{s}\vec{\sigma }+a^{(1)}_{s}\tau _{3}\vec{\sigma }+a^{(0)}_
{p}[Y_{2}\otimes \vec{\sigma }]_{1}+a^{(1)}_{p}\tau _{3}[Y_{2}\otimes
\vec{\sigma }]_{1}).
\end{equation}
As the average excitation energy is measured in units of
$\hbar \omega$, the bare couplings $a_{(l,s,p)}^{(0,1)}$
are dimensionless.  We then evaluate matrix elements of
this one-body operator and of the full polarization sum (chosing DDH 
``best-value'' meson-nucleon couplings) in a single-particle model for a variety of nuclei in the
Pb and Sn regions, fitting the coefficients of the one-body
operator to reproduce the polarization results.  The results
for Tl (Pb region) are presented in a series of three tables,
Tables \ref{Tab: Pb data}, \ref{Tab: Pb fit}, and \ref{Tab: Pb s.p.},
giving, respectively, the comparison of the calculated s.p.
polarization results with those generated by the effective 
operator, the best fit values found for the coefficients of the effective operator,
and the matrix elements of the various terms in the effective
one-body operator.  The following three tables give the analogous  
results for Cs (Sn region).
Calculations were done with no spin-orbit potential as well as
with a spin-orbit potential of strength -0.1$\hbar \omega 
\vec{\sigma} \cdot \vec{l}$: the results show little sensitivity
to the spin-orbit contribution.

The tables show that the orbital
contributions to the effective operator are neglible: the
dominant terms are the spin and spin-tensor 
operators, with the former (folding the results of Tables
\ref{Tab: Pb fit} and \ref{Tab: Pb s.p.} and of Tables \ref{Tab: Sn fit} and 
\ref{Tab: Sn s.p.})
accounting typically for about 70\% of the AM strength.  
Furthermore the spin isoscalar and spin isovector operators
contribute with the same relative sign, with the isovector
contribution larger.  It follows for $^{205}$Tl, 
where the single particle assignment is $3s_{1/2}$, eliminating
both the spin-tensor and orbital contributions, that the 
effective AM operator is very similar to the magnetic moment
operator, and thus should be renormalized in a very similar
way.  From Table \ref{Tab: M1 Moment} one concludes that our
SM estimates are not sufficiently quenched, overestimating 
the Tl AM by about a factor 1.6.  The consequence of 
this would be to broaden the allowed Tl band (only partially
shown) in Fig. \ref{Fig: Comparison} proportionately.

The case of $^{133}$Cs is more difficult in that the spin-tensor
operator now plays a significant role: the s.p. assignment
is $1g_{7/2}$.  This operator does not arise as a bare operator
in Gamow-Teller, $M1$, or other familiar responses.   
Our approach is somewhat unsatisfactory, but perhaps of some help.
In Table \ref{Tab: Light Nu. mag. Mom.} we compare s.p. and 
full $1p-$ and $2s1d-$shell SM calculations of magnetic moments
with the experimental values
for a series of light nuclei.  This seems to establish that, in
these nuclei, the bulk of the needed renormalization of 
s.p. estimates does come from the SM (sweeping under the rug
issues like exchange currents, etc.).  In Table \ref{Tab: Light Nu. Ren.}
we make a similar comparison of s.p. and SM AM operator
matrix elements.  The pattern of significant quenching of 
spin matrix elements again emerges from this purely theoretical
comparison.  In the case of the spin-tensor operator, the
renormalizations do not seem very large, nor do they appear
to follow a simple pattern.  While there are cases of modest 
spin-tensor matrix element enhancement when the full-shell
correlations are turned on, these enhancements are smaller
than the quenching that occurs in the spin matrix elements.
The overall tendancy of the correlations is to suppress the
AM prediction.

While these arguments are of a hand-waving nature, they 
favor the conclusion that better SM calculations will produce
a somewhat smaller, not larger, predicted Cs AM.  The dominant 
missing physics appears to be insufficient quenching of the
spin matrix elements.  This will clearly exacerbate the
discrepancies apparent in Fig. \ref{Fig: Comparison}.
As a full-shell calculation for $^{133}$Cs will likely 
become feasible within the next few years, there may soon be
an opportunity to demonstrate that improved calculations
will produce a smaller AM.

\begin{table}

\caption{Comparison of calculated s.p. polarization anapole moments \protect\( \langle ||\vec{a}||\rangle /e\protect \)
in the \protect\( ^{208}\protect \)Pb region 
with results for a fitted phenomenological effective operator.}

\label{Tab: Pb data}

\begin{center}

\begin{tabular}{lrrrr}
\multicolumn{1}{c}{Nucleus}&
\multicolumn{2}{c}{No s.o.}&
\multicolumn{2}{c}{With s.o.}\\
\cline{2-3} \cline{4-5}
&
Calc.&
Fit&
Calc.&
Fit\\
\hline
\( ^{207} \)Tl\( (3s^{-1}_{\frac{1}{2}}) \)&
-578&
-593&
-542&
-536\\
\( ^{207} \)Tl\( (2d^{-1}_{\frac{3}{2}}) \)&
759&
763&
699&
692\\
\( ^{207} \)Tl\( (2d^{-1}_{\frac{5}{2}}) \)&
-780&
-889&
-691&
-825\\
\( ^{207} \)Pb\( (3p^{-1}_{\frac{1}{2}}) \)&
-131&
-122&
-132&
-123\\
\( ^{207} \)Pb\( (3p^{-1}_{\frac{3}{2}}) \)&
161&
154&
158&
151\\
\( ^{207} \)Pb\( (2f^{-1}_{\frac{5}{2}}) \)&
-180&
-190&
-184&
-194\\
\( ^{209} \)Bi\( (2f_{\frac{5}{2}}) \)&
970&
924&
881&
830\\
\( ^{209} \)Bi\( (2f_{\frac{7}{2}}) \)&
-919&
-1012&
-821&
-949\\
\( ^{209} \)Bi\( (1h_{\frac{9}{2}}) \)&
1154&
1198&
990&
1057\\
\( ^{209} \)Pb\( (2g_{\frac{9}{2}}) \)&
224&
232&
212&
220\\
\end{tabular}

\end{center}
\end{table}

\begin{table}

\caption{The fitted parameters in \protect\( ^{208}\protect \)Pb region.}

\label{Tab: Pb fit}

\begin{center}

\begin{tabular}{crrrrrr}
&
\multicolumn{1}{c}{\( a^{(0)}_{l} \)}&
\multicolumn{1}{c}{\( a^{(1)}_{l} \)}&
\multicolumn{1}{c}{\( a^{(0)}_{s} \)}&
\multicolumn{1}{c}{\( a^{(1)}_{s} \)}&
\multicolumn{1}{c}{\( a^{(0)}_{p} \)}&
\multicolumn{1}{c}{\( a^{(1)}_{p} \)}\\
\hline
no s.o.&
0.990&
1.458&
-95.838&
-146.159&
-243.094&
-366.696\\
with s.o.&
-0.721&
0.432&
-84.580&
-134.308&
-224.986&
-348.570\\
\end{tabular}

\end{center}
\end{table}

\begin{table}
[p]

\caption{The single particle reduced matrix elements used for \protect\(
^{208}\protect \)Pb
region fits.}

\label{Tab: Pb s.p.}

\begin{center}

\begin{tabular}{lrrrrrr}
&
\( \vec{l} \)&
\( \tau _{3}\vec{l} \)&
\( \vec{\sigma } \)&
\( \tau _{3}\vec{\sigma } \)&
\( [Y_{2}\otimes \vec{\sigma }]_{1} \)&
\( \tau _{3}[Y_{2}\otimes \vec{\sigma }]_{1} \)\\
\hline
\( ^{207} \)Tl\( (3s^{-1}_{\frac{1}{2}}) \)&
0.000&
0.000&
2.449&
2.449&
0.000&
0.000\\
\( ^{207} \)Tl\( (2d^{-1}_{\frac{3}{2}}) \)&
4.648&
4.648&
-1.549&
-1.549&
-0.618&
-0.618\\
\( ^{207} \)Tl\( (2d^{-1}_{\frac{5}{2}}) \)&
5.797&
5.797&
2.898&
2.898&
0.330&
0.330\\
\( ^{207} \)Pb\( (3p^{-1}_{\frac{1}{2}}) \)&
1.633&
-1.633&
-0.816&
0.816&
-0.651&
0.651\\
\( ^{207} \)Pb\( (3p^{-1}_{\frac{3}{2}}) \)&
2.582&
-2.582&
2.582&
-2.582&
0.206&
-0.206\\
\( ^{207} \)Pb\( (2f^{-1}_{\frac{5}{2}}) \)&
8.281&
-8.281&
-2.070&
2.070&
-0.661&
0.661\\
\( ^{209} \)Bi\( (2f_{\frac{5}{2}}) \)&
8.281&
8.281&
-2.070&
-2.070&
-0.661&
-0.661\\
\( ^{209} \)Bi\( (2f_{\frac{7}{2}}) \)&
9.621&
9.621&
3.207&
3.207&
0.426&
0.426\\
\( ^{209} \)Bi\( (1h_{\frac{9}{2}}) \)&
17.162&
17.162&
-2.860&
-2.860&
-0.761&
-0.761\\
\( ^{209} \)Pb\( (2g_{\frac{9}{2}}) \)&
13.984&
-13.984&
3.496&
-3.496&
0.507&
-0.507\\
\end{tabular}

\end{center}
\end{table}

\begin{table}

\caption{Comparison of calculated s.p. polarization anapole moments \protect\( \langle ||\vec{a}||\rangle /e\protect \)
in the \protect\( ^{132}\protect \)Sn region 
with results for a fitted phenomenological effective operator.}

\label{Tab: Sn data}

\begin{center}

\begin{tabular}{lrrrr}
\multicolumn{1}{c}{Nucleus}&
\multicolumn{2}{c}{No s.o.}&
\multicolumn{2}{c}{With s.o.}\\
\cline{2-3} \cline{4-5}
&
Calc.&
Fit&
Calc.&
Fit\\
\hline
\( ^{131} \)In\( (2p^{-1}_{\frac{1}{2}}) \)&
433&
422&
409&
429\\
\( ^{131} \)In\( (1f^{-1}_{\frac{5}{2}}) \)&
641&
644&
567&
636\\
\( ^{131} \)In\( (2p^{-1}_{\frac{3}{2}}) \)&
-484&
-450&
-440&
-449\\
\( ^{131} \)Sn\( (3s^{-1}_{\frac{1}{2}}) \)&
103&
93&
102&
89\\
\( ^{131} \)Sn\( (2d^{-1}_{\frac{3}{2}}) \)&
-124&
-109&
-125&
-106\\
\( ^{131} \)Sn\( (2d^{-1}_{\frac{5}{2}}) \)&
137&
144&
127&
136\\
\( ^{133} \)Sb\( (1g_{\frac{7}{2}}) \)&
788&
751&
684&
736\\
\( ^{133} \)Sb\( (2d_{\frac{5}{2}}) \)&
-610&
-534&
-549&
-539\\
\( ^{133} \)Sn\( (2f_{\frac{7}{2}}) \)&
169&
168&
158&
157\\
\( ^{133} \)Sn\( (1h_{\frac{9}{2}}) \)&
-169&
-159&
-171&
-160\\
\end{tabular}

\end{center}
\end{table}

\begin{table}

\caption{The fitted parameters in \protect\( ^{132}\protect \)Sn region.}

\label{Tab: Sn fit}

\begin{center}

\begin{tabular}{crrrrrr}
&
\multicolumn{1}{c}{\( a^{(0)}_{l} \)}&
\multicolumn{1}{c}{\( a^{(1)}_{l} \)}&
\multicolumn{1}{c}{\( a^{(0)}_{s} \)}&
\multicolumn{1}{c}{\( a^{(1)}_{s} \)}&
\multicolumn{1}{c}{\( a^{(0)}_{p} \)}&
\multicolumn{1}{c}{\( a^{(1)}_{p} \)}\\
\hline
no s.o.&
3.284&
2.327&
-52.990&
-0.790&
-182.832&
-271.131\\
with s.o.&
1.807&
1.315&
-44.608&
-80.981&
-176.206&
-260.346\\
\end{tabular}

\end{center}
\end{table}

\begin{table}

\caption{The single particle reduced matrix elements used for \protect\(
^{132}\protect \)Sn
region fits.}

\label{Tab: Sn s.p.}

\begin{center}

\begin{tabular}{lrrrrrr}
&
\( \vec{l} \)&
\( \tau _{3}\vec{l} \)&
\( \vec{\sigma } \)&
\( \tau _{3}\vec{\sigma } \)&
\( [Y_{2}\otimes \vec{\sigma }]_{1} \)&
\( \tau _{3}[Y_{2}\otimes \vec{\sigma }]_{1} \)\\
\hline
\( ^{131} \)In\( (2p^{-1}_{\frac{1}{2}}) \)&
1.633&
1.633&
-0.817&
-0.817&
-0.652&
-0.652\\
\( ^{131} \)In\( (1f^{-1}_{\frac{5}{2}}) \)&
8.281&
8.281&
-2.070&
-2.070&
-0.661&
-0.661\\
\( ^{131} \)In\( (2p^{-1}_{\frac{3}{2}}) \)&
2.582&
2.582&
2.582&
2.582&
0.206&
0.206\\
\( ^{131} \)Sn\( (3s^{-1}_{\frac{1}{2}}) \)&
0.000&
0.000&
2.449&
-2.449&
0.000&
0.000\\
\( ^{131} \)Sn\( (2d^{-1}_{\frac{3}{2}}) \)&
4.648&
-4.648&
-1.549&
1.549&
-0.618&
0.618\\
\( ^{131} \)Sn\( (2d^{-1}_{\frac{5}{2}}) \)&
5.797&
-5.797&
2.898&
-2.898&
0.330&
-0.330\\
\( ^{133} \)Sb\( (1g_{\frac{7}{2}}) \)&
12.470&
12.470&
-2.494&
-2.494&
-0.711&
-0.711\\
\( ^{133} \)Sb\( (2d_{\frac{5}{2}}) \)&
5.797&
5.797&
2.898&
2.898&
0.330&
0.330\\
\( ^{133} \)Sn\( (2f_{\frac{7}{2}}) \)&
9.621&
-9.621&
3.207&
-3.207&
0.427&
-0.427\\
\( ^{133} \)Sn\( (1h_{\frac{9}{2}}) \)&
17.160&
-17.160&
-2.860&
2.860&
-0.761&
0.761\\
\end{tabular}

\end{center}
\end{table}

\begin{table}

\caption{Magnetic moments of light odd-A nuclei.}

\label{Tab: Light Nu. mag. Mom.}

\begin{center}

\begin{tabular}{lrrc}
&
s.p.&
SM&
exp.\\
\hline
\( ^{11} \)B&
3.790&
2.872&
2.689\\
\( ^{13} \)N&
-0.263&
-0.307&
-0.322\\
\( ^{27} \)Al&
4.790&
4.207&
3.642\\
\( ^{29} \)P&
2.790&
1.088&
1.235\\
\( ^{31} \)P&
2.790&
1.252&
1.132\\
\( ^{33} \)Cl&
0.126&
0.634&
0.752\\
\end{tabular}

\end{center}
\end{table}

\begin{table}

\caption{The renormalization of single particle matrix elements in light odd-A
nuclei.}

\label{Tab: Light Nu. Ren.}

\begin{center}

\begin{tabular}{llrrrrrr}
&
&
\multicolumn{1}{c}{\( \vec{l} \)}&
\multicolumn{1}{c}{\( \tau _{3}\vec{l} \)}&
\multicolumn{1}{c}{\( \vec{\sigma } \)}&
\multicolumn{1}{c}{\( \tau _{3}\vec{\sigma } \)}&
\multicolumn{1}{c}{\( [Y_{2}\otimes \vec{\sigma }]_{1} \)}&
\multicolumn{1}{c}{\( \tau _{3}[Y_{2}\otimes \vec{\sigma }]_{1} \)}\\
\hline
\( ^{11} \)B&
s.p.&
2.582&
2.582&
1.291&
1.291&
0.206&
0.206\\
&
SM&
3.100&
3.100&
0.773&
0.773&
0.309&
0.309\\
\( ^{13} \)N&
s.p&
1.632&
1.632&
-0.408&
-0.408&
-0.651&
-0.651\\
&
SM&
1.657&
1.657&
-0.432&
-0.432&
-0.598&
-0.598\\
\( ^{27} \)Al&
s.p.&
5.787&
5.787&
1.449&
1.449&
0.330&
0.330\\
&
SM&
6.164&
6.164&
1.080&
1.080&
0.321&
0.321\\
\( ^{29} \)P&
s.p&
0.000&
0.000&
1.225&
1.225&
0.000&
0.000\\
&
SM&
0.910&
0.910&
0.314&
0.314&
0.246&
0.246\\
\( ^{31} \)P&
s.p.&
0.000&
0.000&
1.225&
1.225&
0.000&
0.000\\
&
SM&
0.822&
0.822&
0.402&
0.402&
0.190&
0.190\\
\( ^{33} \)Cl&
s.p.&
4.648&
4.648&
-0.775&
-0.775&
-0.618&
-0.618\\
&
SM&
4.361&
4.361&
-0.488&
-0.488&
-0.695&
-0.695\\
\end{tabular}

\end{center}
\end{table}

\section{Conclusions}

Recent atomic PNC measurements in $^{133}$Cs reached a new
level of precision that led, for the first time, to detection
of the hyperfine dependence of the signal.  New measurements in
Tl have also imposed important constraints on nuclear-spin-dependent
atomic PNC.  This progress has inspired the calculations reported
here.  In our work we employ a PNC
nucleon-nucleon interaction derived from a \( \pi  \)-, \( \rho  \)-,
and \( \omega \)-meson exchange model, providing sufficient degrees of
freedom to describe fully the five independent $S-P$ amplitudes.
The single-nucleon, exchange current, and nuclear polarization 
AM contributions are then evaluated with this choice of potential.
The end result is an analysis of AM constraints that is fully
consistent with the existing analysis of $A_L(\vec{p}+p)$ and
other hadronic tests of PNC.
  
Our results show
that the weak meson-nucleon couplings favored by nuclear experiments
are not compatible with the large AM value extracted from the
Cs measurement.  The Tl AM limit also favors a sign disfavored 
by theory.  Our qualitative arguments about the effects of
correlations missing from the SM calculations suggest that
improvements in the nuclear structure are likely to lead to
smaller values for the predicted Cs AM, exacerbating the current
discrepancy.

The nuclear constraints favor a small value for $f_\pi$ and 
isoscalar PNC couplings near the DDH ``best values.''  This
pattern is puzzling, and suggests that strong interactions 
modify the isospin of weak-meson nucleon couplings in a 
nontrivial way.  The Cs AM result now has produced a more
confusing situation, one where no one solution satisfies all
constraints.
Hopefully new experiments will provide the redundancy needed
to resolve the conflict.
In the next few years results are expected for
the spin-rotation of polarized slow neutrons in liquid helium
\cite{spin-rot}
and the asymmetry in polarized neutron capture \( \vec{n}+p\rightarrow
d+\gamma  \) \cite{np}.

New AM measurements could also help clarify matters.
A more accurate Tl AM measurement could define the sign of this
quantity: while the current band includes zero, it favors a
sign opposite that predicted by theory.
New AM measurements in odd-neutron nuclei would have great
impact, defining a band in the weak meson-nucleon coupling plane
roughly perpendicular to the Cs and Tl bands.
There are proposals for AM measurements on Dy, Fr, and Ba$^+$.

The accuracy of the Cs AM results sets it apart from any other
atomic PNC result: it has produced a constraint on a weak radiative
correction that, when translated into meson-nucleon weak couplings,
is as accurate as any direct probe of hadronic PNC.  
Thus the challenge of understanding this special measurement
should motivate more theoretical work.  Furthermore,
the implications of this measurement are not necessarily limited
to the issues discussed in this paper.  Our understanding of
V($e$)-A($N$) interactions also affects the interpretation of
electron-nucleus scattering experiments like SAMPLE \cite{SAMPLE},
where a similar discrepancy between theory and experiment
exists and where theoretical predictions also depend on a
proper treatment of the hadronic weak interaction.  Unravelling
the puzzles presented by these measurements constitutes an
important challenge to both theory and experiment.

\begin{acknowledgments}
This work was supported in part by the US Department of Energy and
by the National Science Foundation.
WH thanks the Guggenheim and Miller Foundations for
support during completion of this work, and UC Berkeley for its
hospitality.

\appendix
\end{acknowledgments}

\section{Two-body Exchange Charge and Current Operators in Momentum
Space\label{App: Full 2-body}}

The total Lagrangian density we are considering is

\begin{equation}
{\mathcal L}={\mathcal L}_{\it Free}+{\mathcal L}_{\it PC}+{\mathcal
L}_{\it PNC}+{\mathcal L}_{\it EM}\, ,
\end{equation}
with

\begin{eqnarray}
{\mathcal L}_{\it Free} & = & \overline{N}^{\, \, '}(i\partial
\! \! \! /-M)N+\frac{1}{2}(\partial _{\mu }\vec{\pi })\cdot (\partial
^{\mu }\vec{\pi })-\frac{1}{2}m^{2}_{\pi }\vec{\pi
}^{2}-\frac{1}{4}\vec{F}^{(\rho )}_{\mu \nu }\cdot \vec{F}^{(\rho )\mu \nu
}+\frac{1}{2}m^{2}_{\rho }\vec{\rho }_{\mu }\cdot \vec{\rho }^{\mu
}\nonumber \\
 &-& \frac{1}{4}F^{(\omega )}_{\mu \nu }F^{(\omega )\mu \nu
}+\frac{1}{2}m^{2}_{\omega }\omega _{\mu }\omega ^{\mu },\\
{\mathcal L}_{\it PC} & = & ig_{\pi NN}\overline{N}^{\, \, '}\gamma
_{5}\vec{\tau }\cdot \vec{\pi }N-g_{\rho NN}\overline{N}^{\, \, '}(\gamma
_{\mu }-i\frac{\mu _{v}}{2M}\sigma _{\mu \nu }q^{\nu })\vec{\tau
}\cdot \vec{\rho }^{\mu }N\nonumber \\
 &-& g_{\omega NN}\overline{N}^{\, \, '}(\gamma _{\mu }-i\frac{\mu
_{s}}{2M}\sigma _{\mu \nu }q^{\nu })\omega ^{\mu }N,\\
{\mathcal L}_{\it PNC} & = & -\frac{f_{\pi }}{\sqrt{2}}\overline{N}^{\, \,
'}(\vec{\tau }\times \vec{\pi })_{3}N+\overline{N}^{\, \, '}\Big
(h^{0}_{\rho }\vec{\tau }\cdot \vec{\rho }^{\mu }+h_{\rho }^{1}\rho ^{\mu
}_{3}+\frac{h_{\rho }^{2}}{2\sqrt{6}}(3\tau _{3}\rho ^{\mu }_{3}-\vec{\tau
}\cdot \vec{\rho }^{\mu })\Big )\gamma _{\mu }\gamma _{5}N\nonumber \\
 &+& \overline{N}^{\, \, '}(h^{0}_{\omega }\omega ^{\mu }+h^{1}_{\omega
}\tau _{3}\omega ^{\mu })\gamma _{\mu }\gamma _{5}N+O(h^{1'}_{\rho }),\\
{\mathcal L}_{\it EM} & = & -e\overline{N}^{\, \, '}[\gamma _{\mu
}(F^{(S)}_{1}\frac{1}{2}+F^{(V)}_{1}\frac{\tau
_{3}}{2})-i\frac{1}{2M}\sigma _{\mu \nu }k^{\nu
}(F^{(S)}_{2}\frac{1}{2}+F^{(V)}_{2}\frac{\tau _{3}}{2})]NA^{\mu
}\nonumber \\
 &-& e(\vec{\pi }\times \partial _{\mu }\vec{\pi })_{3}A^{\mu
}-e(\vec{\rho }^{\nu }\times \vec{F}^{(\rho )}_{\nu \mu })_{3}A^{\mu
}\nonumber \\
 &-& e\frac{g_{\rho \pi \gamma }}{2M}\epsilon _{\alpha \beta \gamma
\delta }F^{(\gamma )\alpha \beta }(\vec{\rho }^{\gamma }\cdot \partial
^{\delta }\vec{\pi })-e\frac{g_{\omega \pi \gamma }}{2M}\epsilon
_{\alpha \beta \gamma \delta }F^{(\gamma )\alpha \beta }(\omega ^{\gamma
}\partial ^{\delta }\pi _{3}).
\end{eqnarray}
Note that we use the Bjorken-Drell \cite{BD64} metric exclusively and the DDH definition
of weak couplings. In these expressions \( \vec{\pi } \), \( \vec{\rho
}^{\mu } \),
\( \omega ^{\mu } \), \( A^{\mu } \) denote the pion, rho meson, omega
meson, and photon fields; \( F^{(\rho ,\, \omega ,\, \gamma )}_{\mu \nu }
\)
is the field tensor for the designated field; and
\( q^{\mu } \) and \( k^{\mu } \) are the 4-momenta carried by
the outgoing meson and photon. \( F^{(S,V)}_{1,2} \) denotes the isoscalar
or isovector EM form factors, with \( F^{(S)}_{1}(0)=F^{(V)}_{1}=1 \),
\( F^{(S)}_{2}(0)=\mu _{s}=-0.12 \); \( F^{(V)}_{2}(0)=\mu _{v}=3.70 \). 

After applying the procedure described in Sec. \ref{Fig: 2-body},
we obtain the following results.

\subsection{Pair Currents }

Pair current diagrams are generated by \( \pi \),
\( \rho  \), or \( \omega  \) exchange, and the nucleon coupling to the photon
has either a PC or PNC meson-nucleon coupling.  Thus there are six
cases.  For charge densities to \( O(1/M^{2}) \) we obtain

\begin{mathletters}\begin{eqnarray}
\rho ^{\pi \, \, pair}_{\gamma -PC} & = & \frac{-ieg_{\pi NN}f_{\pi
}}{4\sqrt{2}M^{2}}(1+\mu _{s})(\vec{\tau }(1)\times \vec{\tau
}(2))_{3} \vec{\sigma }(1)\cdot \vec{k}\frac{(2\pi )^{3}\delta
^{(3)}(\cdots )}{(\vec{p}^{\, \, '}_{2}-\vec{p}_{2})^{2}+m^{2}_{\pi
}}+(1\leftrightarrow 2),\\
\rho ^{\pi \, \, pair}_{\gamma -PNC} & = & 0,\\
\rho ^{\rho \, \, pair}_{\gamma -PC} & = & \frac{ieg_{\rho
NN}}{4M^{2}}\Bigg \{\bigg [(1+\mu _{s})\Big (h^{0}_{\rho }\vec{\tau
}(1)\cdot \vec{\tau }(2)+h^{1}_{\rho }\tau _{3}(1)+\frac{h^{2}_{\rho
}}{2\sqrt{6}}(3\tau _{3}(1)\tau _{3}(2)-\vec{\tau }(1)\cdot \vec{\tau
}(2))\Big )\nonumber \\
&+& (1+\mu _{v})\Big (h^{0}_{\rho }\tau _{3}(2)+h^{1}_{\rho
}+\frac{h^{2}_{\rho }}{\sqrt{6}}\tau _{3}(2)\Big )\bigg ] \bigg
[\vec{k}\cdot \vec{\sigma }(1)\times \vec{\sigma }(2)\nonumber \\
&+& \frac{\vec{\sigma }(2)\cdot (\vec{p}^{\, \,
'}_{2}-\vec{p}_{2})}{m^{2}_{\rho }}\vec{k}\cdot \vec{\sigma }(1)\times
(\vec{p}^{\, \, '}_{2}-\vec{p}_{2})\bigg ]+(1+\mu _{v})(h^{0}_{\rho
}-\frac{h^{2}_{\rho }}{2\sqrt{6}})(\vec{\tau }(1)\times \vec{\tau
}(2))_{3}\nonumber \\
&\times&  \bigg [\vec{k}\cdot \vec{\sigma }(2)+\frac{\vec{\sigma
}(2)\cdot (\vec{p}^{\, \, '}_{2}-\vec{p}_{2})}{m^{2}_{\rho
}}\vec{k}\cdot (\vec{p}^{\, \, '}_{2}-\vec{p}_{2})\bigg ]\Bigg
\}\frac{(2\pi )^{3}\delta ^{(3)}(\cdots )}{(\vec{p}^{\, \,
'}_{2}-\vec{p}_{2})^{2}+m^{2}_{\rho }}+(1\leftrightarrow 2),\\
\rho ^{\rho \, \, pair}_{\gamma -PNC} & = & \frac{-ieg_{\rho
NN}}{4M^{2}}(1+\mu _{v})(h^{0}_{\rho }-\frac{h^{2}_{\rho
}}{2\sqrt{6}})(\vec{\tau }(1)\times \vec{\tau }(2))_{3}
 \vec{\sigma }(1)\cdot \vec{k}\frac{(2\pi )^{3}\delta
^{(3)}(\cdots )}{(\vec{p}^{\, \, '}_{2}-\vec{p}_{2})^{2}+m^{2}_{\rho
}}\nonumber \\ &+& (1\leftrightarrow 2),\\
\rho ^{\omega \, \, pair}_{\gamma -PC} & = & \frac{ieg_{\omega
NN}}{4M^{2}}(h^{0}_{\omega }+h^{1}_{\omega }\tau _{3}(2))[(1+\mu
_{s})+(1+\mu _{v})\tau _{3}(1)] \bigg [\vec{k}\cdot \vec{\sigma
}(1)\times \vec{\sigma }(2)\nonumber \\
&+& \frac{\vec{\sigma }(2)\cdot (\vec{p}^{\, \,
'}_{2}-\vec{p}_{2})}{m^{2}_{\omega }}\vec{k}\cdot \vec{\sigma }(1)\times
(\vec{p}^{\, \, '}_{2}-\vec{p}_{2})\bigg ]\frac{(2\pi )^{3}\delta
^{(3)}(\cdots )}{(\vec{p}^{\, \, '}_{2}-\vec{p}_{2})^{2}+m^{2}_{\omega
}}+(1\leftrightarrow 2),\\
\rho ^{\omega \, \, pair}_{\gamma -PNC} & = & 0.
\end{eqnarray}
\end{mathletters}
For current densities to \( O(1/M^{2}) \) we obtain

\begin{mathletters}\begin{eqnarray}
\vec{j}_{\gamma -PC}^{\pi \, \, pair} & = & \frac{-eg_{\pi NN}f_{\pi
}}{2\sqrt{2}M}[\vec{\tau }(1)\cdot \vec{\tau }(2)-\tau _{3}(1)\tau
_{3}(2)] \vec{\sigma }(1)\frac{(2\pi )^{3}\delta ^{(3)}(\cdots
)}{(\vec{p}^{\, \, '}_{2}-\vec{p}_{2})^{2}+m^{2}_{\pi }}+(1\leftrightarrow
2),\\
\vec{j}_{\gamma -PNC}^{\pi \, \, pair} & = & 0,\\
\vec{j}_{\gamma -PC}^{\rho \, \, pair} & = & \frac{eg_{\rho
NN}}{2M}\Bigg \{\bigg [h^{0}_{\rho }(\vec{\tau }(1)\cdot
\vec{\tau }(2)+\tau _{3}(2))+h^{1}_{\rho }(1+\tau
_{3}(1))+\frac{h^{2}_{\rho }}{2\sqrt{6}}(3\tau _{3}(1)\tau
_{3}(2)\nonumber \\
&-& \vec{\tau }(1)\cdot \vec{\tau }(2)+2\tau _{3}(2))\bigg ]
\bigg [\vec{\sigma }(2)+\frac{\vec{\sigma }(2)\cdot (\vec{p}^{\, \,
'}_{2}-\vec{p}_{2})}{m^{2}_{\rho }}(\vec{p}^{\, \,
'}_{2}-\vec{p}_{2})\bigg ]\nonumber \\
 &-& (h^{0}_{\rho }-\frac{h^{2}_{\rho }}{2\sqrt{6}})(\vec{\tau
}(1)\times \vec{\tau }(2))_{3} \bigg [\vec{\sigma }(1)\times
\vec{\sigma }(2)\nonumber \\
 &+& \frac{\vec{\sigma }(2)\cdot (\vec{p}^{\, \,
'}_{2}-\vec{p}_{2})}{m^{2}_{\rho }}\vec{\sigma }(1)\times (\vec{p}^{\, \,
'}_{2}-\vec{p}_{2})\bigg ]\Bigg \}\frac{(2\pi )^{3}\delta ^{(3)}(\cdots
)}{(\vec{p}^{\, \, '}_{2}-\vec{p}_{2})^{2}+m^{2}_{\rho
}}+(1\leftrightarrow 2),\\
\vec{j}_{\gamma -PNC}^{\rho \, \, pair} & = & \frac{-eg_{\rho
NN}}{2M}\bigg [h^{0}_{\rho }(\vec{\tau }(1)\cdot \vec{\tau }(2)+\tau
_{3}(2))+h^{1}_{\rho }(\tau _{3}(2)+\tau _{3}(1)\tau
_{3}(2))+\frac{h^{2}_{\rho }}{2\sqrt{6}}(3\tau _{3}(1)\tau
_{3}(2)\nonumber \\
 &-& \vec{\tau }(1)\cdot \vec{\tau }(2)+2\tau _{3}(2))\bigg ]
\vec{\sigma }(1)\frac{(2\pi )^{3}\delta ^{(3)}(\cdots )}{(\vec{p}^{\, \,
'}_{2}-\vec{p}_{2})^{2}+m^{2}_{\rho }}+(1\leftrightarrow 2),\\
\vec{j}_{\gamma -PC}^{\omega \, \, pair} & = & \frac{eg_{\omega
NN}}{2M}(h^{0}_{\omega }+h^{1}_{\omega }\tau _{3}(2))(1+\tau
_{3}(1))\nonumber \\
 &\times&  \bigg [\vec{\sigma }(2)+\frac{\vec{\sigma }(2)\cdot
(\vec{p}^{\, \, '}_{2}-\vec{p}_{2})}{m^{2}_{\omega }}(\vec{p}^{\, \,
'}_{2}-\vec{p}_{2})\bigg ]\frac{(2\pi )^{3}\delta ^{(3)}(\cdots
)}{(\vec{p}^{\, \, '}_{2}-\vec{p}_{2})^{2}+m^{2}_{\omega
}}+(1\leftrightarrow 2),\\
\vec{j}_{\gamma -PNC}^{\omega \, \, pair} & = & \frac{-eg_{\omega
NN}}{2M}(h^{0}_{\omega }+h^{1}_{\omega })(1+\tau _{3}(1))
\vec{\sigma }(1)\frac{(2\pi )^{3}\delta ^{(3)}(\cdots )}{(\vec{p}^{\, \,
'}_{2}-\vec{p}_{2})^{2}+m^{2}_{\omega }}+(1\leftrightarrow 2).
\end{eqnarray}

\end{mathletters}

\subsection{Transition Currents }

The transition currents can have a \( \pi \pi \gamma  \), \( \rho
\rho \gamma  \),
\( \rho \pi \gamma  \), or \( \omega \pi \gamma  \) vertex. In
the last two cases, the heavier mesons \( \rho  \) and \( \omega  \)
can have either a PC or PNC coupling.  Thus there are six possibilities.
For charge densities to \( O(1/M^{2}) \) we obtain

\begin{mathletters}

\begin{eqnarray}
\rho ^{\pi \pi \gamma } & = & \frac{eg_{\pi NN}f_{\pi
}}{2\sqrt{2}M}[\tau _{3}(1)\tau _{3}(2)-\vec{\tau }(1)\cdot \vec{\tau
}(2)] (E^{\, \, '}_{2}-E_{2}-(E^{\, \, '}_{1}-E_{1}))
\vec{\sigma }(1)\cdot (\vec{p}^{\, \,
'}_{1}-\vec{p}_{1}) \nonumber \\
&\times& \frac{(2\pi )^{3}\delta ^{(3)}(\cdots )}{((\vec{p}^{\,
\, '}_{1}+\vec{p}_{1})^{2}+m^{2}_{\pi })((\vec{p}^{\, \,
'}_{2}-\vec{p}_{2})^{2}+m^{2}_{\pi })}+(1\leftrightarrow 2),\\
\rho ^{\rho \rho \gamma } & = & ieg_{\rho NN}\Big (h^{0}_{\rho
}-\frac{h^{2}_{\rho }}{2\sqrt{6}}\Big )(\vec{\tau }(1)\times \vec{\tau
}(2))_{3} \bigg [\vec{\sigma }(2)\cdot (\vec{p}^{\, \,
'}_{1}-\vec{p}_{1}) \nonumber \\ &+&\frac{\vec{\sigma }(2)\cdot (\vec{p}^{\, \,
'}_{2}-\vec{p}_{2})}{m^{2}_{\rho }}
(\vec{p}^{\, \, '}_{2}-\vec{p}_{2})\cdot (\vec{p}^{\, \,
'}_{1}-\vec{p}_{1})\bigg ]\frac{(2\pi )^{3}\delta ^{(3)}(\cdots
)}{((\vec{p}^{\, \, '}_{1}-\vec{p}_{1})^{2}+m^{2}_{\rho })((\vec{p}^{\, \,
'}_{2}-\vec{p}_{2})^{2}+m^{2}_{\rho })}\nonumber \\
&+&  (1\leftrightarrow 2), \\
\rho _{\rho -PC}^{\rho \pi \gamma } & = & \frac{eg_{\rho NN}f_{\pi
}g_{\rho \pi \gamma }}{2\sqrt{2}Mm_{\rho }}(\vec{\tau }(1)\times
\vec{\tau }(2))_{3} \Big [(\vec{p}^{\, \, '}_{1}-\vec{p}_{1})\cdot
(\vec{p}^{\, \, '}_{2}-\vec{p}_{2})\times (\vec{p}^{\, \,
'}_{1}+\vec{p}_{1})\nonumber \\
 &+& i(1+\mu _{v})\big (\vec{\sigma }(1)\cdot (\vec{p}^{\, \,
'}_{1}-\vec{p}_{1})(\vec{p}^{\, \, '}_{2}-\vec{p}_{2})\cdot (\vec{p}^{\,
\, '}_{1}-\vec{p}_{1})-\vec{\sigma }(1)\cdot (\vec{p}^{\, \,
'}_{2}-\vec{p}_{2})(\vec{p}^{\, \, '}_{1}-\vec{p}_{1})^{2}\big )\Big
]\nonumber \\
 &\times& \frac{(2\pi )^{3}\delta ^{(3)}(\cdots )}{((\vec{p}^{\, \,
'}_{1}-\vec{p}_{1})^{2}+m^{2}_{\rho })((\vec{p}^{\, \,
'}_{2}-\vec{p}_{2})^{2}+m^{2}_{\pi })}+(1\leftrightarrow 2),\\
\rho _{\rho -PNC}^{\rho \pi \gamma } & = & \frac{ieg_{\rho NN}g_{\rho \pi
\gamma }}{2Mm_{\rho }}\big (h^{0}_{\rho }\vec{\tau }(1)\cdot \vec{\tau
}(2)+h^{1}_{\rho }\tau _{3}(2)+\frac{h^{2}_{\rho }}{2\sqrt{6}}(3\tau
_{3}(1)\tau _{3}(2)-\vec{\tau }(1)\cdot \vec{\tau }(2))\big )\nonumber \\
 &\times&  \vec{\sigma }(2)\cdot (\vec{p}^{\, \,
'}_{2}-\vec{p}_{2})\vec{\sigma }(1)\cdot (\vec{p}^{\, \,
'}_{1}-\vec{p}_{1})\times (\vec{p}^{\, \, '}_{2}-\vec{p}_{2})
\nonumber \\
 &\times&  \frac{(2\pi )^{3}\delta ^{(3)}(\cdots )}{((\vec{p}^{\, \,
'}_{1}-\vec{p}_{1})^{2}+m^{2}_{\rho })((\vec{p}^{\, \,
'}_{2}-\vec{p}_{2})^{2}-m^{2}_{\pi })}+(1\leftrightarrow 2),\\
\rho _{\omega -PC}^{\omega \pi \gamma } & = & 0,\\
\rho _{\omega -PNC}^{\omega \pi \gamma } & = & \frac{ieg_{\omega
NN}g_{\omega \pi \gamma }}{2Mm_{\omega }}(h^{0}_{\omega }\tau
_{3}(2)+h^{1}_{\omega }\tau _{3}(1)\tau _{3}(2))\vec{\sigma
}(2)\cdot (\vec{p}^{\, \, '}_{2}-\vec{p}_{2})\nonumber \\
 &\times&  \vec{\sigma }(1)\cdot (\vec{p}^{\, \, '}_{1}-\vec{p}_{1})\times
(\vec{p}^{\, \, '}_{2}-\vec{p}_{2})\frac{(2\pi )^{3}\delta ^{(3)}(\cdots
)}{((\vec{p}^{\, \, '}_{1}-\vec{p}_{1})^{2}+m^{2}_{\omega })((\vec{p}^{\,
\, '}_{2}-\vec{p}_{2})^{2}+m^{2}_{\pi })}+(1\leftrightarrow 2).
\end{eqnarray}
\end{mathletters}
For current densities to \( O(1/M^{2}) \) we obtain

\begin{mathletters}

\begin{eqnarray}
\vec{j}^{\pi \pi \gamma } & = & \frac{eg_{\pi NN}f_{\pi
}}{2\sqrt{2}M}[\tau _{3}(1)\tau _{3}(2)-\vec{\tau }(1)\cdot \vec{\tau
}(2)]((\vec{p}^{\, \, '}_{2}-\vec{p}_{2})-(\vec{p}^{\, \,
'}_{1}-\vec{p}_{1}))\vec{\sigma }(1)\cdot (\vec{p}^{\, \,
'}_{1}-\vec{p}_{1})\nonumber \\
 &\times&  \frac{(2\pi )^{3}\delta ^{(3)}(\cdots )}{((\vec{p}^{\, \,
'}_{1}-\vec{p}_{1})^{2}+m^{2}_{\pi })((\vec{p}^{\, \,
'}_{2}-\vec{p}_{2})^{2}+m^{2}_{\pi })}+(1\leftrightarrow 2),\\
\vec{j}^{\rho \rho \gamma } & = & \frac{ieg_{\rho NN}}{2M}\Big
(h^{0}_{\rho }-\frac{h^{2}_{\rho }}{2\sqrt{6}}\Big )(\vec{\tau }(1)\times
\vec{\tau }(2))_{3}\Bigg \{\Big [(\vec{p}^{\, \,
'}_{1}-\vec{p}_{1})-(\vec{p}^{\, \, '}_{2}-\vec{p}_{2})\Big ]\Big
(\vec{\sigma }(2)\nonumber \\ &+& \frac{\vec{\sigma }(2)\cdot (\vec{p}^{\, \,
'}_{2}-\vec{p}_{2})}{m^{2}_{\rho }}
 (\vec{p}^{\, \, '}_{2}-\vec{p}_{2})\Big )\cdot \Big ((\vec{p}^{\, \,
'}_{2}+\vec{p}_{2})-(\vec{p}^{\, \, '}_{1}+\vec{p}_{1})-i(1+\mu
_{v})\vec{\sigma }(1)\times (\vec{p}^{\, \, '}_{1}-\vec{p}_{1})\Big )\nonumber \\ &+& \Big
[(\vec{p}^{\, \, '}_{1}+\vec{p}_{1})
 + i(1+\mu _{v})\vec{\sigma }(1)\times (\vec{p}^{\, \,
'}_{1}-\vec{p}_{1})\Big ](\vec{p}^{\, \, '}_{1}-\vec{p}_{1})\cdot \Big
(\vec{\sigma }(2)+\frac{\vec{\sigma }(2)\cdot (\vec{p}^{\, \,
'}_{2}-\vec{p}_{2})}{m^{2}_{\rho }}(\vec{p}^{\, \, '}_{2}-\vec{p}_{2})\Big
)\nonumber \\
 &+& \bigg [\vec{\sigma }(2)+\frac{\vec{\sigma }(2)\cdot (\vec{p}^{\, \,
'}_{2}-\vec{p}_{2})}{m^{2}_{\rho }}(\vec{p}^{\, \,
'}_{2}-\vec{p}_{2})\bigg ](\vec{p}^{\, \, '}_{2}-\vec{p}_{2})\cdot \Big
((\vec{p}^{\, \, '}_{2}+\vec{p}_{2})-(\vec{p}^{\, \,
'}_{1}+\vec{p}_{1})\nonumber \\ &-& i(1+\mu _{v})
 \vec{\sigma }(1)\times (\vec{p}^{\, \, '}_{1}-\vec{p}_{1})\Big
)\Bigg \}\frac{(2\pi )^{3}\delta ^{(3)}(\cdots )}{((\vec{p}^{\, \,
'}_{1}-\vec{p}_{1})^{2}+m^{2}_{\rho })((\vec{p}^{\, \,
'}_{2}-\vec{p}_{2})^{2}+m^{2}_{\rho })}+(1\leftrightarrow 2),\\
\vec{j}_{\rho -PC}^{\rho \pi \gamma } & = & \frac{eg_{\rho NN}f_{\pi
}g_{\rho \pi \gamma }}{\sqrt{2}m_{\rho }}(\vec{\tau }(1)\times \vec{\tau
}(2))_{3}(\vec{p}^{\, \, '}_{1}-\vec{p}_{1})\times (\vec{p}^{\, \,
'}_{2}-\vec{p}_{2})\nonumber \\
 &\times& \frac{(2\pi )^{3}\delta ^{(3)}(\cdots )}{((\vec{p}^{\, \,
'}_{1}-\vec{p}_{1})^{2}+m^{2}_{\rho })((\vec{p}^{\, \,
'}_{2}-\vec{p}_{2})^{2}+m^{2}_{\pi })}+(1\leftrightarrow 2),\\
\vec{j}_{\rho -PNC}^{\rho \pi \gamma } & = & \frac{-ieg_{\rho NN}g_{\rho
\pi \gamma }}{4M^{2}m_{\rho }}\Big (h^{0}_{\rho }\vec{\tau }(1)\cdot
\vec{\tau }(2)+h^{1}_{\rho }\tau _{3}(2)+\frac{h^{2}_{\rho
}}{2\sqrt{6}}(3\tau _{3}(1)\tau _{3}(2)-\vec{\tau }(1)\cdot \vec{\tau
}(2))\Big )\nonumber \\
 &\times& \vec{\sigma}(2)\cdot (\vec{p}^{\, \, '}_{2}-\vec{p}_{2})]\bigg \{\Big [(\vec{p}^{\,
\, '2}_{2}-\vec{p}^{2}_{2})(\vec{p}^{\, \,
'}_{1}-\vec{p}_{1})-(\vec{p}^{\, \, '2}_{1}-\vec{p}^{2}_{1})(\vec{p}^{\,
\, '}_{2}-\vec{p}_{2})\Big ]\times \vec{\sigma }(1)\nonumber \\
&-& \vec{\sigma}(1) \cdot
(\vec{p}^{\, \, '}_{1}+\vec{p}_{1})
(\vec{p}^{\, \, '}_{1}-\vec{p}_{1})\times (\vec{p}^{\, \,
'}_{2}-\vec{p}_{2})\bigg \} \nonumber \\
&\times& \frac{(2\pi )^{3}\delta ^{(3)}(\cdots
)}{((\vec{p}^{\, \, '}_{1}-\vec{p}_{1})^{2}+m^{2}_{\rho })((\vec{p}^{\, \,
'}_{2}-\vec{p}_{2})^{2}+m^{2}_{\pi })}+(1\leftrightarrow 2),\\
\vec{j}_{\omega -PC}^{\omega \pi \gamma } & = & 0,\\
\vec{j}_{\omega -PNC}^{\omega \pi \gamma } & = & \frac{-ieg_{\omega
NN}g_{\omega \pi \gamma }}{4M^{2}m_{\omega }}(h^{0}_{\omega }\tau
_{3}(2)+h^{1}_{\omega }\tau _{3}(1)\tau _{3}(2)) \vec{\sigma
}(2)\cdot (\vec{p}^{\, \, '}_{2}-\vec{p}_{2})\bigg \{\Big [(\vec{p}^{\,
\, '2}_{2}-\vec{p}^{2}_{2})(\vec{p}^{\, \, '}_{1}-\vec{p}_{1})\nonumber \\
 &-& (\vec{p}^{\, \, '2}_{1}-\vec{p}^{2}_{1})(\vec{p}^{\, \,
'}_{2}-\vec{p}_{2})\Big ]\times \vec{\sigma }(1)-\vec{\sigma }(1)\cdot
(\vec{p}^{\, \, '}_{1}+\vec{p}_{1})(\vec{p}^{\, \,
'}_{1}-\vec{p}_{1})\times (\vec{p}^{\, \, '}_{2}-\vec{p}_{2})\bigg
\}\nonumber \\
 &\times& \frac{(2\pi )^{3}\delta ^{(3)}(\cdots )}{((\vec{p}^{\, \,
'}_{1}-\vec{p}_{1})^{2}+m^{2}_{\omega })((\vec{p}^{\, \,
'}_{2}-\vec{p}_{2})^{2}-m^{2}_{\pi })}+(1\leftrightarrow 2).
\end{eqnarray}

\end{mathletters}

\section{Two-body Exchange Current Operators in Position Space to Order of
\protect\( 1/M\protect \)}

Only the three-current operators are needed for the AM calculation.
We keep terms to $O(1/M)$.  The following results follow from
Fourier transformations of selected terms in Appendix A.

\begin{eqnarray}
\vec{j}^{\pi \, \, pair}_{\gamma -PC} & = & \frac{-eg_{\pi NN}f_{\pi
}}{8\sqrt{2}\pi M}(\vec{\tau }(1)\cdot \vec{\tau }(2)-\tau _{3}(1)\tau
_{3}(2))\vec{\sigma }(1)\delta ^{(3)}(\vec{x}-\vec{x}_{1})\frac{e^{-m_{\pi
}r}}{r}+(1\leftrightarrow 2),\\
\vec{j}^{\rho \, \, pair}_{\gamma -PC} & = & \frac{eg_{\rho NN}}{12\pi
M}\bigg \{\Big [h^{0}_{\rho }(\vec{\tau }(1)\cdot \vec{\tau }(2)+\tau
_{3}(2))+h^{1}_{\rho }(1+\tau _{3}(1))+\frac{h^{2}_{\rho
}}{2\sqrt{6}}(3\tau _{3}(1)\tau _{3}(2)-\vec{\tau }(1)\cdot \vec{\tau
}(2)\nonumber \\
 &+& 2\tau _{3}(2))\Big ]\Big (\vec{\sigma }(2)+\sqrt{2\pi
}(1+\frac{3}{m_{\rho }r}+\frac{3}{(m_{\rho }r)^{2}})[Y_{2}(\Omega
_{r})\otimes \vec{\sigma }(2)]_{1}\Big )\nonumber \\
 &+& (h^{0}_{\rho }-\frac{h^{2}_{\rho }}{2\sqrt{6}})(\vec{\tau
}(1)\times \vec{\tau }(2))_{3}\Big (\vec{\sigma }(1)\times \vec{\sigma
}(2)-\sqrt{\frac{\pi }{2}}(1+\frac{3}{m_{\rho }r}+\frac{3}{(m_{\rho
}r)^{2}})\nonumber \\
 &\times& [Y_{2}(\Omega _{r})\otimes (\vec{\sigma }(1)\times \vec{\sigma
}(2)]_{1}+i\sqrt{3\pi }(1+\frac{3}{m_{\rho }r}+\frac{3}{(m_{\rho
}r)^{2}})[Y_{2}(\Omega _{r})\otimes [\vec{\sigma }(1)\otimes \vec{\sigma
}(2)]_{2}]_{1}\Big )\bigg \}\nonumber \\
 &\times& \delta ^{(3)}(\vec{x}-\vec{x}_{1})\frac{e^{-m_{\rho
}r}}{r}+(1\leftrightarrow 2),\\
\vec{j}^{\rho \, \, pair}_{\gamma -PNC} & = & \frac{-eg_{\rho NN}}{8\pi
M}\Big [h^{0}_{\rho }(\vec{\tau }(1)\cdot \vec{\tau }(2)+\tau
_{3}(2))+h^{1}_{\rho }(\tau _{3}(2)+\tau _{3}(1)\tau
_{3}(2))+\frac{h^{2}_{\rho }}{2\sqrt{6}}(3\tau _{3}(1)\tau
_{3}(2)\nonumber \\
 &-& \vec{\tau }(1)\cdot \vec{\tau }(2)+2\tau _{3}(2))\Big ]\vec{\sigma
}(1)\delta ^{(3)}(\vec{x}-\vec{x}_{1})\frac{e^{-m_{\rho
}r}}{r}+(1\leftrightarrow 2),\\
\vec{j}^{\omega \, \, pair}_{\gamma -PC} & = & \frac{eg_{\omega NN}}{12\pi
M}(h^{0}_{\omega }+h^{1}_{\omega }\tau _{3}(2))(1+\tau _{3}(1))\Big
(\vec{\sigma }(2)+\sqrt{2\pi }(1+\frac{3}{m_{\omega
}r}+\frac{3}{(m_{\omega }r)^{2}})\nonumber \\
 &\times& [Y_{2}(\Omega _{r})\otimes \vec{\sigma }(2)]_{1}\Big )\delta
^{(3)}(\vec{x}-\vec{x}_{1})\frac{e^{-m_{\omega }r}}{r}+(1\leftrightarrow
2),\\
\vec{j}^{\omega \, \, pair}_{\gamma -PNC} & = & \frac{-eg_{\omega NN}}{4\pi
M}(h^{0}_{\omega }+h^{1}_{\omega })(1+\tau _{3}(1))\vec{\sigma
}(1)\delta ^{(3)}(\vec{x}-\vec{x}_{1})\frac{e^{-m_{\omega
}r}}{r}+(1\leftrightarrow 2),\\
\vec{j}^{\pi \pi \gamma } & = & \frac{-eg_{\pi NN}f_{\pi }}{16\sqrt{2}\pi
M}(\vec{\tau }(1)\cdot \vec{\tau }(2)-\tau _{3}(1)\tau
_{3}(2))\vec{\sigma }(1)\cdot \vec{\nabla }_{1}(\vec{\nabla
}_{1}-\vec{\nabla }_{2}) \nonumber \\
&\times& \int \frac{d^{3}k}{(2\pi )^{3}}e^{i\vec{k}\cdot
(\vec{x}-\vec{R})}
 \int ^{1/2}_{-1/2}d\alpha e^{i\alpha \vec{k}\cdot
\vec{r}}\frac{e^{-L_{\pi }r}}{L_{\pi }}+(1\leftrightarrow 2),\\
\vec{j}^{\rho \rho \gamma } & = & \frac{ieg_{\rho NN}}{16\pi
M}(h^{0}_{\rho }-\frac{h^{2}_{\rho }}{2\sqrt{6}})(\vec{\tau }(1)\times
\vec{\tau }(2))_{3}\bigg [(\vec{\sigma }(2)-\frac{\vec{\sigma }(2)\cdot
\vec{\nabla }_{2}}{m^{2}_{\rho }}\vec{\nabla }_{2})\cdot (\lrope
_{1}-\lrope _{2}\nonumber \\
&+& i(1+\mu _{v})
 \vec{\sigma }(1)\times \vec{\nabla }_{1})(\vec{\nabla
}_{1}-\vec{\nabla }_{2})-\vec{\nabla }_{1}\cdot (\vec{\sigma
}(2)-\frac{\vec{\sigma }(2)\cdot \vec{\nabla }_{2}}{m^{2}_{\rho
}}\vec{\nabla }_{2})(\lrope _{1} \nonumber \\ &+&i(1+\mu _{v})\vec{\sigma }(1)\times
\vec{\nabla }_{1})
+\vec{\nabla }_{2}\cdot (\lrope _{1}-\lrope _{2}+i(1+\mu
_{v})\vec{\sigma }(1)\times \vec{\nabla }_{1})(\vec{\sigma
}(2)-\frac{\vec{\sigma }(2)\cdot \vec{\nabla }_{2}}{m^{2}_{\rho
}}\vec{\nabla }_{2})\bigg ] \nonumber \\
&\times& \int \frac{d^{3}k}{(2\pi )^{3}}e^{i\vec{k}\cdot
(\vec{x}-\vec{R})}
 \int ^{1/2}_{-1/2}d\alpha e^{i\alpha \vec{k}\cdot
\vec{r}}\frac{e^{-L_{\rho }r}}{L_{\rho }}+(1\leftrightarrow 2),\\
\vec{j}_{\rho -PC}^{\rho \pi \gamma } & = & \frac{-ieg_{\rho NN}f_{\pi
}g_{\rho \pi \gamma }}{8\sqrt{2}\pi m_{\rho }}(\vec{\tau }(1)\times
\vec{\tau }(2))_{3}(\vec{\nabla }_{1}\times \vec{\nabla }_{2})\int
\frac{d^{3}k}{(2\pi )^{3}}e^{i\vec{k}\cdot (\vec{x}-\vec{R})}\int
^{1/2}_{-1/2}d\alpha e^{i\alpha \vec{k}\cdot \vec{r}}\frac{e^{-L_{\rho \pi
}r}}{L_{\rho \pi }}\nonumber \\
 &+& (1\leftrightarrow 2),
\end{eqnarray}
where \( \vec{R}=(\vec{x}_{1}+\vec{x}_{2})/2, \) \(
\vec{r}=\vec{x}_{1}-\vec{x}_{2} \),
\( r=|\vec{r}| \), \( L_{\pi (\rho )}=(m^{2}_{\pi (\rho
)}+\vec{k}^{2}(\frac{1}{4}-\alpha ^{2}))^{1/2} \),
\( L_{\rho \pi }=(\frac{m^{2}_{\rho }+m^{2}_{\pi }}{2}+\alpha (m^{2}_{\rho
}-m^{2}_{\pi })+\vec{k}^{2}(\frac{1}{4}-\alpha ^{2}))^{1/2} \),
and the operation of \( \lrope  \) should be understood as \( \lrope
F(\cdots )=F(\cdots )\vec{\nabla }-\nabla
^{^{^{\! \! \! \! \! \! \! \leftarrow }}}F(...) \)
. For transition currents, the full Fourier transformation is not easily
evaluated in the general case, so we leave the integration undone.

\section{An Example of Fermi Gas One-body Averages\label{App: 1-body Ave}}

Here we describe how effective operators are obtained in the Fermi
gas model by one-body averages.  Most of the discussion is general,
though we use the simple $\pi$ pair current when specific examples
are needed.

An effective one-body operator 
is obtained by performing a mean-field-like sum 
over the direct and exchange terms

\begin{equation}
<\alpha |O^{(1)}|\beta >\equiv \sum _{\gamma }<\alpha \gamma
|O^{(2)}|\beta \gamma >-<\alpha \gamma |O^{(2)}|\gamma \beta >,
\end{equation}
where the sum extends over occupied core states.
In the nonrelativistic Fermi gas each s.p. state is a direct product
of space, spin, and isospin components

\begin{equation}
|\alpha \rangle =|\vec{p}(\alpha )\rangle \otimes |\frac{1}{2}m_{s}(\alpha
)\rangle \otimes |\frac{1}{2}m_{t}(\alpha )\rangle .
\end{equation}
Thus the wave function factors, allowing space, spin, and isospin
sum to be performed independently.

The spin and isospin averages for common operators are easily
done.  The results are displayed in Table \ref{Tab: 1-body S}
and \ref{Tab: 1-body T}.

\begin{table}

\caption{One-body averaged spin operators.}

\label{Tab: 1-body S}

\begin{center}

\begin{tabular}{ccc}
2-body&
1-body direct&
1-body exchange\\
\hline
\( 1 \)&
\( 2 \)&
\( 1 \)\\
\( \vec{\sigma }(1)+\vec{\sigma }(2) \)&
\( 2\vec{\sigma } \)&
\( 2\vec{\sigma } \)\\
\( \vec{\sigma }(1)-\vec{\sigma }(2) \)&
\( 2\vec{\sigma } \)&
\( 0 \)\\
\( \vec{\sigma }(1)\times \vec{\sigma }(2) \)&
\( 0 \)&
\( 2i\vec{\sigma } \)\\
\end{tabular}
\end{center}
\end{table}

\begin{figure}
{\centering \resizebox*{0.5\columnwidth}{!}{\includegraphics{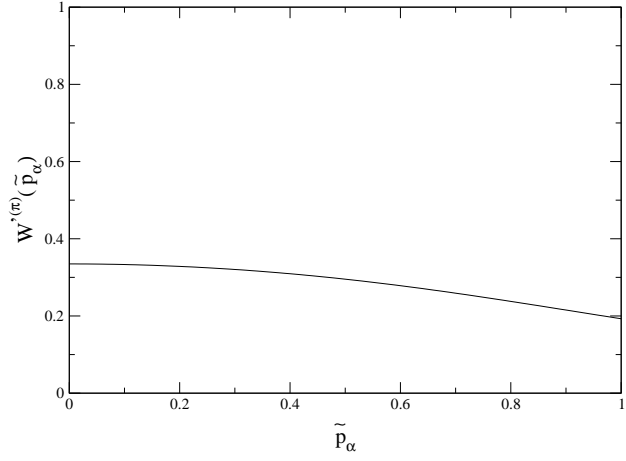}}
\par}

\caption{The smoothness of \protect\( W^{'(\pi )}\protect \) as a function
of \protect\( \tilde{p}_{\alpha }\protect \)\label{Fig: W Function}}
\end{figure}

\begin{table}

\caption{One-body averaged isospin operators.}

\label{Tab: 1-body T}

\begin{center}

\begin{tabular}{ccc}
2-body&
1-body direct&
1-body exchange\\
\hline
\( 1 \)&
\( (\theta _{p}+\theta _{n}) \)&
\( (\theta _{p}+\theta _{n})\frac{1}{2}+(\theta _{p}-\theta
_{n})\frac{\tau _{3}}{2} \)\\
\( \vec{\tau }(1)\cdot \vec{\tau }(2) \)&
\( (\theta _{p}-\theta _{n})\tau _{3} \)&
\( (\theta _{p}+\theta _{n})\frac{3}{2}-(\theta _{p}-\theta
_{n})\frac{\tau _{3}}{2} \)\\
\( \tau _{3}(1)\tau _{3}(2) \)&
\( (\theta _{p}-\theta _{n})\tau _{3} \)&
\( (\theta _{p}+\theta _{n})\frac{1}{2}+(\theta _{p}-\theta
_{n})\frac{\tau _{3}}{2} \)\\
\( \tau _{3}(1)+\tau _{3}(2) \)&
\( (\theta _{p}-\theta _{n})+(\theta _{p}+\theta _{n})\tau _{3} \)&
\( (\theta _{p}-\theta _{n})+(\theta _{p}+\theta _{n})\tau _{3} \)\\
\( \tau _{3}(1)-\tau _{3}(2) \)&
\( -(\theta _{p}-\theta _{n})+(\theta _{p}+\theta _{n})\tau _{3} \)&
\( 0 \)\\
\( (\vec{\tau }(1)\times \vec{\tau }(2))_{3} \)&
\( 0 \)&
\( -i(\theta _{p}-\theta _{n})+i(\theta _{p}+\theta _{n})\tau _{3} \)\\
\( 3\tau _{3}(1)\tau _{3}(2)-\vec{\tau }(1)\cdot \vec{\tau }(2) \)&
\( 2(\theta _{p}-\theta _{n})\tau _{3} \)&
\( 2(\theta _{p}-\theta _{n})\tau _{3} \)\\
\end{tabular}
\end{center}
\end{table}
Turning to spatial averages, we first consider pair currents.
The spatial parts of these operators take one of the generic forms:
i) \( f(r)\delta ^{(3)}(\vec{x}-\vec{x}_{1}) \) or ii) \( f(r)\delta
^{(3)}(\vec{x}-\vec{x}_{2}) \),
where \( f(r) \) is a function of \( r\equiv
|\vec{r}|=|\vec{x}_{1}-\vec{x}_{2}| \).
Therefore, the direct average is

\begin{eqnarray}
 &  & \sum _{\vec{p}_{\gamma }}\langle \vec{p}_{\alpha },\vec{p}_{\gamma
}|f(r)\delta ^{(3)}(\vec{x}-\vec{x}_{1})|\vec{p}_{\beta },\vec{p}_{\gamma
}\rangle =e^{-i(\vec{p}_{\alpha }-\vec{p}_{\beta })\cdot \vec{x}}\sum
_{\vec{p}_{\gamma }}\, \int d^{3}r\, f(r),\\
 &  & \sum _{\vec{p}_{\gamma }}\langle \vec{p}_{\alpha },\vec{p}_{\gamma
}|f(r)\delta ^{(3)}(\vec{x}-\vec{x}_{2})|\vec{p}_{\beta },\vec{p}_{\gamma
}\rangle =e^{-i(\vec{p}_{\alpha }-\vec{p}_{\beta })\cdot \vec{x}}\sum
_{\vec{p}_{\gamma }}\, \int d^{3}r\, e^{-i(\vec{p}_{\alpha
}-\vec{p}_{\beta })\cdot \vec{r}}\, f(r),
\end{eqnarray}
and the exchange average 

\begin{eqnarray}
 &  & \sum _{\vec{p}_{\gamma }}\langle \vec{p}_{\alpha },\vec{p}_{\gamma
}|f(r)\delta ^{(3)}(\vec{x}-\vec{x}_{2})|\vec{p}_{\gamma },\vec{p}_{\beta
}\rangle =e^{-i(\vec{p}_{\alpha }-\vec{p}_{\beta })\cdot \vec{x}}\sum
_{\vec{p}_{\gamma }}\, \int d^{3}r\, e^{-i(\vec{p}_{\beta
}-\vec{p}_{\gamma })\cdot \vec{r}}\, f(r),\\
 &  & \sum _{\vec{p}_{\gamma }}\langle \vec{p}_{\alpha },\vec{p}_{\gamma
}|f(r)\delta ^{(3)}(\vec{x}-\vec{x}_{2})|\vec{p}_{\gamma },\vec{p}_{\alpha
}\rangle =e^{-i(\vec{p}_{\alpha }-\vec{p}_{\beta })\cdot \vec{x}}\sum
_{\vec{p}_{\gamma }}\, \int d^{3}r\, e^{-i(\vec{p}_{\alpha
}-\vec{p}_{\gamma })\cdot \vec{r}}\, f(r).
\end{eqnarray}
Now \( e^{-i(\vec{p}_{\alpha }-\vec{p}_{\beta })\cdot \vec{x}} \)
gives a \( \delta ^{(3)}(\vec{x}-\vec{x}_{i}) \) for a first
quantized operator in position space, i.e., \( \langle \vec{p}_{\alpha
}|\delta ^{(3)}(\vec{x}-\vec{x}_{i})|\vec{p}_{\beta }\rangle
=e^{-i(\vec{p}_{\alpha }-\vec{p}_{\beta })\cdot \vec{x}} \).
We specialize the remaining integration to the $\pi$
pair current where \( f(r) \) has the Yukawa form \( e^{-m_{\pi
}}/r \),
\begin{eqnarray}
 &  & \int d^{3}r\, \frac{e^{-m_{\pi }}}{r}=\frac{4\pi }{m^{2}_{\pi }},\\
 &  & \int d^{3}r\, e^{-i\vec{p}\cdot \vec{r}}\frac{e^{-m_{\pi
}}}{r}=\frac{4\pi }{\vec{p}^{2}+m^{2}_{\pi }}.
\end{eqnarray}
After performing the sum over the Fermi sphere by using the quasi-continuum
limit, \( \sum _{\vec{p}_{\gamma }}\rightarrow \int ^{p_{F}}_{0}dp_{\gamma
}\, \int ^{4\pi }_{0}d\Omega (p_{\gamma }) \),
we find
\begin{eqnarray}
\sum _{\vec{p}_{\gamma }}1 & = & \frac{p^{3}_{F}}{6\pi ^{2}}=\frac{\rho
}{2},\\
\sum _{\vec{p}_{\gamma }}\frac{1}{(\vec{p}_{\alpha (\beta
)}-\vec{p}_{\gamma })^{2}+m^{2}_{\pi }} & = & \frac{\rho }{2m^{2}_{\pi
}}W^{'}(\tilde{p}_{\alpha (\beta )},\tilde{m}_{\pi }),
\end{eqnarray}
where the \( W^{'} \) function represents the full result after
the volume integration,
\begin{eqnarray}
W^{'}(\tilde{p}_{\alpha },\tilde{m}_{\pi }) & = & \frac{3m^{2}_{\pi
}}{4p^{2}_{F}}\Bigg \{2-2\tilde{m}_{\pi }\bigg [\arctan \Big
(\frac{1+\tilde{p}_{\alpha }}{\tilde{m}_{\pi }}\Big )+\arctan \Big
(\frac{1-\tilde{p}_{\alpha }}{\tilde{m}_{\pi }}\Big )\bigg ]\nonumber \\
 &+& \frac{1}{2\tilde{p}_{\alpha }}[1-\tilde{p}^{2}_{\alpha
}+\tilde{m}^{2}_{\pi }]\ln \bigg [\frac{(1+\tilde{p}_{\alpha
})^{2}+\tilde{m}^{2}_{\pi }}{(1-\tilde{p}_{\alpha
})^{2}+\tilde{m}^{2}_{\pi }}\bigg ] \Bigg \}
\end{eqnarray}
with all the tilde quantities normalized by
the Fermi momentum, i.e. \( \tilde{X}\equiv X/p_{F} \). As shown
in Fig. \ref{Fig: W Function}, the \( W' \) function varies very
slowly as \( \tilde{p}_{\alpha } \) runs from 0 to 1. Therefore,
it is reasonable to replace this quantity by its average value \( \langle W^{'(\pi )}\rangle  \).

Combining the spatial result with the spin and isospin factors
from the tables the yields the $\pi$ pair current one-body
averged form
\begin{eqnarray}
j_{\gamma -PC}^{\pi \, \, pair} & = & \frac{eg_{\pi NN}f_{\pi
}}{8\sqrt{2}\pi M}((\theta _{p}+\theta _{n})-(\theta _{p}-\theta
_{n})\tau _{3})\vec{\sigma }\frac{2\pi \rho }{m^{2}_{\pi }}\langle
W^{'(\pi )}\rangle \delta ^{(3)}(\vec{x}-\vec{x}_{i})+(1\leftrightarrow
2)\\
 & = & \frac{eg_{\pi NN}f_{\pi }}{2\sqrt{2}\pi Mm^{2}_{\pi }}((\theta
_{n}+\theta _{n})+(\theta _{n}-\theta _{p})\tau _{3})\times \rho \langle
W^{'(\pi )} \rangle\vec{\sigma } \delta ^{(3)}(\vec{x}-\vec{x}_{i}).
\end{eqnarray}
Other currents are similar, though generally more tedious.

\section{One-body Fermi-Gas Averaged Current Operators}

We list the relevant one-body Fermi gas averaged operators in momentum
space,

\begin{eqnarray}
\vec{j}^{\pi \, \, pair}_{\gamma -PC} & = & \frac{e g_{\pi NN}f_{\pi
}}{2\sqrt{2}Mm^{2}_{\pi }}((\theta _{n}+\theta _{p})+(\theta
_{n}-\theta _{p})\tau _{3})\times \rho \langle W^{'(\pi )}\rangle
\vec{\sigma },\\
\vec{j}^{\rho \, \, pair}_{\gamma -PC} & = & \frac{2 e g_{\rho
NN}}{3Mm^{2}_{\rho }}(h^{1}_{\rho }+(h^{0}_{\rho }+\frac{h^{2}_{\rho
}}{2\sqrt{6}})\tau _{3})\theta _{p} \rho \{\vec{\sigma
}+\sqrt{\frac{15}{4}}\frac{1}{m^{2}_{\rho }}[[\vec{k}\otimes
\vec{k}]_{2}\otimes \vec{\sigma }]_{1}\}\nonumber \\
 &-& \frac{g_{\rho NN}}{6Mm^{2}_{\rho }}\Bigg (\bigg [\Big
((h^{1}_{\rho }+\frac{3h^{2}_{\rho }}{2\sqrt{6}})+(2h^{0}_{\rho
}+h^{1}_{\rho }+\frac{h^{2}_{\rho }}{2\sqrt{6}})\tau _{3}\Big )\theta
_{p}-3(h^{0}_{\rho }-\frac{h^{2}_{\rho }}{2\sqrt{6}})(1+\tau _{3})\theta
_{n}\bigg ]
  2 \rho \langle W^{'(\rho )}\rangle \vec{\sigma } \nonumber \\ &+&\bigg
[(3h^{0}_{\rho }+h^{1}_{\rho })-(h^{0}_{\rho }-h^{1}_{\rho
}-\frac{2h^{2}_{\rho }}{\sqrt{6}})\tau _{3})\bigg ]\theta _{n}
  \rho \sqrt{\frac{15}{4}}\langle W^{''(\rho )}\rangle
\frac{1}{p^{2}_{F}}[[\vec{K}\otimes \vec{K}]_{2}\otimes \vec{\sigma
}]_{1}\Bigg ),\\
\vec{j}^{\rho \, \, pair}_{\gamma -PNC} & = & \frac{e g_{\rho
NN}}{2Mm^{2}_{\rho }}(h^{0}_{\rho }+h^{1}_{\rho }+\frac{h^{2}_{\rho
}}{2\sqrt{6}})(1+\tau _{3})(\theta _{n}-\theta _{p}) \rho
\vec{\sigma }\nonumber \\
 &+& \frac{e g_{\rho NN}}{2Mm^{2}_{\rho }}\bigg [\Big (2h^{0}_{\rho
}+h^{1}_{\rho }+\frac{h^{2}_{\rho }}{2\sqrt{6}}+(h^{1}_{\rho
}+\frac{3h^{2}_{\rho }}{2\sqrt{6}})\tau _{3}\Big )\theta _{p}+(h^{0}_{\rho
}-\frac{h^{2}_{\rho }}{2\sqrt{6}})(1+\tau _{3})\theta _{n}\bigg ]
  \rho \langle W^{'(\rho )}\rangle \vec{\sigma },\\
\vec{j}^{\omega \, \, pair}_{\gamma -PC} & = & \frac{2 e g_{\omega
NN}}{3Mm^{2}_{\omega }}(h^{0}_{\omega }+h^{1}_{\omega }\tau
_{3})\theta _{p} \rho \{\vec{\sigma
}+\sqrt{\frac{15}{4}}\frac{1}{m^{2}_{\omega }}[[\vec{k}\otimes
\vec{k}]_{2}\otimes \vec{\sigma }]_{1}\}\nonumber \\
 &-& \frac{e g_{\omega NN}}{6Mm^{2}_{\omega }}(h^{0}_{\omega
}+h^{1}_{\omega })(1+\tau _{3})\theta _{p}\ \rho \{2\langle
W^{'(\omega )}\rangle \vec{\sigma }\
 - \sqrt{\frac{15}{4}}\langle W^{''(\omega )}\rangle
\frac{1}{p^{2}_{F}}[[\vec{K}\otimes \vec{K}]_{2}\otimes \vec{\sigma
}]_{1}\},\\
\vec{j}^{\omega \, \, pair}_{\gamma -PNC} & = & \frac{-e g_{\omega
NN}}{2Mm^{2}_{\omega }}(h^{0}_{\omega }+h^{1}_{\omega })(1+\tau
_{3})((\theta _{n}+\theta _{p})\ \rho \vec{\sigma }
 - \theta _{p}\ \rho \langle
W^{'(\omega )}\rangle \vec{\sigma }),\\
\vec{j}^{\pi \pi \gamma } & = & \frac{-e g_{\pi NN}f_{\pi
}}{\sqrt{2}Mm^{4}_{\pi }}((\theta _{n}+\theta _{p})+(\theta
_{n}-\theta _{p})\tau _{3})\ \rho \{\langle Y_{1}\rangle (\vec{\sigma
}\cdot \vec{k})\vec{k}+\langle Y_{2}\rangle (\vec{\sigma }\cdot
\vec{K})\vec{K}
 +\langle Y_{3}\rangle (p^{2}_{F})\vec{\sigma }\},\\
\vec{j}^{\rho \rho \gamma } & = & \frac{-e g_{\rho
NN}}{2\sqrt{2}Mm^{4}_{\rho }}(h^{0}_{\rho }-\frac{h^{2}_{\rho
}}{2\sqrt{6}})((\theta _{n}-\theta _{p})+(\theta _{n}+\theta _{p})\tau
_{3})\ \rho \bigg \{\langle I_{1}\rangle (\vec{k}^{2})\vec{\sigma
}+\langle I_{2}\rangle (p^{2}_{F})\vec{\sigma }\nonumber \\
 &+& \langle I_{3}\rangle (\vec{\sigma }\cdot \vec{k})\vec{k}+\langle
I_{4}\rangle (\vec{\sigma }\cdot \vec{K})\vec{K}-i\langle I_{5}\rangle
\vec{k}\times \vec{K}+\frac{1}{m^{2}_{\rho }}\Big (\langle J_{1}\rangle
(\vec{k}^{4})\vec{\sigma }+\langle J_{2}\rangle (\vec{k}\cdot
\vec{K})^{2}\vec{\sigma }\nonumber \\
 &+& \langle J_{3}\rangle (p^{2}_{F}\vec{k}^{2})\vec{\sigma }+\langle
J_{4}\rangle (p^{4}_{F})\vec{\sigma }+\langle J_{5}\rangle (\vec{\sigma
}\cdot \vec{k})(\vec{k}^{2})\vec{k}+\langle J_{6}\rangle (\vec{\sigma
}\cdot \vec{K})(\vec{k}\cdot \vec{K})\vec{k}\nonumber \\
 &+& \langle J_{7}\rangle (p^{2}_{F})(\vec{\sigma }\cdot
\vec{k})\vec{k}+\langle J_{8}\rangle (\vec{\sigma }\cdot
\vec{K})(\vec{k}^{2})\vec{K}+\langle J_{9}\rangle (p^{2}_{F})(\vec{\sigma
}\cdot \vec{K})\vec{K}+\langle J_{10}\rangle (\vec{\sigma }\cdot
\vec{k})(\vec{k}\cdot \vec{K})\vec{K}\nonumber \\
 &-& i\langle J_{11}\rangle (\vec{k}^{2})\vec{k}\times \vec{K}-i\langle
J_{12}\rangle (p^{2}_{F})\vec{k}\times \vec{K}\Big )\bigg \},\\
\vec{j}_{\rho -PC}^{\rho \pi \gamma } & = & \frac{-i2\sqrt{2}e g_{\rho
NN}f_{\pi }g_{\rho \pi \gamma }}{m_{\rho }(m^{2}_{\rho }+m^{2}_{\pi
})^{2}}((\theta _{n}-\theta _{p})+(\theta _{n}+\theta _{p})\tau
_{3})\ \langle Z\rangle  \vec{k}\times \vec{K}.
\end{eqnarray}

\( \theta _{p(n)} \) is a projection operator:
when \( \theta _{p(n)} \) acts on \{\( \rho  \) (nuclear
density), \( p_{F} \) (Fermi momentum), \( \langle X\rangle  \)
(averaged weighting function such as \( \langle W^{'(\pi )}\rangle  \))\},
the results should read \{\( \rho _{p(n)} \), \( p^{p(n)}_{F} \),
\( \langle X_{p(n)}\rangle  \)\}. The conversion rules from momentum
space to position space are simple,
\begin{eqnarray}
1 & \longrightarrow  & \delta ^{(3)}(\vec{x}-\vec{x}_{i}),\\
\vec{k}=\vec{p}_{\beta }-\vec{p}_{\alpha } & \longrightarrow  &
-i[\vec{\nabla },\delta ^{(3)}(\vec{x}-\vec{x}_{i})],\\
\vec{K}=\frac{\vec{p}_{\beta }+\vec{p}_{\alpha }}{2} & \longrightarrow  &
-i\{\vec{\nabla }_{i},\delta ^{(3)}(\vec{x}-\vec{x}_{i})\}_{sym}.
\end{eqnarray}
 For \( ^{133}\)Cs, \( p^{p(n)}_{F}\sim 260(300)\) MeV.
The average weighting factors are given in Table \ref{Tab: Weighting
Functions}.
For \( ^{205}\)Tl, these numbers are almost the same. We do
not distinguish between the $\rho$ and $\omega$ masses in evaluating
weighting functions.

\begin{table}

\caption{Average weighting functions. (Note: First
number
refers to proton part and the the second the neutron
part)}

\label{Tab: Weighting Functions}

\begin{center}

\begin{tabular}{c|c|c|c|c|c|c}
\( \langle W^{'(\pi )}\rangle  \)&
\( \langle W^{'(\rho ,\omega )}\rangle  \)&
\( \langle W^{''(\pi )}\rangle  \)&
\( \langle W^{''(\rho ,\omega )}\rangle  \)&
\( \langle Y_{1}\rangle  \)&
\( \langle Y_{2}\rangle  \)&
\( \langle Y_{3}\rangle  \)\\
\( 0.30/0.26 \)&
\( 0.90/0.88 \)&
\( 0.99/1.02 \)&
\( 0.19/0.23 \)&
\( 0.0039/0.0033 \)&
\( 0.0173/0.0111 \)&
\( 0.0144/0.0103 \)\\
\hline 
\( \langle I_{1}\rangle  \)&
\( \langle I_{2}\rangle  \)&
\( \langle I_{3}\rangle  \)&
\( \langle I_{4}\rangle  \)&
\( \langle I_{5}\rangle  \)&
\( \langle J_{1}\rangle  \)&
\( \langle J_{2}\rangle  \)\\
\( 2.25/2.18 \)&
\( 10.09/9.58 \)&
\( -1.53/-1.46 \)&
\( 21.30/19.61 \)&
\( 10.18/9.56 \)&
\( 0.42/0.39 \)&
\( 1.10/1.02 \)\\
\hline
\( \langle J_{3}\rangle  \)&
\( \langle J_{4}\rangle  \)&
\( \langle J_{5}\rangle  \)&
\( \langle J_{6}\rangle  \)&
\( \langle J_{7}\rangle  \)&
\( \langle J_{8}\rangle  \)&
\( \langle J_{9}\rangle  \)\\
\( 4.11/3.91 \)&
\( 1.43/1.32 \)&
\( -0.83/-0.77 \)&
\( -2.68/-2.48 \)&
\( 2.57/2.40 \)&
\( 4.25/3.89 \)&
\( -8.21/-7.56 \)\\
\hline
\( \langle J_{10}\rangle  \)&
\( \langle J_{11}\rangle  \)&
\( \langle J_{12}\rangle  \)&
\( \langle Z\rangle  \)&
&
&
\\
\( 2.53/2.32 \)&
\( 3.26/3.03 \)&
\( -11.16/-10.42 \)&
\( 1.43/1.15 \)&
&
&
\\
\end{tabular}
\end{center}
\end{table}

\end{document}